\documentclass[12pt]{article}
\usepackage{amsmath}
\usepackage{amstext}
\usepackage{amsthm}
\usepackage{amssymb}
\usepackage{mathrsfs} 
\usepackage[bottom]{footmisc}
\usepackage{indentfirst}
\usepackage{upref}
\usepackage{cite}
\usepackage{units}
\numberwithin{equation}{section}
\renewcommand{\baselinestretch}{1.303}
\begin{document}
\begin{titlepage} 
\renewcommand{\baselinestretch}{1.3}
\small\normalsize
\begin{flushright}
MZ-TH/04-14\footnote{to appear in Phys.\ Rev.\ D}
\end{flushright}

\vspace{0.1cm}

\begin{center}   

{\LARGE \textsc{Running Newton Constant,\\[2.35mm] Improved Gravitational
Actions, and\\
[6.5mm] Galaxy Rotation Curves}}

\vspace{1.4cm}
{\large M.~Reuter and H.~Weyer}\\

\vspace{0.7cm}
\noindent
\textit{Institute of Physics, University of Mainz\\
Staudingerweg 7, D-55099 Mainz, Germany}\\

\end{center}   

\vspace*{0.6cm}
\begin{abstract}
A renormalization group (RG) improvement of the Einstein-Hilbert action
is performed which promotes Newton's constant and the cosmological
constant to scalar functions on spacetime. They arise from solutions
of an exact RG equation by means of a ``cutoff identification'' which 
associates RG scales to the points of spacetime. The resulting modified
Einstein equations for spherically symmetric, static spacetimes are
derived and analyzed in detail. The modifications of the Newtonian limit
due to the RG evolution are obtained for the general case. As an application,
the viability of a scenario is investigated where strong quantum effects
in the infrared cause Newton's constant to grow at large (astrophysical)
distances. For two specific RG trajectories exact vacuum spacetimes
modifying the Schwarzschild metric are obtained by means of a 
solution-generating Weyl transformation. Their possible relevance to the 
problem of the observed approximately flat galaxy rotation curves is
discussed. It is found that a power law running of Newton's constant
with a small exponent of the order $10^{-6}$ would account for their
non-Keplerian behavior without having to postulate the presence of any
dark matter in the galactic halo.

04.60.-m, 95.35.+d, 98.62.Gq, 11.10.Hi 
\end{abstract}
\end{titlepage}
\section{\label{1}Introduction}
Renormalization group improvement is a standard tool of elementary particle
physics. Typically it is used for including higher order quantum corrections
into scattering amplitudes or cross-sections by replacing the classical or
``bare'' parameters on which they depend with scale dependent ``running''
coupling constants and masses obtained by solving an appropriate 
renormalization group equation. A nontrivial step in this procedure is
identifying the physical scale (momentum transfer, etc.) at which the running 
couplings are to be evaluated. While originally employed within perturbation
theory, this idea can also be used in the context of the various ``exact
renormalization group equations'' \cite{bagber} which implement the idea of
the Wilsonian renormalization group in quantum field theory. A particularly 
convenient and powerful formulation is the ``effective average action''
\cite{avact,avactrev} and its associated flow equation. One of its advantages 
is that it can be applied to gauge theories \cite{ym} and also to
gravity \cite{mr}.

In the latter case the effective average action
$\Gamma_{k} \bigl[ g_{\mu \nu} \bigr]$ is a free energy functional which
depends on the metric and a momentum scale $k$ with the
interpretation of a variable infrared (IR) cutoff. The action
$\Gamma_{k}$ is similar to the ordinary effective action $\Gamma$ which
it approaches for $k \to 0$. The main difference is that the path integral
defining $\Gamma_{k}$ extends only over quantum fluctuations with covariant
momenta $p^{2} > k^{2}$. As a result, $\Gamma_{k}$ describes the dynamics
of metrics averaged over spacetime volumes of linear dimension $k^{-1}$.
The functional $\Gamma_{k} \bigl[ g_{\mu \nu} \bigr]$ gives rise to an
effective field theory valid near the scale $k$. Hence, when evaluated
at tree level, $\Gamma_{k}$ correctly describes all quantum gravitational
phenomena, including all loop effects, provided the typical momentum scales
involved are all of order $k$. Considered a function of $k$, $\Gamma_{k}$
describes a renormalization group (RG) trajectory in the space of all action
functionals. The trajectory can be obtained by solving an exact functional
RG equation. In practice one has to resort to approximations. Nonperturbative
approximate solutions can be obtained by truncating the space of action
functionals, i.\,e.\ by projecting the RG flow onto a finite-dimensional
subspace which encapsulates the essential physics. Recently a great deal of
work went into the investigation of such truncated RG flows of quantum
gravity \cite{mr,percadou,oliver1,frank1,oliver2,souma,percacciperini,frank2,litimgrav}.
Most of these studies employ the ``Einstein-Hilbert truncation'' which
approximates $\Gamma_{k}$ by a linear combination of the monomials
$\int \!\!\sqrt{g\,} \, R$ and $\int \!\!\sqrt{g\,}$. Their prefactors contain the
running Newton constant $G (k)$ and the running cosmological constant
$\Lambda (k)$. Their $k$-dependence is governed by a system of two coupled
ordinary differential equations.

Let us assume we found some solution to this system, a RG trajectory
$\bigl( G (k), \Lambda (k) \bigr)$. How can we take advantage of this
information? Sometimes it is possible to ``RG improve'' the predictions of
classical General Relativity by replacing the pertinent classical (bare)
parameters with the running ones, $G (k)$ and $\Lambda (k)$. In a spacetime
formulation, one would try to find a ``cutoff identification'' of the
form \cite{nelson}
\begin{align}
k = k (x)
\label{1.1}
\end{align}
which converts the scale dependence of $G$ and $\Lambda$ to a position
dependence:
\begin{align}
G (x) \equiv G \bigl( k = k (x) \bigr), \quad
\Lambda (x) \equiv \Lambda \bigl( k = k (x) \bigr).
\label{1.2}
\end{align}
Clearly it is not always possible to associate cutoff values to spacetime
points and to find a map $x \mapsto k (x)$. As we discussed in
ref.\ \cite{h1}, a detailed discussion of the possible \textit{physical}
cutoff mechanisms which might be operative in a given system is required here.
However, typically when the problem has a high degree of symmetry, maps of
the kind $k = k (x)$ actually do exist and can be guessed almost 
unambiguously \cite{h1}. For instance, in Robertson-Walker cosmology the
condition of homogeneity and isotropy leads to strong constraints on the
possible form of $k (x)$: it can depend on the cosmological time $t$ only.
In fact, the most natural choice is \cite{cosmo1,cosmo2,elo,esposito}
\begin{align}
k (t) = \widehat \xi / t,
\label{1.3}
\end{align}
with $\widehat \xi$ a constant of order unity. The motivation for \eqref{1.3}
is obvious: By the time the age of the Universe is $t$, no quantum fluctuation
with a frequency smaller than $1/t$ can have played any role yet.
Hence the integrating-out of field modes should be stopped at the IR cutoff
$k \approx 1/t$. (See also \cite{sola,scalfact}.)

Likewise, if the spacetime is spherically symmetric and static, $k$ can depend
on the radial coordinate $r$ only. In ref.\ \cite{bh} the RG improvement
of Schwarzschild black holes has been discussed and it was argued that the 
correct cutoff is 
\begin{align}
k (r) = \xi / d (r),
\label{1.4}
\end{align}
where, outside the horizon, $d (r)$ is the radial proper distance to the
center of the black hole.

Knowing the position dependence of $G (x)$ and $\Lambda (x)$ the actual
RG improvement can be performed in either of the following ways. (1) One
picks a spacetime solving the classical Einstein equation and substitutes
$G \to G (x)$, $\Lambda \to \Lambda (x)$ in this solution. (2) One performs
the substitution one step earlier at the level of the field equations, and
then solves the resulting ``improved Einstein equation''
\begin{align}
G_{\mu \nu} = - \Lambda (x) \, g_{\mu \nu} + 8 \pi \, G (x) \, T_{\mu \nu}.
\label{1.5}
\end{align}
(3) One performs the substitution still another step earlier, at the level of
the action from which Einstein's equation is derived, then derives the field
equations implied by the improved action and solves them.

The three options in this list are ordered according to increasing predictive
power and generality. The improved solution approach \cite{bh} can be used
only when the quantum corrected metric is not very different from the classical
one. The improved equation approach \cite{cosmo2} does allow for large quantum
corrections, but there are also situations where the improved Einstein equation
\eqref{1.5} obviously does not capture the essential quantum effects. The prime
example are vacuum spacetimes for which the cosmological constant is not an
issue, small isolated systems such as black holes, for instance. While we expect
on general grounds that the running of $G$ is important for the structure
of such spacetimes, the improved Einstein equation \eqref{1.5} cannot
``know'' about this running since it does not differ from the 
classical one, $G_{\mu \nu} =0$.

In ref.\ \cite{cosmo2} the improved equation approach has been applied to
cosmology and its relation to the improved solution technique was clarified.
Ref.\ \cite{bh} contains the first application of the effective average action
to black holes. It used the improved solution technique which is insufficient
near the central singularity, for instance, where the quantum effects become
strong. This was one of the motivations for developing the improved action
approach in ref.\ \cite{h1}. In the present paper we are going to take a first
step towards its application to black holes.

The idea is to start from the classical Einstein-Hilbert action
$S_{\text{EH}}= \int \!\text{d}^{4} x~\sqrt{-g\,} \, \mathscr{L}_{\text{EH}}$
with the Lagrangian 
$\mathscr{L}_{\text{EH}} = \left( R - 2 \, \Lambda \right) / 
\left( 16 \pi \, G \right)$ and to promote, in this Lagrangian, $G$ and
$\Lambda$ to scalar fields on spacetime. This leads to the modified 
Einstein-Hilbert (mEH) Lagrangian
\begin{align}
\mathscr{L}_{\text{mEH}} = \frac{1}{16 \pi \, G (x)} \,
\big \{ R - 2 \, \Lambda (x) \big \}.
\label{1.6}
\end{align}
The resulting modified Einstein equations differ from the classical ones not
only by the substitution $G \to G (x)$, $\Lambda \to \Lambda (x)$ but also
by terms containing derivatives of $G (x)$. Because of those terms, the
improvement of the field equations and of the action functional are not
equivalent a priori. In ref.\ \cite{h1} the relation between the two approaches
was discussed in detail. The upshot of this analysis was that, in the improved
action approach, the modified Einstein-Hilbert action
$S_{\text{mEH}}= \int \!\text{d}^{4} x~\sqrt{-g\,} \, \mathscr{L}_{\text{mEH}}$
should be supplemented by an action $S_{\theta} \bigl[ g_{\mu \nu}, G, \Lambda
\bigr]$ whose associated energy-momentum tensor
\begin{align}
\theta^{\mu \nu} = \frac{2}{\sqrt{-g\,}\,} \,
\frac{\delta S_{\theta}}{\delta g_{\mu \nu}}
\label{1.7}
\end{align}
describes the 4-momentum carried by the scalar fields $G (x)$ and 
$\Lambda (x)$. In the approach of improving field equations those scalars do
not carry gravitating 4-momentum, but they do so in the approach based upon
$\mathscr{L}_{\text{mEH}}$. The energy-momentum tensor $\theta^{\mu \nu}$
plays a double role: In physical situations where both approaches are
applicable it renders their respective results equivalent, at least at a
qualitative level, and at the same time it helps in making the modified
field equations mathematically consistent. In fact, the modified Einstein
equation has the form $G_{\mu \nu} = \mathcal{E}_{\mu \nu}$ where
$\mathcal{E}_{\mu \nu}$ contains the energy-momentum tensor of the matter
system, $T_{\mu \nu}$, the scalars $G (x)$ and $\Lambda (x)$, and their
derivatives. Since $D^{\mu} G_{\mu \nu}=0$ by Bianchi's identity, consistency
requires that $D^{\mu} \mathcal{E}_{\mu \nu}=0$. In classical General
Relativity this condition is satisfied if $T_{\mu \nu}$ is conserved. In the
RG improved setting additional conditions on $\Lambda (x)$ and $G (x)$
arise, henceforth referred to as ``consistency conditions''. As a rule, the
consistency conditions become much less restrictive when an appropriate
$\theta$-tensor is included.

Before continuing, let us recall some properties of the RG trajectories of 
Quantum Einstein Gravity \cite{frank1,oliver2}. At the exact level, they
consist of infinitely many generalized couplings $G (k)$, 
$\Lambda (k)$, $\cdots$
needed in order to specify a generic action function, including higher
derivative and nonlocal invariants \cite{frank2,cwnonloc}. Rather than in
terms of the dimensionful quantities $G (k)$, $\Lambda (k)$, $\cdots$
the flow equations are usually written in terms of their dimensionless
counterparts
$g (k) \equiv k^{2} \, G (k)$, $\lambda (k) \equiv \Lambda (k) / k^{2}$,
$\cdots$.
Within the Einstein-Hilbert truncation, the RG equations for $g (k)$ and
$\lambda (k)$ were solved numerically in \cite{frank1}. The RG flow on the
$g$-$\lambda$-plane is dominated by two fixed points
$\bigl( g_{*}, \lambda_{*} \bigr)$: a Gaussian fixed point at
$g_{*} = \lambda_{*} =0$, and a non-Gaussian one with $g_{*} >0$ and
$\lambda_{*} >0$. The high-energy, i.\,e.\ short-distance behavior of
Quantum Einstein Gravity is governed by the non-Gaussian fixed point:
for $k \to \infty$, all RG trajectories run into this fixed point. If it is
present in the exact flow equation, it can by used to construct a fundamental,
microscopic quantum theory of gravity by taking the limit of infinite UV cutoff
along one of the trajectories running into the fixed point. The theory is said
to be ``nonperturbatively renormalizable'' then
\cite{wein,souma,oliver2,max}. At the non-Gaussian fixed point the 
dimensionless couplings assume constant values so that the dimensionful ones
display a power law running:
\begin{align}
G (k) = g_{*} / k^{2}, \quad
\Lambda (k) = \lambda_{*} \, k^{2}.
\label{1.8}
\end{align}
Newton's constant $G (k)$ is an asymptotically free coupling: it vanishes for
$k \to \infty$. Conversely, lowering $k$, one finds that $G (k)$ increases
monotonically all the way down from the ultraviolet (UV) to the IR.
Interpreting $1/k$ as a distance scale, we see that $G$ \textit{in}creases
with distance: quantum gravity is \textit{anti}screening, a property it shares
with QCD, but not with QED, for instance. For the RG
trajectory ending at $\lambda (k=0) =0$ the scale dependence of $G$ can be
approximated by
\begin{align}
G (k) = \frac{\overline{G}}{1 + \omega \, \overline{G} \, k^{2}}
\label{1.9}
\end{align}
where $\omega$ is a constant. Here $\overline{G} \equiv G (k=0)$ is the
renormalized Newton constant. If the behavior \eqref{1.9} applies, this is
the value we measure in the laboratory because the difference between
$G$ at $k=0$ and at a typical laboratory scale,
$k=\left( \unit[1]{meter} \right)^{-1}$ say, is utterly negligible.
According to \eqref{1.9}, $G (k)$ is practically constant once $k$ is smaller
than the Planck mass $m_{\text{Pl}} \equiv 1 / \sqrt{\,\overline{G}\,}$.

It is known \cite{frank1} that the Einstein-Hilbert approximation performs
better in the UV than in the IR. In particular it seems clear that for
$k \to 0$ it should be supplemented by more complicated (nonlocal) invariants.
One might wonder what their impact on the RG evolution in the IR could be.
It is an exciting conjecture that gravity is subject to strong quantum
corrections at large distances \cite{tsamis,mottola} which might lead to
visible renormalization effects at large, i.\,e.\ astrophysical or
cosmological distance scales \cite{bertodm,bertoproc,cosmo2}.
From the theoretical point of view, this conjecture is supported by the
observation that Quantum Einstein Gravity has many features in common with
non-Abelian gauge theories \cite{frank1,frank2}. In Yang-Mills theory,
QCD, say, we know that the large-distance physics is governed by strong
and highly nonperturbative quantum effects which are responsible for color
confinement and chiral symmetry breaking, for instance. Because of the
extreme algebraic complexity of the corresponding flow equations it has not
yet been possible to explore the IR of Quantum Einstein Gravity with realistic
nonlocal truncations; see \cite{frank2}, however, for a first step in this
direction.

From the phenomenological point of view, the astrophysical and cosmological
implications of running gravitational coupling constants were explored in
\cite{bertodm,bertoproc} in the framework of a higher-derivative theory
of gravity; in particular, primordial nucleosynthesis, structure
formation, and galaxy rotation curves were discussed. While the observed
flatness of galaxy rotation curves $v (r)$ \cite{combbook} is usually
considered evidence for the existence of dark matter, it was argued that
macroscopic quantum effects could mimic the presence of dark matter.

Similarly, the phenomenological theory of the MOdified Newtonian Dynamics
called ``MOND'' \cite{mond} is designed so as to explain the galaxy rotation
curves without the need for dark matter. For a point mass $M$, say, its
modified (nonrelativistic) field equation reproduces the familiar 
$- \overline{G} \, M / r$-potential at small distances; at large distances it
yields instead
\begin{align}
\phi_{\text{MOND}} (r) = \sqrt{a_{0} \, \overline{G} \, M\,} \, \ln r
\label{1.10} 
\end{align}
which implies a flat rotation curve.
Here $a_{0} \approx \unit[10^{-8}]{cm\,sec^{-2}}$ is a universal constant.
If it exists, the $\ln r$-term in \eqref{1.10} overwhelms the classical
$1/r$-potential at large distances. In quantum gravity the perturbative
corrections to Newton's potential \cite{donoghue,bohr,burgess,ward}
vanish for $r \to \infty$, so a $\ln r$-behavior would require strong
nonperturbative effects.

A fully relativistic theory which also tries to explain the rotation curves
as due to a non-Newtonian force is the Weyl theory put forward by
Mannheim \cite{mannheim}. Its Lagrangian is the square of the Weyl tensor,
leading to a potential of the type $1/r + r$. The resulting rotation curves
are not flat, but seem to be in accord with the observations.

For a review of further work on non-standard gravity at galactic scales and
a comprehensive list of references we refer to \cite{aguirre}. In this context
we also mention the possibility that the flatness of the rotation curves is
caused by the quintessence field \cite{cwgal} originally introduced for
cosmological reasons \cite{cwquint,quint}.

As for strong (quantum or classical \cite{carfora}) renormalization effects
at cosmological scales, an \textit{infrared fixed point model} was proposed in
ref.\ \cite{cosmo2}. Its basic assumption is that, for $k \to 0$,
the RG trajectory of the Universe is attracted to a second fixed point
$\bigl( g_{*}^{\text{IR}}, \lambda_{*}^{\text{IR}} \bigr)$,
implying $G (k) = g_{*}^{\text{IR}} / k^{2}$ and
$\Lambda (k) = \lambda_{*}^{\text{IR}} \, k^{2}$ or, with $k \propto 1/t$,
$G \propto t^{2}$ and $\Lambda \propto 1 / t^{2}$ at late times
($t \to \infty$). It was not yet possible to derive this IR fixed point from a
flow equation. Its main motivation is the interesting phenomenology it
entails \cite{cosmo2,elo}. In particular it explains without any finetuning
why in the present Universe the matter energy density equals (approximately)
the vacuum energy density. At the same time it performs with respect to the
standard tests (supernovae, compact radio sources, etc.) as well as the
best-fit Friedmann model \cite{elo}.

One of the central questions we are going to address in the present paper is
whether the flat galaxy rotation curves can be explained as due to the onset
of the cosmological fixed point running at galactic scales. We shall employ a
model trajectory of the form
\begin{align}
G (k) = \overline{G} + g_{*}^{\text{IR}} / k^{2}
\label{1.11}
\end{align}
which interpolates between a constant $\overline{G}$ at large $k$ (small
distances) and the $1/k^{2}$-running at small $k$ (large distances).
Since the cosmological constant is not expected to be essential at the scale
of galaxies we set $\Lambda (k) \equiv 0$ right from the start. The ansatz
\eqref{1.11} is assumed to be valid at distances larger than laboratory
scales, $\overline{G}$ being the measured value of Newton's constant. This
ansatz neglects the UV renormalization effects of \eqref{1.9} which are
completely irrelevant for galaxies.

Within the improved action approach, we shall find that the $G (k)$ of
\eqref{1.11} has indeed the right qualitative properties. The running of $G$
does not only prevent the rotation curves from falling off at large distances,
they show a significant growth even. While many of the observed rotation
curves are indeed gently rising \cite{combbook}, it appears that the
growth implied by \eqref{1.11} is too strong and the relevant velocities
$v (r)$ are too small.

A second model trajectory which we are going to analyze assumes
a power law
\begin{align}
G (k) \propto k^{-q}, \quad 0 < q \ll 1,
\label{1.12}
\end{align}
over a wide range of scales. We shall see that if the observed flatness of the
rotation curves is indeed a running-$G$ effect, the ansatz \eqref{1.12}
should come very close to the truth about the underlying trajectory. At
sufficiently large distances it gives rise to perfectly flat rotation
curves.
\vspace*{2\baselinestretch pt}

The purpose of the present paper is twofold. On the theoretical side,
we study the application of the improved action approach of
ref.\ \cite{h1} to static isotropic (spherically symmetric) spacetimes
for an arbitrary RG trajectory. We derive the corresponding modified Einstein
equations, analyze their general structure, and construct a cutoff which is
selfconsistent in the following sense. Employing a cutoff identification
$k (r) \propto 1 / d (r)$ one has to specify a metric by means of which the
radial proper distance $d (r)$ is to be computed. When the quantum effects
are weak one can use a fixed background metric, but in general one would
like to compute $d (r)$ using the actual metric, i.\,e.\ the solution of 
the modified Einstein equation. Clearly the problem is that by the time one
writes down the cutoff and the resulting modified field equation one does not
know its solution yet. We shall solve this problem by a kind of ``bootstrap''
which allows us to \textit{compute} the relation between $k$ and the
coordinate $r$ in such a way that $k (r) \propto 1 / d_{g} (r)$ where
$d_{g}$ is the proper distance with respect to the solution $g_{\mu \nu}$.

These results will be important for a later application of the improved action
approach to black holes, for instance, with a $G (k)$ of the type
\eqref{1.9}.

In the present paper, on the applied side, we are going to use these methods
in order to explore the impact a scale dependent Newton constant could have
on the properties of (idealized) galaxies. We shall derive exact results for the model
trajectories \eqref{1.11} and \eqref{1.12} and allow for an arbitrary
$G (k)$ in the (modified) Newtonian approximation.

With regard to the underlying RG trajectories the application of
\eqref{1.11} and \eqref{1.12} to galaxies is certainly more speculative than
the use of \eqref{1.9} in black hole physics since the latter $G (k)$ has
actually been computed within the Einstein-Hilbert truncation, while the
former ones have a phenomenological motivation only. The reason why we
nevertheless start with the galaxies as a first application is that here
the essential new physics can be understood in a quasi-Newtonian language. In
fact, a large portion of the present paper (Section \ref{7}) will be devoted
to the modified Newtonian limit of the improved action approach. This part
of the paper is completely general, no special symmetries or
trajectories are assumed.

The remaining sections of this paper are organized as follows. In Section
\ref{2} we review some properties of the improved action approach which will
be needed later on. In Section \ref{3} we discuss the modified Einstein
equations in the static isotropic case, and in Section \ref{4} we use
Weyl transformations in order to generate a large class of solutions and to
solve the problem of the selfconsistent cutoff identification. Up to this point the
discussion applies to any scale dependence $G (k)$. Then, in Sections
\ref{5} and \ref{6}, we construct and analyze the static isotropic spacetimes
resulting from the running $G$ of the fixed point scenario, eq.\ \eqref{1.11},
and the power law \eqref{1.12}, respectively. In Section \ref{7} we return
to the general theory and explore its modified Newtonian limit. At the end
of this section we describe a simple RG-based model of galactic haloes and
demonstrate explicitly that, with an appropriate RG trajectory, no dark matter
would be needed in order to explain the flatness of the observed rotation curves.
Section \ref{8} contains the conclusions.
%
%
%
%
%
%
\section{\label{2}RG improved gravitational actions}
In this section we briefly review the improved action approach of 
ref.\ \cite{h1} and discuss some new aspects which will be important later on.
We investigate gravity interacting with an arbitrary set of minimally
coupled matter fields, $A (x)$. Their dynamics is governed by the action
$S_{\text{M}} [g,A]$ which gives rise to the energy-momentum tensor
\begin{align}
T^{\mu \nu} = \frac{2}{\sqrt{-g\,}\,} \,
\frac{\delta S_{\text{M}} [g,A]}{\delta g_{\mu \nu} (x)}.
\label{2.1}
\end{align}
Since $S_{\text{M}}$ is diffeomorphism invariant by assumption,
$T^{\mu \nu}$ is conserved when $A$ satisfies its equation of motion,
$\delta S_{\text{M}} / \delta A =0$:
\begin{align}
D_{\mu} T^{\mu \nu} = 0.
\label{2.2}
\end{align}
The total action is taken to be
\begin{align}
S_{\text{tot}} = S_{\text{mEH}} [g,G,\Lambda]
+ S_{\text{M}} [g,A]
+ S_{\theta} [g,G,\Lambda]
\label{2.3}
\end{align}
with the modified Einstein-Hilbert action\footnote{Our curvature conventions
are $R^{\sigma}_{~\rho \mu \nu} = 
- \partial_{\nu} \Gamma_{\mu \rho}^{~~\,\sigma} + \cdots$,
$R_{\mu \nu} = R^{\sigma}_{~\mu \sigma \nu}$. The metric signature is
$\left( -+++ \right)$. We abbreviate $\left( D G \right)^{2} \equiv
g^{\mu \nu} \, D_{\mu} G \, D_{\nu} G$ and
$D^{2} G \equiv g^{\mu \nu} \, D_{\mu} D_{\nu} G$. The labeling of the
Schwarzschild coordinates is $x^{0} = t$, $x^{1} = r$, $x^{2} = \theta$,
$x^{3} = \varphi$.}
\begin{align}
S_{\text{mEH}} [g,G,\Lambda] =
\frac{1}{16 \pi} \, \int \!\! \text{d}^{4} x~
\sqrt{-g\,} \, \bigg \{ \frac{R}{G (x)} 
- 2 \, \frac{\Lambda (x)}{G (x)} \bigg \}.
\label{2.4}
\end{align}
As we mentioned already, the purpose of the term $S_{\theta} [g,G,\Lambda]$
is to generate the energy-momentum tensor $\theta^{\mu \nu}$ of the scalar
fields $G (x)$ and $\Lambda (x)$. A priori $S_{\theta}$ is quite arbitrary
but, as we shall see, the consistency condition will lead to strong
constraints on $S_{\theta}$ and $\theta^{\mu \nu}$. The only general
requirements are that $S_{\theta}$ is independent of $A$, that it is
diffeomorphism invariant if $G$ and $\Lambda$ are transformed as scalars,
and that $\theta_{\mu \nu}$ vanishes for $G, \Lambda = const$
(in which case the scalars carry no momentum).

Varying $S_{\text{tot}}$ with respect to the metric we obtain the modified
Einstein equation
\begin{align}
G_{\mu \nu} = - \Lambda \, g_{\mu \nu}
+ 8 \pi G \, \bigl( T_{\mu \nu} + \Delta T_{\mu \nu}
+ \theta_{\mu \nu} \bigr),
\label{2.5}
\end{align}
or, with the convenient definitions 
$\vartheta_{\mu \nu} \equiv 8 \pi G \, \theta_{\mu \nu}$ and
$\Delta t_{\mu \nu} \equiv 8 \pi G \, \Delta T_{\mu \nu}$,
\begin{align}
G_{\mu \nu} = - \Lambda \, g_{\mu \nu}
+ 8 \pi G \, T_{\mu \nu} + \Delta t_{\mu \nu}
+ \vartheta_{\mu \nu}.
\label{2.6}
\end{align}
The tensors $T_{\mu \nu}$ and $\theta_{\mu \nu}$ obtain
as the functional derivatives of $S_{\text{M}}$ and $S_{\theta}$,
respectively, and $\Delta T_{\mu \nu}$ originates from the $x$-dependence
of Newton's constant in $S_{\text{mEH}}$:
\begin{align}
\begin{split}
\Delta t_{\mu \nu}
& = G (x) \, \bigl( D_{\mu} D_{\nu} - g_{\mu \nu} \, D^{2} \bigr)
\, \frac{1}{G (x)} \\
& = \frac{1}{G^{2}} \, \Big \{
2 \, D_{\mu} G \, D_{\nu} G - G \, D_{\mu} D_{\nu} G
- g_{\mu \nu} \, \bigl[2 \, \left( DG \right)^{2} - G \, D^{2} G \bigr] 
\Big \}.
\end{split}
\label{2.7}
\end{align}
The modified Einstein equation is coupled to the equations of motion of the
matter system,
\begin{align}
\frac{\delta S_{\text{tot}}}{\delta A} = 
\frac{\delta S_{\text{M}}}{\delta A} =0.
\label{2.8}
\end{align}
There is, however, no corresponding field equation for the scalars $G (x)$ and
$\Lambda (x)$. They are ``external fields'' defining the
``background'' for the dynamics of $g_{\mu \nu}$ and $A$.

Not every choice of $G (x)$, $\Lambda (x)$, and $\theta_{\mu \nu}$ leads
to consistent field equations. Since the LHS of \eqref{2.6} has vanishing
covariant divergence, consistency requires the same to hold true for its
RHS. This implies the ``consistency condition''
\begin{align}
D^{\mu} \Delta t_{\mu \nu} + D^{\mu} \vartheta_{\mu \nu}
- D_{\nu} \Lambda
+ 8 \pi \, \bigl( D_{\mu} G \bigr) \, T^{\mu}_{~\nu} =0,
\label{2.9}
\end{align}
or, in terms of $\Delta T_{\mu \nu}$ and $\theta_{\mu \nu}$,
\begin{align}
8 \pi G \, D^{\mu} \bigl[ \Delta T_{\mu \nu} + \theta_{\mu \nu} \bigr]
+ 8 \pi \, \bigl[ T_{\mu \nu} + \Delta T_{\mu \nu} + \theta_{\mu \nu}\bigr]
\, D^{\mu} G - D_{\nu} \Lambda =0.
\label{2.10}
\end{align}
In writing down \eqref{2.9} and \eqref{2.10} we used that
$D^{\mu} T_{\mu \nu} =0$. After working out the covariant divergence of 
$\Delta t_{\mu \nu}$, eq.\ \eqref{2.9} becomes
\begin{align}
G^{-1} \, D^{\mu} G \, \bigl( \Delta t_{\mu \nu} - R_{\mu \nu} \bigr)
+ D^{\mu} \vartheta_{\mu \nu} - D_{\nu} \Lambda 
+ 8 \pi \, \bigl( D_{\mu} G \bigr) \, T^{\mu}_{~\nu} =0.
\label{2.9-1}
\end{align}

Let us fix a constant reference value of Newton's constant, $\overline{G}$,
and let us define the ``total'' energy-momentum tensor 
$T_{\mu \nu}^{\text{ tot}}$ according to
\begin{align}
8 \pi \overline{G} \, T_{\mu \nu}^{\text{ tot}} \equiv
- \Lambda \, g_{\mu \nu}
+ 8 \pi G \, \bigl( T_{\mu \nu} + \Delta T_{\mu \nu}
+ \theta_{\mu \nu} \bigr).
\label{2.11}
\end{align}
Then the modified field equation assumes the form of the classical
equation with $\Lambda =0$:
\begin{align}
G_{\mu \nu} = 8 \pi \overline{G} \, T_{\mu \nu}^{\text{ tot}}.
\label{2.12}
\end{align}
The consistency condition \eqref{2.9} is nothing but the continuity equation
for $T_{\mu \nu}^{\text{ tot}}$ which guarantees the integrability of
\eqref{2.12}:
\begin{align}
D^{\mu} T_{\mu \nu}^{\text{ tot}} =0.
\label{2.13}
\end{align}

Occasionally it is helpful to employ the following alternative form of the
consistency condition:
\begin{align}
\begin{split}
& \frac{3}{2} \, \frac{D_{\nu} G}{G^{3}} \,
\Bigl[ G \, D^{2} G - 2 \, \left( DG \right)^{2} \Bigr]
+ D^{\mu} \vartheta_{\mu \nu} 
- \frac{D^{\mu} G}{G} \, \widetilde \vartheta_{\mu \nu} \\
& \phantom{{=}} + 4 \pi \, T \, D_{\nu} G 
- \frac{1}{G} \, D_{\nu} \left( G \Lambda \right) =0.
\end{split}
\label{2.14}
\end{align}
Here $\widetilde \vartheta_{\mu \nu} \equiv \vartheta_{\mu \nu}
- g_{\mu \nu} \, \vartheta_{\alpha}^{~\alpha} / 2$
and $T \equiv T_{\mu}^{~\mu}$. Eq.\ \eqref{2.14} follows from 
eq.\ \eqref{2.9-1} by making use of the modified Einstein equation. In fact,
eq.\ \eqref{2.14} is equivalent to eq.\ \eqref{2.9} only ``on-shell'',
i.\,e.\ when $g_{\mu \nu}$ satisfies its equation of motion. However,
since the field equations anyhow have to be solved together with the
consistency condition we may equally well use \eqref{2.14} rather
than \eqref{2.9} as the independent equation.

In order to get a first impression of the content of the ``on-shell
consistency condition'' \eqref{2.14} let us assume that $T_{\mu \nu}$ is
traceless ($T=0$), at least when $A$ satisfies its equation of motion, and
that the $x$-dependence of $G$ and $\Lambda$ is such that
$\partial_{\mu} \left( G \Lambda \right) =0$. In this case \eqref{2.14}
boils down to
\begin{align}
D^{\mu} \vartheta_{\mu \nu} 
- \frac{D^{\mu} G}{G} \, \widetilde \vartheta_{\mu \nu}
+ \frac{3}{2} \, \frac{D_{\nu} G}{G^{3}} \,
\Bigl[ G \, D^{2} G - 2 \, \left( DG \right)^{2} \Bigr] =0.
\label{2.15} 
\end{align}
In ref.\ \cite{h1} we showed that there exists a unique tensor
$\vartheta_{\mu \nu} = \vartheta_{\mu \nu}^{\text{ BD}}$ which solves
\eqref{2.15} identically in $G (x)$, i.\,e.\ without assuming any special
properties of this function, and which vanishes for $G = const$. The
corresponding $\theta_{\mu \nu}^{\text{ BD}} \equiv
\vartheta_{\mu \nu}^{\text{ BD}} / 8 \pi G$ reads
\begin{align}
\theta_{\mu \nu}^{\text{ BD}} = \left( - \frac{3}{2} \right) \,
\frac{1}{8 \pi G^{3}} \,
\Bigl[ D_{\mu} G \, D_{\nu} G 
- \tfrac{1}{2} \, g_{\mu \nu}\, \left( DG \right)^{2} \Bigr].
\label{2.16}
\end{align}
The superscript ``BD'' stands for ``Brans-Dicke''. In fact, the tensor
\eqref{2.16} made its first appearance in standard Brans-Dicke theory
\cite{bradi}. For a singular choice of the Brans-Dicke parameter,
$\omega = - 3/2$, \eqref{2.16} is exactly the energy-momentum tensor
of the BD-scalar $\phi \equiv 1/G$. In ref.\ \cite{h1} we described in detail
the relation of our theory to Brans-Dicke theory. This discussion is not
relevant for the purposes of the present paper. As the prototype of an
admissible $\theta$-tensor, $\theta_{\mu \nu}^{\text{ BD}}$ will play
an important role, though. It can be derived from the action
\begin{align}
S_{\theta}^{\text{ BD}} [g,G] =
\frac{3}{32 \pi} \, \int \!\! \text{d}^{4} x~
\sqrt{-g\,}\, \frac{D_{\mu}G \, D^{\mu} G}{G^{3}}.
\label{2.17}
\end{align}

In ref.\ \cite{h1} we identified various classes of solutions which
enjoy special properties. Here a ``solution'' is a set
$\big \{ g_{\mu \nu}, A, G, \Lambda, \theta_{\mu \nu} \big \}$
which satisfies the coupled system of equations consisting of the modified 
Einstein equation, the matter equation of motion, and the consistency
condition. The specification of a class involves (i) a choice of 
$\theta_{\mu \nu}$, (ii) assumptions about the scalar fields $G (x)$
and $\Lambda (x)$, and (iii) assumptions about the matter sector. The defining
properties of the classes considered are:
\begin{subequations} \label{2.18}
\begin{align}
& \quad \text{Class I:}
& & \theta_{\mu \nu} =0
\label{2.18a}
\\
& \quad \text{Class II:}
& & \theta_{\mu \nu} = \theta_{\mu \nu}^{\text{ BD}},~
\Lambda = 0,~T=0
\label{2.18b}
\\
& \quad \text{Class IIIa:}
& & \theta_{\mu \nu} = \theta_{\mu \nu}^{\text{ BD}},~
G \Lambda = const \neq 0,~
T=0
\label{2.18c}
\\
& \quad \text{Class IIIb:}
& & \theta_{\mu \nu} = \theta_{\mu \nu}^{\text{ BD}},~
\Lambda \neq 0,~
\partial_{\mu} \left( G \Lambda \right) \neq 0 
\label{2.18d}
\end{align}
\end{subequations}
In Class II and Class IIIa the on-shell consistency condition \eqref{2.14}
is satisfied by construction. In Class IIIb it reduces to the simple
requirement 
$4 \pi G \, T \, \partial_{\mu} G = \partial_{\mu} \left( G \Lambda \right)
\neq 0$. The importance of the solutions in Class IIIa resides in the fact
that, near a RG fixed point, $G (k) \propto 1 / k^{2}$ and
$\Lambda (k) \propto k^{2}$ which, for \textit{every} cutoff identification
$k = k (x)$, implies $G (x) \Lambda (x) = const$.

In the cases where $T=0$ we assume that the tracelessness of $T_{\mu \nu}$
is due to a Weyl-invariance of the matter action, i.\,e., if $\Delta_{A}$
is the Weyl weight of the matter field, 
$S_{\text{M}} \bigl[ \exp \left( 2 \sigma \right) \, g_{\mu \nu},
~\exp \bigl( - 2 \Delta_{A} \, \sigma \bigr) \, A \bigr]
= S_{\text{M}} \bigl[ g_{\mu \nu},A \bigr]$ for any $\sigma (x)$.
This condition insures that $T=0$ if $\delta S_{\text{M}} / \delta A =0$.

In ref.\ \cite{h1} we demonstrated that all solutions 
$\bigl( g_{\mu \nu},A \bigr)$ of Class II can be generated in the following
very efficient way. Let us pick a solution 
$\bigl( \gamma_{\mu \nu}, \mathcal{A} \bigr)$ of the conventional Einstein
equation with $G = const \equiv \overline{G}$, $\Lambda =0$, coupled to the
matter equation of motion:
\begin{subequations} 
\begin{gather}
G_{\mu \nu} (\gamma) = 8 \pi \, \overline{G} \,
T_{\mu \nu} (\mathcal{A}, \gamma),
\label{2.19a}
\\
\delta S_{\text{M}} [\gamma, \mathcal{A} ] / \delta \mathcal{A} =0.
\label{2.19b}
\end{gather}
\end{subequations}
Now we use the prescribed field $G (x)$ in order to perform the following Weyl
transformation of the metric $\gamma_{\mu \nu}$ and the matter field
$\mathcal{A}$:
\begin{subequations} \label{2.20}
\begin{align}
g_{\mu \nu} (x) & = \frac{G (x)}{\,\overline{G}\,} \, 
\gamma_{\mu \nu} (x),
\label{2.20a}
\\
A (x) & = \left[ \frac{G (x)}{\,\overline{G}\,} \right]^{-\Delta_{A}}
\, \mathcal{A} (x)
\label{2.20b}.
\end{align}
\end{subequations}
It can be shown that the pair $\bigl( g_{\mu \nu},A \bigr)$ thus
obtained is a Class II solution of the RG improved system, with a position
dependent Newton constant $G = G (x)$, and that \textit{all} solutions of
Class II can be obtained in this way.

The proof makes essential use of the fact that, for $T=0$ and $\Lambda =0$,
the Brans-Dicke action $S_{\theta}^{\text{ BD}}$ can be absorbed into
$S_{\text{mEH}}$ by Weyl-rescaling the metric according to \eqref{2.20a}.
A Weyl invariant matter action is ``blind'' to this rescaling, explaining
why the case $T=0$ is special. (See \cite{h1} for further details.)

Generically a nonzero cosmological constant with an arbitrary $x$-dependence
will spoil the simplicity of this method. An exception is the case when 
$\Lambda$ varies according to
\begin{align}
\Lambda (x) = \overline{\Lambda} ~ \overline{G} / G (x),
\quad \overline{\Lambda} = const.
\label{2.21}
\end{align}
Hence $\Lambda G = const$ and we are in Class IIIa. It turns out that all 
solutions $\bigl( g_{\mu \nu},A \bigr)$ of Class IIIa can be obtained as the
Weyl transform according to \eqref{2.20} of some solution
$\bigl( \gamma_{\mu \nu}, \mathcal{A} \bigr)$ of the classical equations
\begin{subequations} \label{2.22}
\begin{gather}
G_{\mu \nu} (\gamma) = - \overline{\Lambda} \, \gamma_{\mu \nu}
+ 8 \pi \, \overline{G} \, T_{\mu \nu} (\mathcal{A}, \gamma),
\label{2.22a}
\\
\delta S_{\text{M}} [\gamma, \mathcal{A}] / \delta \mathcal{A} =0.
\label{2.22b}
\end{gather}
\end{subequations}
For a proof and further details about the use of Weyl rescalings as a
solution-generating transformation the reader is referred to ref.\ \cite{h1}.

Very often one would like to find metrics $g_{\mu \nu}$ which enjoy special
symmetry properties. For a symmetry transformation generated by the Killing
vector field $K^{\mu}$ the condition is that the Lie-derivative of
$g_{\mu\nu}$
with respect to $K^{\mu}$ vanishes: $\mathscr{L}_{K} g_{\mu \nu} =0$.
If \eqref{2.20a} applies, i.\,e.\ in Classes II and IIIa, this leads to the
requirement
$G \, \mathscr{L}_{K} \gamma_{\mu \nu} + 
\gamma_{\mu\nu} \,  \mathscr{L}_{K} G =0$.
For this requirement to be met it is sufficient, but not necessary, that $K$
is a Killing vector of $\gamma_{\mu \nu}$, too, 
$\mathscr{L}_{K} \gamma_{\mu \nu} =0$, and that the ``external field''
$G (x)$ is chosen so as to respect the symmetry:
$\mathscr{L}_{K} G  \equiv K^{\mu} \, \partial_{\mu} G =0$.
In the present paper we shall always generate symmetric $g_{\mu \nu}$'s
from $\gamma_{\mu \nu}$'s
and $G$'s possessing the same symmetry. (In principle a more general situation
is conceivable where the non-invariance of $G (x)$ compensates for a 
non-invariance of $\gamma_{\mu \nu}$.)
%
%
%
%
%
%
%
%
\section{\label{3}Static isotropic solutions}
\subsection{\label{s3.1}Symmetry properties}
In this section we focus on solutions to the modified Einstein equations 
with special symmetry properties. We search for spacetimes which are
\textit{static} and \textit{isotropic}, i.\,e.\ spherically symmetric.
The corresponding $g_{\mu \nu}$ admits 4 Killing vectors: one related to time
translation, and 3 to spatial rotations. In Schwarzschild-type coordinates
$x^{\mu} = (t,r,\theta,\varphi)$ the most general metric of this kind can
be brought to the ``standard form'' \cite{weinbook}
\begin{subequations} \label{3.1}
\begin{align}
\text{d} s^{2} \equiv g_{\mu \nu} \, \text{d} x^{\mu} \, \text{d} x^{\nu} =
- B (r) \, \text{d} t^{2} + A (r) \, \text{d} r^{2}
+ r^{2} \, \text{d} \sigma^{2},
\label{3.1a}
\end{align}
where
\begin{align}
\text{d} \sigma^{2} \equiv \text{d} \theta^{2} + \sin^{2} \theta \,
\text{d} \varphi^{2}
\label{3.1b}
\end{align}
\end{subequations}
denotes the line element on the unit 2-sphere. For the spacetimes discussed
in the present paper the functions $A (r)$ and $B (r)$ are strictly positive
and horizons are not an issue. Hence $t$ and $r$ have the interpretation of
the time and the radial coordinate, respectively.

In the following the ``external fields'' $G (x)$ and $\Lambda (x)$ are chosen
in accord with the above symmetry requirements. We demand that
$K^{\mu} \, \partial_{\mu} G = K^{\mu} \, \partial_{\mu} \Lambda =0$
for all four Killing vectors $K^{\mu}$ which implies that $G$ and $\Lambda$
can depend on the radial coordinate $r$ only:
\begin{align}
G = G (r), \quad \Lambda = \Lambda (r).
\label{3.2}
\end{align}
Likewise the matter field configuration is assumed to respect all symmetries.
\subsection{\label{s3.2}Decomposition of the field equations}
We model the matter system by a perfect fluid in hydrostatic equilibrium.
In the metric \eqref{3.1} its energy-momentum tensor 
$T_{\mu}^{~\nu} = p \, g_{\mu}^{~\nu} + \left( p + \rho \right)
u_{\mu} u^{\nu}$ is diagonal,
\begin{align}
T_{\mu}^{~\nu} = \text{diag} [-\rho,p,p,p].
\label{3.3}
\end{align}
The energy density $\rho$ and the pressure $p$ depend on 
$r$ only. To fully characterize the fluid one has to specify an equation 
of state
\begin{align}
p = p (\rho).
\label{3.4}
\end{align}
For the metric \eqref{3.1} the continuity equation 
$D^{\mu} T_{\mu}^{~\nu} =0$ amounts to a single condition:
\begin{align}
\frac{\text{d}}{\text{d} r} \, p (r)
+ \frac{B^{\prime} (r)}{2 \, B(r)} \,
\bigl( \rho (r) + p (r) \bigr) =0.
\label{3.5}
\end{align}
(Here and in the following a prime always denotes the derivative with
respect to the argument.)

Next let us evaluate the tensor $\Delta T_{\mu \nu}$ for metrics of the
``$AB$-form'' \eqref{3.1}. Working out the covariant derivatives in
\eqref{2.7} one finds that $\Delta T_{\mu}^{~\nu}$ is diagonal:
\begin{align}
\Delta T_{\mu}^{~\nu} \equiv \Delta t_{\mu}^{~\nu} / \left( 8 \pi \, G \right)
= \text{diag} \bigl[ -\Delta \rho,
 \Delta p_{r}, \Delta p_{\perp}, \Delta p_{\perp} \bigr].
\label{3.6}
\end{align}
Here $\Delta \rho = - \Delta T_{t}^{~t}$ is an energy density which is induced
by the position dependence of Newton's constant:
\begin{subequations} \label{3.7}
\begin{align}
\Delta \rho =
\frac{1}{8 \pi G} \left[
- \frac{2}{r \, A} \, \left( \frac{G^{\prime}}{G} \right) 
+ \frac{A^{\prime}}{2 \, A^{2}} \, \left( \frac{G^{\prime}}{G} \right)
+ \frac{2}{A} \, \left( \frac{G^{\prime}}{G} \right)^{2}
- \frac{1}{A} \, \left( \frac{G^{\prime \prime}}{G} \right)
\right].
\label{3.7a}
\end{align}
Likewise there is a contribution to the radial pressure,
\begin{align}
\Delta p_{r} 
& = \Delta T_{r}^{~r}
=
\frac{1}{8 \pi G} \left[
\frac{B^{\prime}}{2 \, A \, B} \, \left( \frac{G^{\prime}}{G} \right)
+ \frac{2}{r \, A} \, \left( \frac{G^{\prime}}{G} \right)
\right],
\label{3.7b}
\end{align}
and to the tangential pressure:
\begin{align}
\begin{split}
\Delta p_{\perp} 
= \Delta T_{\theta}^{~\theta} = \Delta T_{\varphi}^{~\varphi}
& =
\frac{1}{8 \pi G} \left[
\frac{1}{r \, A} \, \left( \frac{G^{\prime}}{G} \right)
- \frac{A^{\prime}}{2 \, A^{2}} \, \left( \frac{G^{\prime}}{G} \right)
+ \frac{B^{\prime}}{2 \, A \, B} \, \left( \frac{G^{\prime}}{G} \right)
\right.
\\
& \phantom{{==} \frac{1}{8 \pi G} \biggl[ \biggr.} \left.
- \frac{2}{A} \, \left( \frac{G^{\prime}}{G} \right)^{2}
+ \frac{1}{A} \, \left( \frac{G^{\prime \prime}}{G} \right)
\right].
\end{split}
\label{3.7c}
\end{align}
\end{subequations}
As expected on symmetry grounds, the components $\Delta T_{\theta}^{~\theta}$
and $\Delta T_{\varphi}^{~\varphi}$ are found to be equal. However, in general
the radial and the tangential pressures are different so that 
$\Delta T_{\mu}^{~\nu}$ is not of the perfect fluid type.

The precise form of the $\theta$-tensor will be left open for the time being.
It is enough to know that it has the structure
\begin{align}
\theta_{\mu}^{~\nu} & \equiv
\vartheta_{\mu}^{~\nu} / \left( 8 \pi G \right) =
\text{diag} \bigl[ - \rho_{\theta}, \, p_{\theta, r}, \,
p_{\theta, \perp}, \, p_{\theta, \perp} \bigr].
\label{3.8}
\end{align}
It supplies additional contributions $\rho_{\theta}$, $p_{\theta,r}$,
and $p_{\theta,\perp}$ to the energy density and the radial and tangential
pressure, respectively, which are functionally dependent on $A (r)$ and
$B (r)$.

A further source term appearing on the RHS of Einstein's equation is the one
containing the cosmological constant. Setting 
${T^{\Lambda}}_{\mu}^{~\nu} \equiv - \left( \Lambda / 8 \pi G \right) \,
g_{\mu}^{~\nu} = \text{diag} \bigl[ - \rho_{\Lambda}, p_{\Lambda},
p_{\Lambda}, p_{\Lambda} \bigr]$, the vacuum energy density and pressure
due to the cosmological constant are, respectively, 
$\rho_{\Lambda} = \Lambda / \left( 8 \pi G \right)$ and
$p_{\Lambda} = - \Lambda / \left( 8 \pi G \right)$. In this notation, the
total energy-momentum tensor defined by eq.\ \eqref{2.11} reads
${T^{\text{ tot}}}_{\mu}^{~\nu} =
\text{diag} \bigl[ - \rho^{\text{ tot}}, p_{r}^{\text{ tot}}, 
p_{\perp}^{\text{ tot}}, p_{\perp}^{\text{ tot}} \bigr]$ with the entries
\begin{align}
\begin{split}
\rho^{\text{ tot}} & =
\left[ G (r) / \, \overline{G} \, \right] \,
\bigl( \rho + \rho_{\Lambda} + \Delta \rho + \rho_{\theta} \bigr),
\\
p_{r}^{\text{ tot}} & =
\left[ G (r) / \, \overline{G} \, \right] \,
\bigl( p + p_{\Lambda} + \Delta p_{r} + p_{\theta,r} \bigr),
\\
p_{\perp}^{\text{ tot}} & =
\left[ G (r) / \, \overline{G} \, \right] \,
\bigl( p + p_{\Lambda} + \Delta p_{\perp} + p_{\theta,\perp} \bigr).
\end{split}
\label{3.9}
\end{align}

For the metric \eqref{3.1} the only nonvanishing components of the Einstein 
tensor $G_{\mu}^{~\nu}$ are $G_{t}^{~t}$, $G_{r}^{~r}$, and 
$G_{\theta}^{~\theta} = G_{\varphi}^{~\varphi}$. As a result, the modified
Einstein equation $G_{\mu}^{~\nu} = 8 \pi \, \overline{G} \, 
{T^{\text{ tot}}}_{\mu}^{~\nu}$ decomposes into the following 3 equations:
the $tt$-component,
\begin{subequations} \label{3.10}
\begin{align}
G_{t}^{~t} & \equiv \frac{1}{r^{2} \, A} - \frac{1}{r^{2}} 
- \frac{A^{\prime}}{r \, A^{2}}
=
- 8 \pi G \, \bigl( \rho + \rho_{\Lambda} + \Delta \rho + \rho_{\theta} \bigr),
\label{3.10a}
\end{align}
the $rr$-component,
\begin{align}
G_{r}^{~r} & \equiv \frac{1}{r^{2} \, A} - \frac{1}{r^{2}}
+ \frac{B^{\prime}}{r \, A \, B} 
=
8 \pi G \, \bigl( p + p_{\Lambda} + \Delta p_{r} + p_{\theta,r} \bigr),
\label{3.10b}
\end{align}
and the $\theta \theta$- or $\varphi \varphi$-component:
\begin{align}
\begin{split}
G_{\theta}^{~\theta} = G_{\varphi}^{~\varphi}
& \equiv - \frac{A^{\prime}}{2 \, r \, A^{2}}
+ \frac{B^{\prime}}{2 \, r \, A \, B}
- \frac{A^{\prime} \, B^{\prime}}{4 \, A^{2} \, B}
- \frac{{B^{\prime}}^{2}}{4 \, A \, B^{2}}
+ \frac{B^{\prime \prime}}{2 \, A \, B}
\\
& =
8 \pi G \, \bigl( p + p_{\Lambda} + \Delta p_{\perp} + p_{\theta,\perp} \bigr).
\end{split}
\label{3.10c}
\end{align}
\end{subequations}

In the same notation, the consistency condition assumes the form
$D^{\mu} {T^{\text{ tot}}}_{\mu}^{~\nu} =0$. Working out the covariant
derivative leads to
\begin{align}
\frac{\text{d}}{\text{d} r} \, p_{r}^{\text{ tot}}
+ \frac{B^{\prime}}{2 \, B} \, \bigl( \rho^{\text{ tot}} 
+ p_{r}^{\text{ tot}} \bigr)
+ \frac{2}{r} \, \bigl( p_{r}^{\text{ tot}}
- p_{\perp}^{\text{ tot}} \bigr) =0
\label{3.einschub}
\end{align}
whence
\begin{align}
\begin{split}
& G^{\prime} \, \bigl( p + p_{\Lambda} + \Delta p_{r} + 
p_{\theta,r} \bigr)
+ G \, \frac{\text{d}}{\text{d} r} \, \bigl( p_{\Lambda} + \Delta p_{r} + 
p_{\theta,r} \bigr)
\\
& \phantom{{=}}
+ \frac{B^{\prime}}{2 \, B} \, G \, 
\bigl( \Delta \rho + \Delta p_{r}
+ \rho_{\theta} + p_{\theta,r} \bigr)
+ \frac{2}{r} \, G \, 
\bigl( \Delta p_{r} - \Delta p_{\perp}
+ p_{\theta,r} - p_{\theta,\perp} \bigr) 
=0.
\end{split}
\label{3.11}
\end{align}
In deriving \eqref{3.11} we exploited that $T_{\mu \nu}$ obeys the ordinary
continuity equation \eqref{3.5}.

As in the classical case, Einstein's equations are not independent of the
conservation law for the energy-momentum tensor. Applying the standard
discussion to 
$G_{\mu \nu} = 8 \pi \, \overline{G} \, T_{\mu \nu}^{\text{ tot}}$
we learn that, provided $D^{\mu} T_{\mu \nu}^{\text{ tot}} =0$ is obeyed,
the validity of the $tt$- and $rr$-component equations (\ref{3.10}\,a,b)
implies the validity of the $\theta \theta$- or
$\varphi \varphi$-equation \eqref{3.10c}. In the following we
replace this latter equation, \eqref{3.10c}, with
$D^{\mu} T_{\mu \nu}^{\text{ tot}} =0$ as the independent condition. It is
satisfied provided the consistency condition \eqref{3.11} and the ordinary
continuity equation \eqref{3.5} are valid.

Let us summarize. After fixing the form of $\theta_{\mu \nu}$ and picking
a background configuration $G (r)$, $\Lambda (r)$, a system of independent
(differential) equations consists of
\begin{align}
\begin{split}
& 
\quad \bullet~~ \text{the $tt$-component \eqref{3.10a}, }
\\
& 
\quad \bullet~~ \text{the $rr$-component \eqref{3.10b}, }
\\
&
\quad \bullet~~ \text{the consistency condition \eqref{3.11}, }
\\
&
\quad \bullet~~ \text{the continuity equation \eqref{3.5}, }
\\
&
\quad \bullet~~ \text{the equation of state \eqref{3.4}. }
\end{split}
\label{3.set}
\end{align}
These are 5 equations for the 4 unknown functions $A (r)$, $B (r)$,
$\rho (r)$, and $p (r)$. Even though the system appears to be overdetermined
at first sight it has indeed nontrivial consistent solutions. An obvious example
is provided by the Classes II and IIIa. From the general discussion we know 
that every solution of the classical Einstein equation 
induces a solution of the
modified one. As a consequence, every static isotropic solution of the
analogous classical equations generates a consistent solution to the above 5
equations. The solution-generating algorithm based upon Weyl transformations
requires the choice $\theta_{\mu \nu} = \theta_{\mu \nu}^{\text{ BD}}$.
Based on our experience with cosmological spacetimes \cite{h1} we would
expect that the coupled system has solutions also for other choices of the
$\theta$-tensor, $\theta_{\mu \nu} =0$ for instance. (However, an
``extreme'' choice such as $\theta_{\mu \nu} =0$ might be in conflict with
the positivity properties of $T_{\mu \nu}$ expected for ordinary matter and
might necessitate a reinterpretation of $p$ and $\rho$, see ref.\ \cite{h1}.)
The weak field approximation of the system \eqref{3.set} is discussed
in Appendix \ref{A}.

Since the Brans-Dicke-type $\theta$-tensor plays a distinguished role 
\cite{h1} we shall adopt the following choice for $\theta_{\mu \nu}$ in the
rest of this paper:
\begin{align}
\theta_{\mu \nu} = \varepsilon \, \theta_{\mu \nu}^{\text{ BD}}.
\label{3.12}
\end{align}
Here $\varepsilon$ is an arbitrary real constant. Particularly interesting
values include $\varepsilon =0$ (Class I) and $\varepsilon =1$ (Classes II
and IIIa). From \eqref{3.12} with \eqref{2.16} we obtain a diagonal
$\theta_{\mu}^{~\nu}$ of the type \eqref{3.8} with the entries
\begin{subequations} \label{3.13}
\begin{align}
\rho_{\theta} 
 = - \theta_{t}^{~t}
& =
- \frac{3 \, \varepsilon}{32 \pi G} \,
\frac{1}{A} \, \left( \frac{G^{\prime}}{G} \right)^{2},
\label{3.13a}
\\
p_{\theta, r} 
= \theta_{r}^{~r}
& =
- \frac{3 \, \varepsilon}{32 \pi G} \,
\frac{1}{A} \, \left( \frac{G^{\prime}}{G} \right)^{2},
\label{3.13b}
\\
 p_{\theta, \perp} 
= \theta_{\theta}^{~\theta}
&  =
\frac{3\, \varepsilon}{32 \pi G} \,
\frac{1}{A} \, \left( \frac{G^{\prime}}{G} \right)^{2}.
\label{3.13c}
\end{align}
\end{subequations}
We shall see that the essential conclusions of this paper are actually independent of 
the choice for $\theta_{\mu \nu}$.

With the choice \eqref{3.12} we can now write down the differential equations
in explicit form. We find the $tt$-component of Einstein's equation
\begin{subequations} \label{3.14}
%
\begin{align}
\begin{split}
& \frac{1}{r^{2} \, A} - \frac{1}{r^{2}} 
- \frac{A^{\prime}}{r \, A^{2}}
\\
& = - \Lambda - 8 \pi G \, \rho
+ \left[
\frac{2}{r \, A} - \frac{A^{\prime}}{2 \, A^{2}}
\right] \,\left( \frac{G^{\prime}}{G} \right)
- \frac{1}{4 \, A} \, \left( 8 - 3 \, \varepsilon \right) \,
\left( \frac{G^{\prime}}{G} \right)^{2}
 + \frac{1}{A} \, \left( \frac{G^{\prime \prime}}{G} \right),
\end{split}
\label{3.14a}
\end{align}
the $rr$-component
%
\begin{align}
\frac{1}{r^{2} \, A} - \frac{1}{r^{2}}
+ \frac{B^{\prime}}{r \, A \, B} 
= - \Lambda + 8 \pi \, G \, p
+ \left[ \frac{B^{\prime}}{2 \, A \, B} + \frac{2}{r \, A} 
\right] \, \left( \frac{G^{\prime}}{G} \right)
- \frac{3 \, \varepsilon}{4 \, A} \, \left( \frac{G^{\prime}}{G} \right)^{2},
\label{3.14b}
\end{align}
the $\theta \theta$- and $\varphi \varphi$-component
%
\begin{align}
\begin{split}
&- \frac{A^{\prime}}{2 \, r \, A^{2}}
+ \frac{B^{\prime}}{2 \, r \, A \, B}
- \frac{A^{\prime} \, B^{\prime}}{4 \, A^{2} \, B}
- \frac{{B^{\prime}}^{2}}{4 \, A \, B^{2}}
+ \frac{B^{\prime \prime}}{2 \, A \, B}
\\
& = 
- \Lambda + 8 \pi \, G \, p
+ \left[
\frac{1}{r \, A} - \frac{A^{\prime}}{2 \, A^{2}} 
+ \frac{B^{\prime}}{2 \, A \, B} 
\right] \, \left( \frac{G^{\prime}}{G} \right)
\\
& \phantom{{==}}
- \frac{1}{4 \, A} \, \left( 8 - 3 \, \varepsilon \right) 
\, \left( \frac{G^{\prime}}{G} \right)^{2}
+ \frac{1}{A} \, \left( \frac{G^{\prime \prime}}{G} \right),
\end{split}
\label{3.14c}
\end{align}
\end{subequations}
and the consistency condition
%
\begin{align}
\begin{split}
& - \left[ 
\frac{A^{\prime}}{r \, A^{2}}
+ \frac{A^{\prime} \, B^{\prime}}{4 \, A^{2} \, B}
+ \frac{{B^{\prime}}^{2}}{4 \, A \, B^{2}}
- \frac{B^{\prime \prime}}{2 \, A \, B} \right]
\, \left( \frac{G^{\prime}}{G} \right)
- \frac{3 \, \varepsilon}{2 \, A} 
\, \left( \frac{G^{\prime} G^{\prime \prime}}{G^{2}} \right)
+ \frac{3 \, \varepsilon}{2 \, A}
 \, \left( \frac{G^{\prime}}{G} \right)^{3}
\\
& \phantom{{=}}
+ \left[ \frac{3 \, \varepsilon\, A^{\prime}}{4 \, A^{2}} 
+ \frac{B^{\prime}}{4 \, A \, B} \, \left( 2 - 3 \, \varepsilon \right)
+ \frac{1}{r \, A} \, \left( 2 - 3 \, \varepsilon \right) \right]
\, \left( \frac{G^{\prime}}{G} \right)^{2} 
- \Lambda^{\prime} + 8 \pi \, G^{\prime} \, p
=0.
\end{split}
\label{3.15}
\end{align}
\subsection{\label{s3.3}Selfconsistent cutoff identification}
Up to this point we assumed that $G$ and $\Lambda$ are given to us as
functions of the radial coordinate $r$. In the RG setting this is not the
case, however. The flow equation provides us only with a RG trajectory
$\bigl( G (k), \Lambda (k), \cdots \bigr)$ into which one must insert a cutoff
identification  $k = k (x)$, or here $k = k (r)$. A detailed discussion of 
the physical cutoff mechanism \cite{h1} and its earlier use in the improved
solution approach \cite{bh} suggests an identification 
$k (r) \propto 1 / d (r)$. Here $d (r)$ is the proper length of a radial
curve $\mathcal{C}$, defined by $\text{d} t = \text{d} \theta = \text{d}
\varphi = 0$, connecting a point with coordinate $r$ to the origin
($r=0$). In view of the symmetries of the problem the choice of the curve
$\mathcal{C}$ is certainly the most natural one. The more subtle
question is: which metric should we use for the computation of the proper
length $d (r)$?

If the renormalization effects are weak and $G$ and $\Lambda$ are
approximately constant, one might pick some background metric 
$\gamma_{\mu \nu}$, a solution for the case of exact constancy, and then
compute $d (r) \equiv d_{\gamma} (r)$ using this background metric.
In the weak field approximation, for instance, the Minkowski metric is the
natural choice for $\gamma_{\mu \nu}$. In the Classes II and IIIa the metric
$\gamma_{\mu \nu}$ which generates the actual solution $g_{\mu \nu}$ by a 
Weyl rescaling can serve as a background metric for this purpose. It is
clear, however, that the ``$d_{\gamma}$-improvement'' based upon the
identification $k (r) = \xi / d_{\gamma} (r)$ becomes inapplicable when
the RG effects are strong. Then $g_{\mu \nu}$ differs significantly from
$\gamma_{\mu \nu}$, and since physical distances are measured by $g_{\mu \nu}$
one should compute the proper length $d (r) \equiv d_{g} (r)$ from this
metric.

This leads to the complication that in this manner the identification
$k = k (r)$ becomes functionally dependent on the metric $g_{\mu \nu}$ which
we do not know beforehand but rather would like to compute. In fact,
for metrics of the form \eqref{3.1} the cutoff identification of this
``$d_{g}$-improvement'' involves the function $A (r)$:
\begin{align}
k (r) = \xi / d_{g} (r) 
= \xi \left( \int \limits_{0}^{r} \!\! \text{d} r^{\prime}~
\sqrt{A (r^{\prime})\,}\, \right)^{-1}.
\label{3.17}
\end{align} 
Using \eqref{3.17} we define the $r$-dependent ``constants''
$G (r) \equiv G \bigl( k = k (r) \bigr)$ and
$\Lambda (r) \equiv \Lambda \bigl( k = k (r) \bigr)$ 
and insert them into the system of equations \eqref{3.set} we derived in
the previous subsection. This results in a system of coupled
integro-differential equations for $A (r)$, $B (r)$, $\rho (r)$, and $p (r)$.
The input for this system are the $k$-dependent ``constants'' $G (k)$ and
$\Lambda (k)$. Their $r$-dependent counterparts are obtained as part of the
solution. Given a solution for $A (r)$ and $B (r)$, the metric $g_{\mu \nu}$
and the corresponding distance function $d_{g} (r)$ are known. This in turn
fixes the hitherto unknown $r$-dependence of $G$ and $\Lambda$.
In this way the cutoff \eqref{3.17} determines the relation between $k$ and
$r$ dynamically. We shall refer to it as a \textit{selfconsistent cutoff
identification} therefore.

Solving the system of integro-differential equations\footnote{In principle
one can convert the integro-differential equations to differential equations
by treating $d_{g} (r)$ as an independent function satisfying the additional
differential equation $d_{g}^{\prime} (r) = \sqrt{A (r)\,}$.} appears
to be a formidable task in general. Therefore it comes as a very pleasant
surprise that, even with a selfconsistent cutoff, solutions of Classes II and
IIIa can be found very efficiently in closed form almost. This will 
be the topic of the next section.

Before closing this section a comment might be in order about the possibility
of horizons where the function $A (r)$ turns negative. In
ref.\ \cite{bh} a ``$d_{\gamma}$-improvement'' of the Schwarzschild solution has
been performed for which $A (r) = \left[ 1 - 2 \overline{G} M / r
\right]^{-1}$ is negative inside the horizon, and $r$ plays the role of a time
coordinate for $r < 2 \overline{G} M$. In \cite{bh} detailed
physical arguments were given which indicate that, where $\text{d} s^{2}$ is
negative along $\mathcal{C}$, one should define the ``distance'' function as
the integral over $\big \lvert \text{d} s^{2} \big \rvert = - \text{d} s^{2}$,
rather than $\text{d} s^{2}$ itself. We adopt the same rule here which amounts
to replacing $A$ by $|A|$ in \eqref{3.17}:
\begin{align}
d_{g} (r) & =
\int \limits_{\mathcal{C}} \!\! 
\sqrt{\big \lvert \text{d} s^{2} \big \rvert\,}
= \int \limits_{0}^{r} \!\! \text{d} r^{\prime}~
\sqrt{\big \lvert A (r^{\prime}) \big \rvert\,}.
\label{3.18}
\end{align}
(See \cite{MTW} for a related discussion.) However, for the purposes of the 
present paper the question of how to define $\mathcal{C}$ inside the 
Schwarzschild horizon is completely irrelevant. We are interested in the
physics at $r \gg 2 \overline{G} M$ where the contribution of the
interval $0 < r < 2 \overline{G} M$ to $d_{g} (r)$ is utterly
negligible.
%
%
%
%
%
%
%
%
\section{\label{4}Class II vacuum solutions}
In this section we implement the idea of a selfconsistent cutoff
identification in the framework of the solution-generating Weyl 
transformations. This restricts us to the Classes II and IIIa. Both of them
require that $T = 0$, leading to the ``radiation'' equation 
of state $p = \rho / 3$ in the hydrodynamical model. Since this equation
of state is not appropriate for the physical situations we envisage in the
present paper we satisfy $T=0$ in the trivial way $\rho = p = 0$,
i.\,e.\ we concentrate on vacuum spacetimes. The difference of the Classes II
and IIIa is that in the first case $\Lambda (r)$ vanishes identically, while
it is inversely proportional to $G (r)$ in the second:
$\Lambda (r) = \overline{\Lambda}~\overline{G} / G (r) \neq 0$. Since on
the (galactic) scales we are interested in the cosmological constant is
unlikely to play a decisive role we are not going to analyze the case IIIa
here. Moreover, one should be aware of the following difficulty. If one
actually wants to explore the effect of a $r$-dependent cosmological constant
it is not enough to have solutions of Class IIIa available. They are indeed
realized near a fixed point where $\Lambda (k) \propto 1 / G (k) \propto
k^{2}$ so that $\Lambda (r) \propto 1 / G (r)$, but other relevant
regimes, a ``cross over'' from the fixed point to ordinary
gravity with constant $G$ and $\Lambda$, say, are not in Class IIIa and cannot
be obtained by the Weyl technique.

For these reasons we confine the following discussion to vacuum spacetimes of
Class II, setting $\Lambda (r) =0$, $T_{\mu \nu} =0$, and
$\theta_{\mu \nu} = \theta_{\mu \nu}^{\text{ BD}}$, i.\,e.\ $\varepsilon =1$.
Then solutions $g_{\mu \nu}$ of the modified Einstein equation are obtained by
Weyl-transforming solutions $\gamma_{\mu \nu}$ of
\begin{align}
G_{\mu \nu} (\gamma) =0.
\label{4.1}
\end{align}
According to eq.\ \eqref{2.20a},
$g_{\mu \nu} = \left[ G (x) / \, \overline{G} \, \right] \, \gamma_{\mu \nu}$,
or
\begin{align}
\text{d} s^{2} = \left[ G (x) / \, \overline{G} \, \right] \, 
\text{d} s_{\gamma}^{2}
\label{4.2}
\end{align}
where $\text{d} s^{2}$ is the line element of $g_{\mu \nu}$ and 
$\text{d} s_{\gamma}^{2} \equiv \gamma_{\mu \nu} \, \text{d} x^{\mu} 
\, \text{d} x^{\nu}$. We implement the symmetry requirements by demanding 
that $\gamma_{\mu \nu}$, too, describes a static isotropic spacetime, and
that $G (x)$ is annihilated by all 4 Killing vectors,
$K^{\mu} \, \partial_{\mu} G =0$, implying that it can depend on the
radial coordinate only. Hence, by Birkhoff's theorem, the only eligible
$\gamma_{\mu \nu}$ is the Schwarzschild metric
\begin{subequations} \label{4.3}
\begin{align}
\text{d} s_{\gamma}^{2} =
- f (\rho) \, \text{d} t^{2}
+ f (\rho)^{-1} \, \text{d} \rho^{2}
+ \rho^{2} \, \text{d} \sigma^{2}
\label{4.3a}
\end{align}
with
\begin{align}
f (\rho) \equiv 1 - \frac{2 \, \overline{G} \, M}{\rho}.
\label{4.3b}
\end{align}
\end{subequations}
For a reason which will become clear in a moment we denoted the radial
coordinate $\rho$ rather than $r$. (There should be no confusion
with an energy density.) Note that the factor $\overline{G}$ in \eqref{4.3b}
arises because the coefficient of the $1/\rho$-term is a free constant of
integration, usually written $2 G M$, and to be interpreted as
$2 \overline{G} M$ in the present context; $\overline{G}$ does not appear
in the field equation $G_{\mu \nu} (\gamma) =0$. Minkowski space, the special
case $M=0$, can be used as well for generating solutions.

If $G$ was known as a function of $\rho$, the line element we are after could
be obtained as $\text{d} s^{2} = \left[ G (\rho) / \, \overline{G} \, \right] 
\, \text{d} s_{\gamma}^{2}$ simply. However, as we discussed in Subsection
\ref{s3.3}, what the RG analysis provides us with is the function $G (k)$ 
only. In the following we employ the selfconsistent, hence
$g_{\mu \nu}$-dependent, cutoff identification
\begin{align}
k (\rho) = \xi / d_{g} (\rho)
\label{4.4}
\end{align}
in order to convert $G (k)$ to $G (\rho) = G \bigl( k = k (\rho) \bigr)$.

Given the function $G (k)$, it will prove convenient to define an associated
function $W (y)$ by
\begin{align}
W (y) \equiv G (k = \xi/y) / \, \overline{G}
\label{4.5}
\end{align}
with $\xi$ an arbitrary fixed constant of order unity whose precise value will
not matter. Obviously $W (y)$ with $y = d_{g} (\rho)$ is the Weyl factor
relating the two metrics:
\begin{align}
\text{d} s^{2} & = W \bigl( d_{g} (\rho) \bigr) \,
\Bigl[ - f (\rho) \, \text{d} t^{2}
+ f (\rho)^{-1} \, \text{d} \rho^{2}
+ \rho^{2} \, \text{d} \sigma^{2} \Bigr].
\label{4.6}
\end{align}
Eq.\ \eqref{4.6} is an implicit equation for $g_{\mu \nu}$: on the RHS of 
\eqref{4.6} it makes its appearance in $d_{g} (\rho)$, on the LHS in
$\text{d} s^{2}$. Solving this equation is tantamount to finding $d_{g}$
as a function of $\rho$. In fact, once we know the function
$\rho \mapsto d_{g} (\rho)$ we can combine it with $y \mapsto W (y)$,
known from the renormalization group, and define
\begin{align}
w (\rho) \equiv W \bigl( d_{g} (\rho) \bigr).
\label{4.7}
\end{align}
By definition, $w$ is a function of $\rho$, while the natural argument of $W$
is $y \equiv d_{g} (\rho)$. We have accomplished our task when we know $w$, for
then
\begin{align}
\text{d} s^{2} & = w (\rho) \,
\Bigl[ - f (\rho) \, \text{d} t^{2}
+ f (\rho)^{-1} \, \text{d} \rho^{2}
+ \rho^{2} \, \text{d} \sigma^{2} \Bigr]
\label{4.8}
\end{align}
is an explicit formula for the line element of $g_{\mu \nu}$.

How can we get a handle on $d_{g} (\rho)$? Eq.\ \eqref{4.6} tells us that,
if $\text{d} t = \text{d} \theta = \text{d} \varphi =0$,
\begin{align}
\big \lvert \text{d} s^{2} \big \rvert & =
W \bigl( d_{g} (\rho) \bigr) \, 
\big \lvert f (\rho) \big \rvert^{-1} \, \text{d} \rho^{2}
\label{4.9}
\end{align}
since $G$ and $W$ are positive throughout. The total proper length of 
$\mathcal{C}$ obtains by integrating \eqref{4.9}:
\begin{align}
d_{g} (\rho) = \int \limits_{0}^{\rho} \!\! \text{d} \rho^{\prime}~
\sqrt{W \bigl( d_{g} (\rho^{\prime}) \bigr) \, 
\big \lvert f (\rho^{\prime}) \big \rvert^{-1} \,}\,.
\label{4.10}
\end{align}
Since $W$ and $f$ are known functions \eqref{4.10} is an integral equation
for $d_{g} (\rho)$. Differentiating it yields the equivalent differential
equation
\begin{align}
\frac{\text{d}}{\text{d} \rho} \,  d_{g} (\rho) 
= \sqrt{W \bigl( d_{g} (\rho) \bigr) \, 
\big \lvert f (\rho) \big \rvert^{-1} \,}\,,
\quad d_{g} (0) =0.
\label{4.11}
\end{align}
All we have to do in order to implement the selfconsistent cutoff is to solve
this differential equation for $d_{g} (\rho)$. Then, by \eqref{4.7}, we
know $w (\rho)$ and, by \eqref{4.8}, the solution $g_{\mu \nu}$.

Eq.\ \eqref{4.11} is easily solved by a separation of variables. After
computing the integrals
\begin{align}
I_{W} (y) & \equiv \int \limits_{0}^{y} 
\frac{\text{d} y^{\prime}}{\sqrt{W \bigl( y^{\prime} \bigr)\,}\,},
\label{4.12}
\\ 
I_{f} (\rho) 
& \equiv
\int \limits_{0}^{\rho}
\frac{\text{d} \rho^{\prime}}{\sqrt{\big \lvert 
f (\rho^{\prime}) \big \rvert \,}\,}
\label{4.13}
\end{align}
it remains to solve the equation
$I_{W} \bigl( d_{g} (\rho) \bigr) = I_{f} (\rho)$ for $d_{g} (\rho)$:
\begin{align}
d_{g} (\rho) & = I_{W}^{-1} \bigl( I_{f} (\rho) \bigr).
\label{4.15}
\end{align}

Note that $I_{W}$ depends only on the RG trajectory, while $I_{f}$ requires
information about $\gamma_{\mu \nu}$ only. With \eqref{4.3b} the latter
integral is elementary. For $\rho > 2 \overline{G} M$:
\begin{align}
\begin{split}
I_{f} (\rho) & =
\pi \, \overline{G} M 
+ 2 \overline{G} M \, \ln \left[
\sqrt{\rho / \left( 2 \overline{G} M \right) \,}
+ \sqrt{\rho / \left( 2 \overline{G} M \right) -1 \,} \,\right]
\\
& \phantom{{==}}
+ \sqrt{\rho \, \left( \rho - 2 \overline{G} M \right) \,}.
\end{split}
\label{4.16}
\end{align}
The asymptotic behavior for $\rho$ large is
\begin{align}
I_{f} (\rho) =
\rho + \overline{G} \, M \, \ln \left( \rho / \rho_{0} \right)
+ const + \mathcal{O} \Bigl( \left( \overline{G}\,M \right)^{2} / \rho \Bigr)
\label{4.17}
\end{align}
with an irrelevant constant $\rho_{0}$.

Our final result, the line element \eqref{4.8}, while describing a static
isotropic spacetime is not in the standard ``$AB$-form'' \eqref{3.1}.
For some applications it is advantageous to analyze it in the
$(t,\rho,\theta,\varphi)$-system of coordinates as it stands. For others it
is better to convert it to the $AB$-form by an appropriate change of 
coordinates. It is sufficient to replace $\rho$ with the new radial variable
$r$ defined as
\begin{align}
r (\rho) \equiv \rho \, \sqrt{w (\rho) \,}.
\label{4.18}
\end{align}
Assuming that the relationship $r = r (\rho)$ can be inverted to obtain
$\rho = \rho (r)$, it is straightforward to rewrite \eqref{4.8} in the
$(t,r,\theta,\varphi)$-system:
\begin{align}
\text{d} s^{2} =
- w \, f \, \text{d} t^{2}
+ \frac{\text{d} r^{2}}{f \, \left[
1 + \frac{\rho}{2} \, \partial_{\rho} \, \ln w \right]^{2}\,}
+ r^{2} \, \text{d} \sigma^{2}.
\label{4.19}
\end{align}
The components of the metric \eqref{4.19} are to be evaluated at 
$\rho = \rho (r)$.

Summarizing the results of this section we can say that vacuum solutions of 
Class II, with a selfconsistent cutoff identification, can be obtained by
means
of the following algorithm.
\begin{description}
\item [Step 1: ]Select a trajectory $G (k)$ and write down the associated
function $W (y)$ of \eqref{4.5}.
\item [Step 2: ]Perform the integral \eqref{4.12} to obtain $I_{W}$ and find
the inverse function $I_{W}^{-1}$.
\item [Step 3: ]Write down $d_{g} (\rho)$ according to \eqref{4.15}. Use it
to construct $w (\rho) = W \bigl( d_{g} (\rho) \bigr)$.
\end{description}

After having completed Step 3 we know the final result for the metric in the
$(t,\rho,\theta,\varphi)$-system, eq.\ \eqref{4.8}. Only at this
point we learn what is $G$ as a function of $\rho$. Since $G (\rho)
=G \bigl( k = \xi / d_{g} (\rho) \bigr) = \overline{G} \,
W \bigl( d_{g} (\rho) \bigr)$ we have
\begin{align}
G (\rho) = \overline{G} \, w (\rho).
\label{4.20}
\end{align}
This is the position dependence of Newton's constant which, by means of the
selfconsistent cutoff identification, gets associated to the trajectory
$G = G (k)$ in a dynamical way. As we shall see, the interrelation between
$G (k)$ and $G (\rho)$ can be highly nontrivial and could not be guessed
beforehand.

If one needs the metric in standard form one continues with
\begin{description}
\item [Step 4: ]Define $r (\rho) \equiv \rho \, \sqrt{w (\rho)\,}$
and find the inverse function $\rho = \rho (r)$.
\item [Step 5: ]Read off the components of the $AB$-metric from \eqref{4.19},
\begin{subequations} \label{4.21}
\begin{align}
A (r) & = f (\rho)^{-1} \,
\left[ 1 + \frac{\rho}{2} \, \partial_{\rho} \ln w (\rho) \right]^{-2}
\bigg \rvert_{\rho=\rho (r)},
\label{4.21a}
\\
B (r) & = w \bigl( \rho (r) \bigr) \, f \bigl( \rho (r) \bigr),
\label{4.21b}
\end{align}
\end{subequations}
and express Newton's constant as a function of $r$:
\begin{align}
G (r) = \overline{G} \, w \bigl( \rho (r) \bigr)
= \overline{G} \, \left( \frac{r}{\rho (r)} \right)^{2}.
\label{4.22}
\end{align}
\end{description}

It can be verified explicitly that, for every given RG trajectory $G = G (k)$,
eqs.\ (\ref{4.21}\,a,b) along with \eqref{4.22} do indeed constitute
a solution to all components of Einstein's equation, 
eqs.\ (\ref{3.14}\,a,b,c) 
and the consistency condition \eqref{3.15} for the case
$T_{\mu \nu} =0$, $\Lambda =0$, and 
$\theta_{\mu \nu}=\theta_{\mu \nu}^{\text{ BD}}$. The calculation is quite 
lengthy and tedious. It demonstrates explicitly that, as it was to be
expected \cite{h1}, the algorithm works only for $\varepsilon =1$, i.\,e.\ for
the Brans-Dicke $\theta$-tensor.
%
%
%
%
%
%
%
%
\section{\label{5}The IR fixed point scenario}
\subsection{\label{s5.1}Scale dependence of $\boldsymbol{G}$}
In this section we apply the algorithm for obtaining Class II vacuum solutions
developed in the previous section to the scale dependence
\begin{align}
G (k) = \overline{G} + g_{*}^{\text{IR}} / k^{2}.
\label{5.1}
\end{align}
As we discussed already in the Introduction, it interpolates between
classical gravity $G (k) \equiv \overline{G}$ for large $k$, and the
fixed point behavior $G (k) \propto 1 / k^{2}$ in the IR ($k \to 0$),
the transition taking place near the scale
$k_{\text{tr}}^{2} = g_{*}^{\text{IR}} / \, \overline{G}$ at which the
two terms on the RHS of \eqref{5.1} are exactly equal.
The choice \eqref{5.1} is motivated by its phenomenological success in
cosmology \cite{cosmo2,elo}.
\subsection{\label{s5.2}Class II vacuum solutions with selfconsistent cutoff
identification}
Let us now follow the steps listed at the end of Section \ref{4}.
According to \eqref{4.5}, the trajectory \eqref{5.1} amounts to
\begin{align}
W (y) = 1 + \kappa^{2} \, y^{2} = 1 + \frac{y^{2}}{L^{2}\,}
\label{5.2}
\end{align}
with
\begin{align}
\kappa^{2} \equiv \frac{1}{L^{2}\,}
\equiv \frac{g_{*}^{\text{IR}}}{\xi^{2} \, \overline{G} \,}.
\label{5.3}
\end{align}
Here $L$ is the proper distance $d_{g}$ at which the fixed point behavior sets
in, and $\kappa = L^{-1}$ denotes the corresponding mass scale. For the
function \eqref{5.2} the integral $I_{W}$ of \eqref{4.12} can be performed
in closed form:
\begin{align}
I_{W} (y) = \kappa^{-1} \, \text{arsinh} \left( \kappa \, y \right)
= \kappa^{-1} \, \ln \left[ \kappa \, y +
\sqrt{1+ \left( \kappa \, y \right)^{2} \, } \, \right].
\label{5.4}
\end{align}

The next step consists in finding $d_{g} (\rho)$ from the implicit equation
$I_{W} \bigl( d_{g} (\rho) \bigr) = I_{f} (\rho)$. For \eqref{5.4}
this is easily done:
\begin{align}
d_{g} (\rho) = \kappa^{-1} \, \sinh \bigl( \kappa \, I_{f} (\rho) \bigr).
\label{5.5}
\end{align}
Recalling eq.\ \eqref{4.16} for $I_{f}$, we see that $d_{g} (\rho)$ is known
in completely explicit form now. By \eqref{4.7}, it implies the Weyl factor
$w(\rho) \equiv W \bigl( d_{g} (\rho) \bigr)
= 1 + \bigl[ \kappa \, d_{g} (\rho) \bigr]^{2}$, or,
\begin{align}
w (\rho) = \cosh^{2} \bigl( \kappa \, I_{f} (\rho) \bigr).
\label{5.6}
\end{align}

Thus we are in a position to write down the final result for the metric and 
the position dependence of $G$. From \eqref{4.8} and \eqref{4.20} we obtain,
respectively,
\begin{subequations} \label{5.7}
\begin{align}
\text{d} s^{2} =
\cosh^{2} \bigl( \kappa \, I_{f} (\rho) \bigr) \,
\Bigl[ - f (\rho) \, \text{d} t^{2}
+ f (\rho) ^{-1} \, \text{d} \rho^{2}
+ \rho^{2} \, \text{d} \sigma^{2} \Bigr]
\label{5.7a}
\end{align}
and
\begin{align}
G (\rho) = \overline{G} \, \cosh^{2}  \bigl( \kappa \, I_{f} (\rho) \bigr).
\label{5.7b}
\end{align}
\end{subequations}
Actually \eqref{5.7} describes a one-parameter family of solutions labeled
by the parameter $M$ which enters via $f (\rho) = 1- 2 \overline{G}
M / \rho$.

Our construction includes $M=0$ as a special case. Here $I_{f} (\rho) = \rho$
entails the simplifications
\begin{subequations} \label{5.8}
\begin{align}
d_{g} (\rho) & = \kappa^{-1} \, \sinh \left( \kappa \, \rho \right),
\label{5.8a}
\\
w (\rho) & = \cosh^{2} \left( \kappa \, \rho \right),
\label{5.8b}
\\
G (\rho) & = \overline{G} \, \cosh^{2} \left( \kappa \, \rho \right),
\label{5.8c}
\\
\text{d} s^{2} &= \cosh^{2} \left( \kappa \, \rho \right) \,
\Bigl[ - \text{d} t^{2} + \text{d} \rho^{2} + \rho^{2} \,
\text{d} \sigma^{2} \Bigr].
\label{5.8d} 
\end{align}
\end{subequations}
Eqs.\ \eqref{5.8} describe a curved, but conformally flat, stationary and
isotropic spacetime. At the origin ($\rho=0$), Newton's constant has the
value $\overline{G}$, and at distances $\rho \gtrsim L$ it starts growing
very rapidly in an exponential fashion. Since the conformal factor $w$ grows
likewise without any bound, the spacetime is not asymptotically flat.

Eq.\ \eqref{5.8a} shows that for $\rho \gtrsim L$ the proper length $d_{g}$
differs significantly from the coordinate ``distance'' $\rho$. This explains the
a priori completely unexpected $\rho$-dependence of $G$, eq.\ \eqref{5.8c}.
It is the result of the selfconsistent cutoff identification which determines
$k = k (\rho)$ dynamically. With a naive cutoff employing the distance
function $d_{\gamma} (\rho) = \rho$ of the Minkowski metric $\gamma_{\mu \nu}$
we would obtain $G (\rho) = \overline{G} \, \left[ 1 + \kappa^{2} \, 
\rho^{2} \right]$ instead. This coincides with \eqref{5.8c} for $\rho \ll L$,
but differs substantially for $\rho \gg L$. In the latter regime it would
make no sense to use the naive cutoff because the renormalization effects
are strong, $w (\rho)$ is large, and $d_{\gamma} (\rho)$ no longer has
anything to do with the physical distance.

For $M>0$ the solutions \eqref{5.7} have a similar structure, with an 
additional distortion due to the curvature of the Schwarzschild spacetime.
It becomes strong for $\rho \to 0$, but in this regime
our model is anyhow 
inapplicable because the trajectory \eqref{5.1} does not ``know'' about
the UV running of $G (k)$. In the following we shall analyze the solution
\eqref{5.7}
only far away from the horizon ($\rho \gg 2 \overline{G} M$). Typical
question about black hole physics (structure of horizons, nature of the
central singularity) must involve a more complete description of $G (k)$ and
will be addressed  elsewhere.

Let us return to the general solution for $M \geq 0$. In \eqref{5.7} it is
written in the $(t,\rho,\theta,\varphi)$-system of coordinates. Introducing
the new radial coordinate $r = \rho \, \sqrt{w (\rho) \,}$ we can transform
it to the standard $AB$-form, but unfortunately it is not possible to
obtain a closed formula for $\rho$ as a function of $r$.
Eqs.\ \eqref{4.20}, \eqref{4.21} yield
\begin{subequations} \label{5.9}
\begin{align}
A (r) & =
\left[ \sqrt{1 - \frac{2 \, \overline{G} \, M}{\rho (r)}\,}
+ \kappa \, \rho (r) \, \sqrt{1 - \left( \frac{\rho (r)}{r} \right)^{2}\,}\,
\right]^{-2},
\label{5.9a}
\\
B (r) & =
\left[ 1 - \frac{2 \, \overline{G} \, M}{\rho (r)} \right] \,
\left( \frac{r}{\rho (r)} \right)^{2},
\label{5.9b}
\\
G (r) & =
\overline{G} \, \left( \frac{r}{\rho (r)} \right)^{2}.
\label{5.9c}
\end{align}
The function $\rho (r)$ is defined by the transcendental
equation
\begin{align}
r = \rho (r) \, \cosh \Bigl[ \kappa \, I_{f} \bigl( \rho (r) \bigr) \Bigr].
\label{5.9d}
\end{align}
\end{subequations}

In order to get a first understanding of the physics behind \eqref{5.9}
we are now going to analyze the regimes $2 \overline{G} M \ll \rho
\ll L$ and $2 \overline{G} M \ll L \ll \rho$ in turn.
\subsection{\label{s5.3}The modified Newtonian regime: $\boldsymbol{2 \, \overline{G} \, 
M \ll \rho \ll L}$}
We consider the portion of spacetime where $\rho$ is much larger than its
value at the Schwarzschild horizon, $2 \overline{G} M$, and at the same
time much smaller than the characteristic scale $L$. In the application to
galaxies which we have in mind, $2 \overline{G} M$ is basically negligible
compared to $L$ so that $\rho$ can vary over many orders of magnitude
without violating $2 \overline{G} M \ll \rho \ll L$. Because here
$2 \overline{G} M / \rho \ll 1$, as in the Newtonian limit of the
Schwarzschild metric, we call this regime ``modified Newtonian''. What we are
interested in are the modifications due to the running of $G$.

Let us expand the solution \eqref{5.9}, with $I_{f} (\rho)$ given by
\eqref{4.17}, for small values of
\begin{align}
\frac{\overline{G} \,M}{\rho},~
\frac{\overline{G} \, M}{\rho} \, \ln \rho,~
\left( \frac{\rho}{L} \right)^{2}
\ll 1.
\label{5.10}
\end{align}
We retain only terms linear in the quantities \eqref{5.10} and neglect higher
powers and cross terms. In leading order one has 
$\rho (r) = r \, \left[ 1 - \tfrac{1}{2} \, \kappa^{2} \, r^{2} + \cdots
\right]$, and \eqref{5.9} yields
\begin{subequations} \label{5.11}
\begin{align}
A (r) & = 1 + \frac{2 \, \overline{G} \, M}{r} - 2 \, \kappa^{2} \, r^{2},
\label{5.11a}
\\
B (r) & = 1 - \frac{2 \, \overline{G} \, M}{r} + \kappa^{2} \, r^{2},
\label{5.11b}
\\
G (r) & = \overline{G} \, \Bigl[ 1 + \kappa^{2} \, r^{2} \Bigr].
\label{5.11c}
\end{align}
\end{subequations}
Since $\rho$, $d_{g}$, and $r$ differ only slightly, $r$ is basically the same
as the physical distance so that \eqref{5.11} is valid for
$2 \overline{G} M \ll r \ll L$.

In order to interpret the above result and similar metrics which we
shall encounter later on let us recall a well-known fact about the 
Newtonian limit \cite{weinbook}. Assume we are given a metric of the form
\begin{align}
\text{d} s^{2} = - \left( 1 + 2 \, \phi \right) \, \text{d} t^{2}
+ \bigl( \delta_{ij} + h_{ij} \bigr) \, \text{d} x^{i} \, \text{d} x^{j},
\label{5.12}
\end{align}
valid to first order in $\phi$ and $h_{ij}$. The timelike geodesics
of test particles are obtained from the mechanical action
$S_{\text{m}} = - m \, \int \!\! \sqrt{- \text{d} s^{2} \,}
= - m \, \int \!\! \text{d} t~ \left[ 1 + 2 \, \phi - \left( \delta_{ij}
+ h_{ij} \right) \, \dot x^{i} \, \dot x^{j} \right]^{1/2}$
where $\dot x^{i} \equiv \text{d} x^{i} / \text{d} t$. 
In the Newton limit one assumes that the particle moves at nonrelativistic
velocities $v^{2} \equiv \delta_{ij} \, \dot x^{i} \, \dot x^{j} \ll 1$,
and that $\phi$, $h_{ij}$, and $v^{2}$ are small of the same order. Expanding
$S_{\text{m}}$ to first order in those quantities leads to
$S_{\text{m}} = \int \!\! \text{d} t~\big \{ \tfrac{1}{2} \, m \, v^{2}
- m \, \phi \big \}$. This shows that the function $\phi$ in \eqref{5.12}
is indeed the Newtonian potential, and that the Newton limit is insensitive
to the spatial components $h_{ij}$ since $h_{ij} \, \dot x^{i} \, \dot x^{j}$
is a second-order term.

By \eqref{5.11}, the Newtonian limit of our static isotropic spacetimes is
described by the line element
\begin{align}
\text{d} s^{2} 
& =
- \left[ 1 - \frac{2 \, \overline{G} \, M}{r} + \kappa^{2} \, r^{2} \right]
\, \text{d} t^{2}
+ \left[ 1 + \frac{2 \, \overline{G} \, M}{r} - 2 \, \kappa^{2} \, r^{2} 
\right] \, \text{d} t^{2}
+ r^{2} \, \text{d} \sigma^{2}.
\label{5.13}
\end{align}
According to the above discussion, it implies the following modified
Newtonian (``mN'') potential:
\begin{align}
\phi_{\text{mN}} (r) = - \frac{\overline{G} \, M}{r} 
+ \tfrac{1}{2} \, \kappa^{2} \, r^{2}
\label{5.14}
\end{align}
While $\phi_{\text{mN}}$ shows the familiar $1/r$-behavior at small distances,
it becomes positive and grows proportional to $r^{2}$ at large distances.
The term $\propto r^{2}$ is a direct consequence, and in fact the only one
in the Newtonian limit, of the IR fixed point running of $G$. It has a
rather dramatic consequence: at sufficiently large distances, the force two
masses exert on each other grows linearly with their distance. This behavior
is reminiscent of the confinement forces among quarks in QCD.

What is the range of validity of the (modified) Newtonian approximation? Is
it really applicable to a regime of radii where the $r^{2}$-behavior is already fully
developed, overriding the classical $1/r$-term in \eqref{5.14}? The answer
is indeed in the affirmative. The reason is the simple but very important
observation that the nonclassical (``nc'') behavior already sets in at a 
radius $r_{\text{nc}}$ which is much smaller than the scale $L$ (where the
approximation breaks down).

Let us define $r_{\text{nc}}$ by the condition
that the two terms on the RHS of \eqref{5.14} are exactly equal there:
$\overline{G} \, M / r_{\text{nc}} = \kappa^{2} \, r_{\text{nc}}^{2} / 2$.
For $r \lesssim r_{\text{nc}}$ the classical term is dominant, while the
$r^{2}$-term prevails for $r \gtrsim r_{\text{nc}}$. Writing
$r_{\text{S}} \equiv 2 \overline{G} M$ for the Schwarzschild radius,
the transition to the nonclassical $r^{2}$-regime takes place at
\begin{align}
r_{\text{nc}} = \bigl( r_{\text{S}} \, L^{2} \bigr)^{1/3}.
\label{5.15}
\end{align}
As a consequence, the ratio $r_{\text{nc}} / L$ is
\begin{align}
\frac{r_{\text{nc}}}{L} =
\left( \frac{r_{\text{S}}}{L} \right)^{1/3} \ll 1.
\label{5.16}
\end{align}
In fact, the modified Newtonian limit applies when
$r_{\text{S}} \ll r \ll L$, so that $r_{\text{S}} \lll L$ in particular.
Thus we conclude that there is a wide range of distances,
$r_{\text{nc}} \lesssim r \ll L$, where the nonclassical behavior dominates
$\phi_{\text{mN}}$ without being in conflict with the approximations made.

The natural emergence of a ``hierarchy'' $r_{\text{nc}} \ll L$ is a very
important feature. For instance, when we apply this theory to galaxies
later on $r_{\text{nc}}$ should be a typical galactic scale while $L$ is of
the order of the Hubble radius.
\subsection{\label{s5.4}The fixed point regime: $\boldsymbol{2 \, \overline{G} \, M
\ll L \ll \rho}$}
In this section we analyze the solution \eqref{5.7} or, equivalently,
\eqref{5.9} at asymptotically large distances from the center.
As the Schwarzschild metric is asymptotically flat, we can approximate
$I_{f} (\rho) \approx \rho$ for $\rho \to \infty$. As a result, all terms
containing $M$ drop out, and it is sufficient to study the $\kappa \, \rho
\to \infty$-behavior of the $M=0$-solution, eqs.\ \eqref{5.8}.
In this limit we can easily transform the metric to the standard $AB$-form.
Since $\cosh \left( \kappa \, \rho \right) \approx \exp \left( \kappa \,
\rho \right) / 2$ for $\kappa \, \rho \to \infty$, the implicit equation
for $\rho = \rho (r)$, eq.\ \eqref{5.9d}, becomes
$2 \, r = \rho (r) \, \exp \bigl[ \kappa \, \rho (r) \bigr]$.
It can be solved in terms of the Lambert W-function\footnote{By definition
\cite{corless}, the Lambert W-function satisfies $\mathcal{W}_{\text{L}} (x) \,
\exp \bigl[ \mathcal{W}_{\text{L}} (x) \bigr] = x$. In our notation, 
$\mathcal{W}_{\text{L}}$ stands for its real branch analytic at $x=0$, denoted
$W$ or $W_{0}$ in ref.\ \cite{corless}.} $\mathcal{W}_{\text{L}}$:
\begin{align}
\rho (r) = \kappa^{-1} \, \mathcal{W}_{\text{L}} \left( 2 \, \kappa \, r \right),
\quad \kappa \, r \gg 1.
\label{5.18}
\end{align}
Hence the asymptotic form of the solution, in the 
$(t,r,\theta,\varphi)$-system, reads
\begin{subequations} \label{5.19}
\begin{align}
A (r) & =
\bigl[ 1 + \mathcal{W}_{\text{L}} \left( 2 \, \kappa \, r \right) \bigr]^{-2},
\label{5.19a}
\\
B (r) & =
\left[ \frac{\kappa \, r}{\mathcal{W}_{\text{L}} \left( 2 \, \kappa \, r \right)}
\right]^{2},
\label{5.19b}
\\
G (r) & =
\overline{G} \, \left[ \frac{\kappa \, r}{\mathcal{W}_{\text{L}} 
\left( 2 \, \kappa \, r \right)} \right]^{2}.
\label{5.19c}
\end{align}
\end{subequations}
For $x \to \infty$ there exists an asymptotic expansion of 
$\mathcal{W}_{\text{L}}$ in terms of iterated logs \cite{corless}:
\begin{align}
\mathcal{W}_{\text{L}} (x) = \ln x - \ln \ln x + \frac{\ln \ln x}{\ln x}
+ \cdots.
\label{5.20}
\end{align}
Therefore, at \textit{very} large distances, $B (r)$ grows proportional
to $r^{2} / \ln^{2} \left( 2 \, \kappa \, r \right)$. The asymptotic spacetime
geometry is strongly curved therefore. This fairly exotic regime will not be relevant
in the context of galaxy haloes.
\subsection{\label{s5.5}Geodesics}
In order to gain a first understanding of the dynamics implied by the metric
$g_{\mu \nu}$ of eqs.\ \eqref{5.9} it is instructive to investigate its
timelike geodesics, the trajectories of massive test particles. Here we
concentrate on periodic circular orbits at constant radius $r$ or $\rho$.

In the Newtonian limit the analysis is very simple. The modified Newtonian
potential $\phi_{\text{mN}}$ allows for circular orbits at any radius $r$.
On those orbits the centrifugal force exactly balances the gravitational
attraction provided the velocity has the time independent value
\begin{align}
v^{2} (r) = r \, \phi_{\text{mN}}^{\prime} (r)
= \frac{\overline{G} \, M}{r} + \kappa^{2} \, r^{2}.
\label{5.21}
\end{align}
At small radii $r \lesssim r_{\text{nc}}$, the classical term
$\overline{G} \, M / r$ dominates, yielding $v (r) \propto 1 / \sqrt{r\,}$
for the velocity and Kepler's $T (r) \propto  r^{3/2}$ for the period of
the orbit, $T (r) \equiv 2 \pi \, r / v (r)$. Instead, for
$r \gtrsim r_{\text{nc}}$, the nonclassical term becomes dominant, giving rise
to a linearly increasing velocity $v (r) = \kappa \, r$ and
a period $T (r) = 2 \pi / \kappa = const$ which is independent of $r$.

Note that there exist periodic orbits even if $M=0$. In this case they owe
their existence entirely to the $r$-dependence of $G$ since the
solution-generating metric $\gamma_{\mu \nu}$ with $M=0$ (Minkowski metric) 
does not give rise to periodic orbits.

Going beyond the Newtonian limit, the analysis is most easily performed using
the radial coordinate $\rho$ rather than $r$. In Appendix \ref{B} we derive the
geodesic equation for a metric of the type \eqref{4.8} with arbitrary functions
$w (\rho)$ and $f (\rho)$. We find that it admits a circular orbit at
coordinate radius $\rho$ provided the quantity
\begin{align}
\left( \frac{T (\rho)}{2 \pi \, \rho} \right)^{2} =
\frac{\frac{2}{\rho} + \frac{w^{\prime}}{w}}{f^{\prime}
+ \frac{w^{\prime}}{w} \, f}
\label{5.22}
\end{align}
is positive. If so, $T (\rho)$ is the period of the orbit, measured in units
of the coordinate time $t$.

In the IR fixed point scenario the Weyl factor $w$ of the exact
metric is given by \eqref{5.6}
and $f$ is of the Schwarzschild form \eqref{4.3b}. Inserting these functions
into the equation \eqref{5.22}
one finds that its RHS is indeed positive for any $\rho > 0$. Hence the exact
solution admits  a circular orbit at every radius.

A simple, closed-form expression for the period $T (\rho)$ can be obtained for
$M=0$ only. In this case, according to \eqref{5.8}, $f \equiv 1$ and
$w (\rho) = \cosh^{2} \left( \kappa \, \rho \right)$ which yields when 
inserted into \eqref{5.22}:
\begin{align}
T (\rho) = 2 \pi \, \rho \, \left[ 1 + \frac{\coth \left( \kappa \, \rho
\right)}{\left( \kappa \, \rho \right)} \right]^{1/2}.
\label{5.23}
\end{align}

The $M=0$-spacetime \eqref{5.8d} is regular at $\rho =0$, and the Weyl factor
equals unity there. Therefore the coordinate time $t$ equals the time
indicated by a clock which is at rest at $\rho =0$. Correspondingly,
$T (\rho)$ is the period as measured by this specific clock. We saw already
that for $\rho \to \infty$ the spacetime is strongly curved and that the
proper distance to the origin, $d_{g} (\rho) \equiv y$, is exponentially
large compared to the coordinate distance $\rho$. In fact, by \eqref{5.8a},
$\kappa \, y = \sinh \left( \kappa \, \rho \right)$. It is more physical,
therefore, to express $T$ in terms of $y$ rather than $\rho$:
\begin{align}
T (y) & =
\frac{2 \pi}{\kappa} \,
\left[ \text{arsinh}^{2} \left( \kappa \, y \right)
+ \text{arsinh} \left( \kappa \, y \right) \,
\sqrt{1 + \left( \kappa \, y \right)^{-2}\,}\, \right]^{1/2}.
\label{5.24}
\end{align}
At small distances $y \ll \kappa^{-1} \equiv L$, this function has a power
series expansion 
$T (y) = 2 \pi \, \kappa^{-1} \, \left[ 1 + \tfrac{2}{3} \, \left( \kappa \, y
\right)^{2} + \mathcal{O} \bigl( (\kappa \, y)^{4} \bigr) \right]$ which,
for $y \to 0$, reproduces the constant period found in the Newtonian limit:
$T \approx 2 \pi \, \kappa^{-1}$. Conversely, at extremely large distances
$y \ggg \kappa^{-1}$, one finds a growth of $T (y)$ which is approximately
logarithmic:
\begin{align}
T (y) \approx \frac{2 \pi}{\kappa} \, \ln \left( 2 \, \kappa \, y \right).
\label{5.25}
\end{align}

As we mentioned in Subsection \ref{s5.4} already, all terms containing the
parameter $M$ are negligible in the fixed point regime
$2 \overline{G} M \ll L \ll \rho$. As a consequence, for $y \to \infty$,
also the solutions with $M \neq 0$ give rise to the 
asymptotic behavior \eqref{5.25}.
\subsection{\label{s5.6}Cosmological fixed point running at galactic scales?}
In this subsection we address the question of whether the non-Gaussian
IR fixed point which was invoked in refs.\ \cite{cosmo2,elo} in order to
describe the cosmology of the late Universe might also be responsible for
the non-Keplerian behavior of the observed galaxy rotation curves $v (r)$.
The model trajectory \eqref{5.1} describes ``ordinary''
gravity with $G = \overline{G} = const$ at small distances,
i.\,e.\ at the scale of laboratory or solar system measurement, and the
IR fixed point running at cosmological scales. The typical scale of
galaxies, $\unit[1]{kpc}$ say, is in between those two regimes. The
intriguing question is whether the unexpected features of their rotation 
curves can be interpreted as due to the onset of the cosmological fixed point
running. If so, the parameter $g_{*}^{\text{IR}}$ appearing in our present
ansatz \eqref{5.1} must have the same numerical value as in cosmology
since one and the same RG fixed point would govern both cosmology and
galaxies.

In the extended IR fixed point model proposed in ref.\ \cite{elo} the fixed
point era proper, with $G (k) \propto 1 / k^{2}$, is preceded by a standard
Friedmann-Robertson-Walker epoch with $G = const$, the transition taking
place at a redshift $z_{\text{tr}}$. In this model, $g_{*}^{\text{IR}}$
can be expressed in terms of $z_{\text{tr}}$ and the present Hubble radius
$l_{\text{H}} (t_{0})$\footnote{Eq.\ \eqref{5.26} obtains by inserting 
eqs.\ (17) and (51) of ref.\ \cite{elo} into eq.\ (29) of \cite{elo}.}:
\begin{align}
g_{*}^{\text{IR}} = \frac{9}{16} \, \widehat \xi^{~2}
\, \bigl( 1 + z_{\text{tr}} \bigr)^{3/2} \,
\left( \frac{l_{\text{Pl}}}{l_{\text{H}} (t_{0})} \right)^{2}.
\label{5.26}
\end{align}
Here $l_{\text{Pl}} \equiv \sqrt{\overline{G}\,}$ is the Planck length
defined in terms of the Newton constant $\overline{G}$ measured in the
laboratory. The constant $\widehat \xi$ relates $k$ to the cosmological
cutoff $1/t$ and may be assumed of order unity therefore. Moreover,
confronting the model with various cosmological standard tests \cite{elo}
it turns out that $z_{\text{tr}}$ cannot be much larger than unity.
This information allows us to estimate the scale $L \equiv 1 / \kappa$.
Using \eqref{5.26} in \eqref{5.3} we obtain
\begin{align}
L = \frac{4}{3} \, \frac{\xi}{\,\widehat \xi\,} \,
\bigl( 1 + z_{\text{tr}} \bigr)^{-3/4} \, l_{\text{H}} (t_{0}).
\label{5.27}
\end{align}
As an order of magnitude estimate we can say that $L$ equals approximately
the present Hubble radius since the prefactor of $l_{\text{H}} (t_{0})$
on the RHS of \eqref{5.27}, while not known exactly, is certainly of order
unity. For simplicity we shall assume in the following that
$L = l_{\text{H}} (t_{0})$; hence $\kappa = H_{0}$ is presicely the present
Hubble parameter, $H_{0} \equiv 1 / l_{\text{H}} (t_{0})$.

This result has an important implication when it comes to applying our
solution of the modified Einstein equation to the galaxy problem.
We had analyzed it in the modified Newtonian regime $\rho \ll L$
and for $\rho \gg L$ where the fixed point behavior is already fully
developed. With $L$ equal to the Hubble radius the latter regime is of no
relevance for the structure of galaxies. We conclude that the portion of
the spacetime which is interesting regarding the problem of the
rotation curves lies entirely within the Newtonian regime $\rho \ll L$.

In this regime the modified potential of a point mass $M$ is given by
\eqref{5.14}; its nonclassical term grows proportional to $r^{2}$,
implying a linearly increasing force and
\begin{align}
v (r) = \kappa \, r = H_{0} \, r.
\label{5.27-1}
\end{align}

We mentioned already that the scale $L$ may not be confused with the
distance $r_{\text{nc}}$ at which the nonclassical force $\propto r$ starts
dominating Newton's $1 / r^{2}$-force. Eq.\ \eqref{5.16} tells us that
$r_{\text{nc}} / L = \left( 2 \, \overline{G} \, M \,
H_{0} \right)^{1/3}$. To be explicit, let us consider a point mass with
$M = 10^{11} \, M_{\odot}$, the mass of a typical galaxy. Using
$H_{0} = \unit[70]{km \, sec^{-1} \, Mpc^{-1}}$ we find the ratio
\begin{align}
r_{\text{nc}} / L \approx 1.3 \cdot 10^{-4}
\label{5.28}
\end{align}
which amounts to
\begin{align}
r_{\text{nc}} \approx \unit[560]{kpc}.
\label{5.29}
\end{align}

Several comments are in order here. First of all it is gratifying to see
that our model is indeed capable of producing a hierarchy between the
cosmological scale $L$ and the (would-be) galactic scale $r_{\text{nc}}$.
However, quantitatively this hierarchy is still too small by about
2 orders of magnitude to be phenomenologically acceptable; rather than
\eqref{5.29}, we would like to find a $r_{\text{nc}}$ of the order
$\unit[1]{kpc}$, which is roughly the distance at which the plateau behavior
in the rotation curve of a typical spiral galaxy
sets in. In a real
galaxy the luminous matter has a complex spatial distribution while above
we were studying the field of a point mass. However, it appears unlikely
that this is the origin of the missing 2 orders of magnitude.
We rather take them as a hint indicating that, while qualitatively pointing
in the right direction, the RG trajectory \eqref{5.1} is probably not very
realistic at the kiloparsec scale.

An even stronger argument for rejecting this trajectory comes from the
rotation curve it gives rise to, eq.\ \eqref{5.27-1}. A linearly rising
$v (r)$ for a point mass can hardly be reconciled with the almost constant
velocities observed in galaxy haloes. Moreover, the predicted velocities
are substantially too small. In fact, \eqref{5.27-1} happens to have the
form of Hubble's law, and $H_{0} \, r$ is several orders of magnitude smaller
than the measured plateau velocities of a few hundred
$\unit{km/sec}$, of course.

The results of this section suggest that, at galactic scales, $G (k)$ should
have a milder growth for decreasing $k$ than it has in the IR fixed point
model. A realistic RG trajectory should entail a $\phi_{\text{mN}}$ which
increases with $r$ less steeply and makes $v (r)$ approximately constant
at large distances. At the same time it must produce higher orbital velocities.
%
%
%
%
%
%
%
%
\section{\label{6}Power law running}
\subsection{\label{s.61}The RG trajectory}
The discussion at the end of the previous section has indicated that an
infrared growth of $G (k)$ proportional to $1 / k^{2}$ is probably too strong
at the scale of galaxies. This motivates us to investigate the consequences
of the following model-trajectory with a milder $k$-dependence of $G (k)$.
We assume that $G (k)$ is approximately constant for large $k$, and that
it runs according to the power law $G (k) \propto 1 / k^{q}$ at small
$k$. Here $q$ is an arbitrary positive constant. Applying the model to
galaxies we shall find that the relevant range of $q$-values is 
$0 < q \ll 1$ so that the variation of $G (k)$ is indeed much weaker than
in the IR fixed point scenario.

For the time being we allow for arbitrary
$q>0$ and make the following ansatz:
\begin{align}
G (k) = 
\begin{cases}
\overline{G} & \text{for } k > k_{\text{tr}}
\\
\overline{G} \, \bigl( k_{\text{tr}} / k \bigr)^{q} & \text{for }
k \leq k_{\text{tr}}
\end{cases}
\label{6.1}
\end{align}
Here $\overline{G}$ and the transition scale $k_{\text{tr}}$ are free
constants. Note that $G (k)$ is continuous but not differentiable at
$k_{\text{tr}}$. Recalling that the dimensionful Newton constant evolves
according to $k \, \partial_{k} G = \eta_{\text{N}} \, G$, we see that
the model trajectory \eqref{6.1} amounts to a constant anomalous dimension
$\eta_{\text{N}} = -q$ for $k \leq k_{\text{tr}}$ and to $\eta_{\text{N}}
=0$ at scales $k > k_{\text{tr}}$. A more realistic RG trajectory would
interpolate smoothly between $\eta_{\text{N}} = -q$ and $\eta_{\text{N}}
=0$, but for our present discussion the idealization of a jump at some
$k_{\text{tr}}$ is good enough. Our main conclusions will be insensitive to
the intermediate regime interpolating between standard gravity
and the power law regime.
\subsection{\label{s6.2}Class II vacuum solutions}
Next we apply the Weyl-algorithm of Section \ref{4} to the $G (k)$ of
eq.\ \eqref{6.1} in order to find the vacuum solutions of Class II with a
selfconsistent cutoff identification. The first step consists in identifying
the function $W (y)$. From \eqref{6.1} we obtain
\begin{align}
W (y) =
\begin{cases}
1 & \text{for } y < L
\\
\left( \kappa \, y \right)^{q} & \text{for } y \geq L
\end{cases}
\label{6.2}
\end{align}
For $k_{\text{tr}}$ and $\xi$ fixed, we introduce the length scale $L$ and
the mass scale $\kappa \equiv L^{-1}$ by
\begin{align}
k_{\text{tr}} \equiv \frac{\xi}{L} \equiv \xi \, \kappa.
\label{6.3}
\end{align}
(The quantities $\kappa$ and $L$ are analogous to those of the previous 
section but not mathematically identical.) The cutoff identification 
$k (\rho) = \xi / d_{g} (\rho)$ provides us with the $\rho$-value 
$\rho_{\text{tr}}$ which corresponds to the transition scale $k_{\text{tr}}$;
from $k ( \rho_{\text{tr}}) = k_{\text{tr}}$ we obtain the condition
$d_{g} (\rho_{\text{tr}}) = L$.
Since $\xi$ is of order unity, $\kappa$ is essentially the same as
$k_{\text{tr}}$ and $L$ is its inverse. When measured with the physical 
metric $g_{\mu \nu}$, the length scale $L$ is the \textit{proper} length
corresponding to the \textit{coordinate} value $\rho_{\text{tr}}$.

For $W (y)$ from \eqref{6.2} the integral $I_{W}$ of \eqref{4.12} can be
done easily:
\begin{align}
I_{W} (y) =
\begin{cases}
y & \text{for } y<L
\\
L + \frac{L}{\left( 1 - q/2 \right)} \,
\left[ \left( \frac{y}{L} \right)^{1-q/2} -1 \right] & \text{for } 
y \geq L
\end{cases}
\label{6.5}
\end{align}
According to \eqref{4.15} we obtain $d_{g} (\rho) = I_{W}^{-1} \bigl(
I_{f} (\rho) \bigr)$ by inverting $I_{W}$. The result reads
\begin{align}
d_{g} (\rho) =
\begin{cases}
I_{f} (\rho) & \text{for } \rho < \rho_{\text{tr}}
\\
L \, \left[ \left( 1 - \frac{q}{2} \right) \,
\frac{I_{f} (\rho)}{L} + \frac{q}{2} \right]^{\frac{2}{2-q}} &
\text{for } \rho \geq \rho_{\text{tr}}
\end{cases}
\label{6.6}
\end{align}
Since $d_{g} = I_{f}$ for $\rho < \rho_{\text{tr}}$, the condition 
$d_{g} \bigl( \rho_{\text{tr}} \bigr) = L$
defining $\rho_{\text{tr}}$ simplifies:
\begin{align}
I_{f} (\rho_{\text{tr}}) = L.
\label{6.7}
\end{align}
Finally we can obtain the Weyl factor $w (\rho) = W \bigl( d_{g} (\rho) \bigr)$
by inserting \eqref{6.6} into \eqref{6.2}:
\begin{align}
w (\rho) =
\begin{cases}
1 & \text{for } \rho < \rho_{\text{tr}}
\\
\Bigl[ \left( 1 - \frac{q}{2} \right) \, \kappa \, I_{f} (\rho)
+ \frac{q}{2} \Bigr]^{\frac{2 \, q}{2-q}} & \text{for } \rho \geq
\rho_{\text{tr}}
\end{cases}
\label{6.8}
\end{align}

With the derivation of $w (\rho)$ we have completed the construction of a
one-parameter family of solutions to the modified Einstein equation with
$T_{\mu \nu}=0$ and $\Lambda =0$. The solutions are labeled by the mass $M$.
In the $(t,\rho,\theta,\varphi)$-system of coordinates, the metric is given
by \eqref{4.8} with \eqref{6.8} and the position dependence of Newton's constant by
$G (\rho) = \overline{G} \, w (\rho)$.
\subsection{\label{s6.3}Rotation curves}
In this subsection we study timelike geodesics of the metric with the
Weyl factor $w (\rho)$ of \eqref{6.8}. Again we consider
circular orbits only, being particularly interested in their rotation curves.
As for the very existence of circular orbits, we showed in Appendix \ref{B}
that for this class of metrics there exists an orbit at constant 
$\rho$ provided the RHS of
\begin{align}
\left( \frac{T (\rho)}{2 \pi \, \rho} \right)^{2} =
\frac{1 + Y (\rho)}{\left( \overline{G} M / \rho \right) + f (\rho) \,
Y (\rho)},
\label{6.9}
\end{align}
with $Y (\rho) \equiv \tfrac{1}{2} \, \rho \, w^{\prime} (\rho) / w (\rho)$,
is positive at this value of $\rho$. For $\rho < \rho_{\text{tr}}$ we have
$Y=0$ and the situation is the same as with the Schwarzschild metric: If
$M>0$ there are circular orbits, satisfying Kepler's law
$\left( T / 2 \pi \right)^{2} = \rho^{3} / \left( \overline{G} M \right)$,
and there are none if $M=0$. In the more interesting region $\rho >
\rho_{\text{tr}}$ the quantity $Y (\rho)$ is given by
\begin{align}
Y (\rho) = \frac{q \, \kappa \, \rho}{\bigl[ \left( 2-q \right) \,
\kappa \, I_{f} (\rho) + q \bigr] \, \sqrt{f (\rho) \,}\,}.
\label{6.10}
\end{align}
It can be shown that the RHS of \eqref{6.9} with \eqref{6.10} is a finite
positive quantity, both if $M>0$ and $M=0$. Hence there exist circular orbits
at all radii $\rho > \rho_{\text{tr}}$ and their periods are given by
$T (\rho)$. Such orbits exist even if, for $M=0$, the construction of the
solution $g_{\mu \nu}$ is based upon
the Minkowski metric which has no circular orbits. In this case
the gravitational attraction is entirely due to the gradient of Newton's
constant.

It is instructive to discuss the following special cases which can be dealt
with analytically.
\subsubsection*{(i) Asymptotically large radii}
Let us focus on orbits with very large radii $\rho \to
\infty$. In this limit, $f (\rho) \to 1$ and $I_{f} (\rho) / \rho \to 1$
which implies $Y (\rho) \to q / \left( 2-q \right)$. Using this
asymptotic behavior in \eqref{6.9} we see that
\begin{align}
\lim_{\rho \to \infty} \left[ \frac{T (\rho)}{2 \pi \, \rho} \right]^{2}
= \, \frac{2}{q}.
\label{6.11}
\end{align}

This is a very remarkable result. In fact, let us tentatively assume that
for $\rho \to \infty$ the curvature of spacetime is small and that the
test particles on their orbits move with velocities much smaller
than the speed of light. In this case a quasi-Newtonian interpretation is
possible which neglects the difference between proper and coordinate time
and likewise between the proper circumference of the orbit and the euclidean
$2 \pi \rho$. Therefore $\left( 2 \pi \rho \right) / T (\rho)
\equiv v (\rho)$ is ``the'' velocity of the test particle. In this
quasi-Newtonian limit, eq.\ \eqref{6.11} tells us that for $\rho \to \infty$
the velocity $v (\rho)$ approaches a finite, $\rho$-independent value
$v_{\infty} \equiv \underset{\rho \to \infty}{\lim} v (\rho)$:
\begin{align}
v_{\infty}^{2} = \tfrac{1}{2} \, q.
\label{6.12}
\end{align}

Thus we may conclude that, provided the quasi-Newtonian interpretation is
applicable, the scale dependence $G (k) \propto 1 / k^{q}$ does indeed give
rise to a rotation curve $v (\rho)$ with a plateau behavior asymptotically.
The plateau value $v_{\infty} = \sqrt{q/2\,}$ is a direct measure of the
exponent appearing in the power law. It does not
depend on $M$.

Clearly it is very tempting at this point to speculate that the observed
plateau behavior in typical rotation curves 
of spiral galaxies is due to a power law running
of $G$. For the Newtonian picture to apply it is necessary that
$v_{\infty} \ll 1$, which implies that the exponent $q$ must be much smaller
than unity: $q \ll 1$. It is gratifying to see that this condition is
indeed met by the ``test particles'' in the galactic haloes whose velocity was
measured \cite{combbook}. Typical values of $v_{\infty}$ range from
\unit[100]{} to \unit[300]{km/sec} so that, in units of the speed of
light, $v_{\infty} \approx 10^{-3}$. Hence, ignoring factors of order unity,
we find that the data require an exponent of the order
\begin{align}
q \approx 10^{-6}.
\label{6.13}
\end{align}
Later on in Subsection \ref{s6.4} we shall see that also the second
prerequisite of the Newtonian interpretation, small spacetime curvature,
is satisfied for $q$ as small as $10^{-6}$.
\subsubsection*{(ii) The case $\boldsymbol{M=0}$}
For non-asymptotic radii $\rho$ the rotation curve $v (\rho)$ can be obtained
in closed form only if $M=0$. In this special case we have $f (\rho) =1$,
$I_{f} (\rho) = \rho$, and \eqref{6.9} becomes
\begin{flalign}
& & & & 
\left( \frac{T}{2 \pi \, \rho} \right)^{2} =
\frac{2}{q} + \frac{1}{\kappa \, \rho}
& & \bigl( \rho \geq \rho_{\text{tr}} \bigr) \qquad
\label{6.14}
\end{flalign}
The RHS of this expression is manifestly positive, showing that there exist
circular orbits at any $\rho \geq \rho_{\text{tr}}$, with period
$T (\rho) = 2 \pi \rho \, \sqrt{\left( 2/q \right) + 1 / \left( \kappa \,
\rho \right)\,}$. Interpreting this result \`a la Newton
yields the velocity
\begin{align}
v^{2} (\rho) = \frac{q}{2} \, \frac{\kappa \, \rho}{\kappa \, \rho
+ \left( q / 2 \right)}.
\label{6.15}
\end{align}
For $\rho \to \infty$ it approaches the plateau value $v^{2}_{\infty} =
q/2$ from below.
\subsubsection*{(iii) First order approximation in $\boldsymbol{q}$
and $\boldsymbol{\overline{G} \, M / \rho}$}
In order to obtain an analytic, albeit approximate expression for
$v (\rho)$ in the generic case $M \neq 0$ we shall assume that
$q \ll 1$ and $\overline{G} M / \rho \ll 1$, the second condition
meaning that $\rho$ is much larger than the Schwarzschild radius.
In the halo of a typical galaxy both conditions are well satisfied.
In evaluating \eqref{6.9} we retain only terms of first order in $q$ and
$\overline{G} M / \rho$ and neglect all higher powers and cross terms.
The result is then
\begin{flalign}
& & & & 
v^{2} (\rho) = \frac{\overline{G} \, M}{\rho} + \frac{q}{2}
& & \bigl( \rho \geq \rho_{\text{tr}} \bigr) \qquad
\label{6.16}
\end{flalign}
This $v^{2} (\rho)$ approaches the plateau value $q/2$ from above.

If $\rho$ is small, the first term on the RHS of \eqref{6.16} dominates,
while the ``nonclassical'' contribution $q/2$ prevails for large values
of $\rho$. We introduce a radius $\rho_{\text{nc}}$ which marks the onset
of the nonclassical behavior. We define $\rho_{\text{nc}}$ to be the
radius where the two terms on the RHS of \eqref{6.16} have equal magnitude:
$\overline{G} M / \rho_{\text{nc}} = q/2$. Obviously the ``new physics''
sets in at
\begin{align}
\rho_{\text{nc}} = \rho_{\text{S}} / q
\label{6.17}
\end{align}
where $\rho_{\text{S}} \equiv 2 \overline{G} M$ is the Schwarzschild 
radius pertaining to $\gamma_{\mu \nu}$.

Since \eqref{6.16} holds for
$\rho \geq \rho_{\text{tr}}$ only, the estimate \eqref{6.17} is valid only
provided $\rho_{\text{nc}} \geq \rho_{\text{tr}}$.
However, it is remarkable that, as long as 
$\rho_{\text{tr}} < \rho_{\text{nc}}$, the value of $\rho_{\text{nc}}$ 
is completely independent of $\rho_{\text{tr}} $ and the scale 
$L \equiv d_{g} (\rho_{\text{tr}})$. Thus we can say that, at least when the
quasi-Newtonian interpretation makes sense, the distance scale where
the galactic plateau behavior sets in, typically 
$\unit[1]{kpc}$ in Nature, is quite unrelated to the ``cross-over scale''
$\kappa \equiv L^{-1}$ where the power law running of $G$ begins.
\subsection{\label{s6.4}Properties of the metric}
In order to fully establish the picture sketched in the previous subsection
we must justify the validity of the Newtonian approximation in the
``halo''. This is most easily done using the standard (``$AB$'') form
of the metric. At the end of Subsection \ref{s6.2} we found the 
Class II solutions pertaining to $G (k) \propto 1 / k^{q}$ in the
$\rho$-coordinate system, the Weyl factor which determines both $g_{\mu \nu}$
and $G (\rho)$ being given by eq.\ \eqref{6.8}. In general it is not
possible to express this solution in the $r$-coordinate system in closed
form. However, there are two instructive limiting cases where this is indeed
possible: asymptotically large radii and the first order approximation
in $q$ and $\overline{G} M / r$. Let us discuss those cases in turn.
\subsubsection*{(i) Asymptotically large radii}
Retaining for $\rho \to \infty$ only the leading-order terms one has
$f (\rho) = 1 + \cdots$, $I_{f} (\rho) / \rho = 1 + \cdots$, and from
\eqref{6.8} it follows that
\begin{align}
w (\rho) = \bigl[ \left( 1 - q/2 \right) \, \kappa \, \rho 
\bigr]^{2 \, q / \left( 2 - q \right)}.
\label{6.18}
\end{align}
Using this approximation for the Weyl factor, $r = \rho \, \sqrt{w (\rho)\,}$
yields the following relationship between the radial coordinates $r$ and
$\rho$:
\begin{align}
\rho = \bigl[ \left( 1 - q/2 \right) \, \kappa \bigr]^{-q/2} \,
r^{\left( 2-q \right) /2}.
\label{6.19}
\end{align}
Eqs.\ \eqref{6.18} and \eqref{6.19} imply
$w \bigl( \rho (r) \bigr) = \bigl[ \left( 1 - q/2 \right) \, 
\kappa \, r\bigr]^{q}$ and $\rho \, \partial_{\rho} \ln w =
2 \, q / \left( 2-q \right)$. As a result, \eqref{4.19} gives rise to
the following metric in the $(t,r,\theta,\varphi)$-system of coordinates,
valid for $r \to \infty$:
\begin{subequations} \label{6.20}
\begin{align}
\text{d} s^{2} = 
- \bigl[ \left( 1 - q/2 \right) \kappa \, r \bigr]^{q} \, \text{d} t^{2}
+ \left( 1 - q/2 \right)^{2} \, \text{d} r^{2}
+ r^{2} \, \text{d} \sigma^{2}.
\label{6.20a}
\end{align}
This solution describes the spacetime geometry due to the $r$-dependent
Newton constant
\begin{align}
G (r) = \overline{G} \, \bigl[ \left( 1 - q/2 \right) \kappa \bigr]^{q}
\, r^{q}.
\label{6.20b}
\end{align}
\end{subequations}

Note that in deriving the asymptotic forms (\ref{6.20}\,a,b)
we did not assume that $q$ is small, they are valid for all
$q \in \left( 0,2 \right)$. (As it should be, $r$ increases monotonically
with $\rho$ in this interval.)

The metric \eqref{6.20a} is the universal $r \to \infty$-asymptotics of all
Class II vacuum spacetimes, i.\,e.\ it does not depend on the parameter $M$.
This geometry is not asymptotically flat which, in fact, is not
too surprising since $G (r) \propto r^{q}$ grows unboundedly for
$r \to \infty$. The spatial part of \eqref{6.20a} is the metric of a cone;
the singular tip of the cone ($r=0$) lies outside the domain of validity of this
metric, however.

It is interesting to consider the asymptotic solution \eqref{6.20} for small
values of $q$: $0 < q \ll 1$. Expanding in powers of $q$ and neglecting
terms of order $q^{2}$ and higher yields
\begin{subequations} \label{6.21}
\begin{gather}
\text{d} s^{2} = 
- \bigl[ 1 + q \, \ln (\kappa \, r) \bigr] \, \text{d} t^{2}
+ \left( 1 - q \right) \, \text{d} r^{2}
+ r^{2} \, \text{d} \sigma^{2},
\label{6.21a}
\\
G (r) = \overline{G} \, \bigl[ 1 + q \, \ln (\kappa \, r) \bigr].
\label{6.21b}
\end{gather}
\end{subequations}

One must be careful about the range of validity of \eqref{6.21}. In order
to obtain this approximation we used the expansion
$\left( \kappa \, r \right)^{q} = \exp \bigl[ q \, \ln (\kappa \, r) \bigr] =
1 + q \, \ln ( \kappa \, r ) + 
\mathcal{O} \bigl( q^{2} \, \ln^{2} (\kappa \, r) \bigr)$
so that the actual expansion parameter is $q \, \ln (\kappa \, r)$ and not
$q$ itself. This approximation is reliable provided 
$q \, \ln (\kappa \, r) \ll 1$ or, equivalently, $r \ll L \, \exp (1/q)$.
Since on the other hand $\rho$, or what is the same if $q$ is small, $r$,
should be much larger than $L$, we see that \eqref{6.21} is valid in
the window
\begin{align}
L \ll r \ll L \, \exp (1/q).
\label{6.22}
\end{align}
For galaxies the upper bound on $r$ is of the order
$\exp (10^{6}) \, L$. Unless $L$ is anomalously small it is probably well
satisfied for all practical purposes so that we may regard
\eqref{6.21} as the asymptotic form of the solution.
\subsubsection*{(ii) First order approximation in $\boldsymbol{q}$
and $\boldsymbol{\overline{G} \, M / r}$}
Next we investigate the solution \eqref{6.8} for all
$\rho \geq \rho_{\text{tr}}$, including non-asymptotic radii. We expand
in $q \ll 1$ and $\overline{G} \, M / \rho \ll 1$ up to the first order
discarding higher powers and products of these quantities, as well as terms of
order $q^{2} \, \ln^{2} (\kappa \, \rho)$. In this approximation,
\begin{align}
w (\rho) = 1 + q \, \ln (\kappa \, \rho)
\label{6.23}
\end{align}
so that $r$ and $\rho$ are related by
\begin{align}
\rho = r \, \Bigl[ 1 - \tfrac{1}{2} \, q \, \ln (\kappa \, r) \Bigr].
\label{6.24}
\end{align}
Hence, to the order considered, $\overline{G} M / r \approx 
\overline{G} M / \rho$. Using \eqref{4.19} we obtain the following result
for the metric in the standard $r$-coordinate system:
\begin{subequations} 
\label{6.25}
\begin{align}
\text{d} s^{2} & =
- \left[ 1 - \frac{2 \, \overline{G} \, M}{r} 
+ q \, \ln (\kappa \, r) \right] \, \text{d} t^{2}
+ \left[ 1 + \frac{2 \, \overline{G} \, M}{r} -q \right] \, \text{d} r^{2}
+ r^{2} \, \text{d} \sigma^{2}.
\label{6.25a}
\end{align}
\end{subequations}
The corresponding position dependence of Newton's constant is given by
\begin{align}
G (r) = \overline{G} \, \bigl[ 1 + \mathcal{N} (r) \bigr]
\label{6.26}
\end{align}
with
\begin{align}
\tag{\ref{6.25}b}
\mathcal{N} (r) = q \, \ln (\kappa \, r).
\label{6.25b}
\end{align}

The result (\ref{6.25}\,a,b) is valid for $L,~\overline{G} M
\ll r \ll L \, \exp (1/q)$. To lowest order, for $q=0$, 
eq.\ \eqref{6.25a} equals the linearized Schwarzschild metric. The terms
$\propto q$ are exactly those which we found in Appendix \ref{A} by
explicitly solving the linearized field equations. In fact,
\eqref{6.25a} is precisely the special case of \eqref{A.22} which follows
from the $\mathcal{N}$-function \eqref{6.25b}.

Within its domain of validity, the line element \eqref{6.25a} is close
to the Minkowski metric. It is the counterpart of \eqref{5.13} which we had
derived for the fixed point running $G (k) \propto 1 / k^{2}$. Like there
we can read off a potential from
$g_{tt} \equiv - \bigl[ 1 + 2 \, \phi_{\text{mN}} \bigr]$
which governs the modified Newtonian limit:
\begin{align}
\phi_{\text{mN}} (r) = - \frac{\overline{G} \, M}{r}
+ \frac{q}{2} \, \ln (\kappa \, r).
\label{6.27}
\end{align}
While in the fixed point scenario the extra term in $\phi_{\text{mN}}$
was growing $\propto r^{2}$, we observe a much milder logarithmic growth
in the power law model with a small $q$.

The time-like geodesics of massive test particles at nonrelativistic
velocities can be found from \eqref{6.27} using the Newtonian equation of
motion $\ddot{\mathbf{x}} = - \nabla \phi_{\text{mN}}$. It implies
$v^{2} = r \, \phi_{\text{mN}}^{\prime} (r)$ for the Newtonian velocity
$v \equiv r \left( \text{d} \varphi / \text{d} t \right)$ along circular
orbits. For \eqref{6.27} we obtain
\begin{align}
v^{2} (r) = \frac{\overline{G} \, M}{r} + \frac{q}{2},
\label{6.28}
\end{align}
displaying the limiting velocity $v^{2} (r \to \infty) = q/2 \equiv
v^{2}_{\infty}$. Eq.\ \eqref{6.28} coincides with our earlier result
\eqref{6.16} since $\overline{G} M / \rho$ differs from
$\overline{G} M / r$ only by a term of second order which we neglect.

When the metric \eqref{6.25a} is valid the curvature of spacetime is small
and an essentially euclidean interpretation applies, with $t$ playing the 
role of ``the'' time, and $r$ differing only slightly from the radial
proper distance. The assumptions which led us to \eqref{6.25} guarantee
that the modified Newtonian potential is always much smaller than unity,
$\phi_{\text{mN}} \ll 1$. One of the consequences is that the proper time
$\tau$ measured by a clock at rest, i.\,e.\ with constant coordinates
$(r,\theta,\varphi)$, differs from the coordinate time $t$ only
by a factor $\text{d} \tau / \text{d} t = 1 + \phi_{\text{mN}} (r)$
which is very close to unity. Thus, within the precision of our calculation,
the velocity $v^{2}$ of \eqref{6.28} coincides with $2 \pi r$ divided by
the period $\tau$ of the orbit as measured by a clock at rest:
$v \equiv 2 \pi r / T = \left( 2 \pi r / \tau \right) \,
\bigl[ 1 + \mathcal{O} \left( \phi_{\text{mN}} \right) \bigr]$.
Note that using the standard form of the metric the proper circumference
of a circle of radius $r$ equals precisely $2 \pi r$. Note also that,
according to \eqref{6.24}, $r$ differs from $\rho$ by a term of order
$\phi_{\text{mN}}$ only. Hence, within the present approximation, we may
identify $v$ with $2 \pi r / T$, or with $2 \pi r / \tau$, or
also with $2 \pi \rho / T$. This justifies our use of the Newtonian
language when we interpreted the rotation curves in Subsection \ref{s6.3}.
\subsection{\label{s6.5}Galactic scales revisited}
The results of the previous subsections indicate that the scale-free power law
$G \propto k^{-q}$ could be able to explain at least the gross features
of the ``non-Keplerian'' dynamics of galaxies without the need for dark
matter. For a point mass it generates a modified Newtonian potential with
a $\ln (r)$-term, implying a flat rotation curve.

Contrary to the IR fixed point scenario of Section \ref{5}, the
$k^{-q}$-model trajectory has no direct relation to the RG-improved cosmology
of \cite{cosmo2,elo}; in particular, $L \equiv \kappa^{-1}$ and $q$ are free
parameters, so far unconstrained by cosmological observations.

Nevertheless the general physical picture is still the same as sketched earlier:
Between laboratory and cosmological scales, $G (k)$ ``crosses over'' from a
$G = const$-regime to the fixed point regime $G (k) \propto 1 / k^{2}$.
In between, at the scale of galaxies, $G (k)$ already shows some
nontrivial running, it increases with increasing distance, but the precise
form of its $k$- or $r$-dependence we do not know yet. The interpolating
ansatz \eqref{5.1} seems problematic, while \eqref{6.1} works much better.

For explaining the rotation curves it suffices that the $k^{-q}$-model 
trajectory \eqref{6.1} is approximately correct near galactic scales only.
We may think of it as part of a trajectory 
$G (k) \propto k^{\eta_{\text{N}}}$ which interpolates between
$G = const$ and $G (k) \propto k^{-2}$ by ``adiabatically'' shifting the
anomalous dimension $\eta_{\text{N}}$ from $\eta_{\text{N}} =0$ to its value
at the fixed point, $\eta_{\text{N}} = -2$. All we need for our above derivation
to apply is a range of $k$-values, comprising 2 or 3 orders of magnitude
say, for which $\eta_{\text{N}} \equiv -q$ is approximately constant,
implying $v (r) \approx const$ between \unit[1]{kpc} and 
\unit[100 or 1000]{kpc}, say.

In principle the numerical value of $q$ can be calculated
by solving the RG equation and identifying the specific trajectory
which is realized in Nature. (As in QED, for instance, this can be done
by fixing certain ``renormalized'' parameters according to their
experimental values.) If the trajectory Nature selects displays an
intermediate $k^{-q}$-behavior with $q$ of the right order of magnitude,
we would predict that the
$\phi_{\text{mN}}$ of any point mass contains a nonclassical term
$q \, \ln r$ where the prefactor $q$ is universal, i.\,e.\ independent
of the mass $M$. Hence the predicted plateau velocity
$v_{\infty} = \sqrt{q/2\,}$ is the same for all masses.

Within a set of typical spiral galaxies the measured $v_{\infty}$-values
differ by a factor of $2$ or $3$, say. While the plateau velocity of real
galaxies is not perfectly universal it does not vary by orders of
magnitude on the other hand, so our prediction of a universal $v_{\infty}$ might be a sensible
first approximation. In reality there exists a correlation between the
maximum rotation velocity $v_{\text{max}}$ and the luminosity $L$ of a galaxy,
the Tully-Fisher relation $v_{\text{max}}^{4} \propto L$. With 
$v_{\text{max}} = v_{\infty}$, and since $M/L$ is approximately constant
for spirals, this means that $v_{\infty}$ has a certain mass dependence:
$v_{\infty} \propto M^{-1/4}$.

The term $q \, \ln r$ which we derived from the $k^{-q}$-running is similar
to the MOND potential \eqref{1.10}. The only difference is its prefactor
$\propto \sqrt{M\,}$ which replaces our constant $q$ and produces the 
desired $v_{\infty} \propto M^{-1/4}$. In fact, the (nonrelativistic!)
field equation of the MOND approach \cite{mond} is designed in precisely
such a way that its solution for the potential of a point mass contains a
logarithm with a prefactor proportional to $\sqrt{M\,}$.

Since in many applications MOND is a successful phenomenological theory
it is plausible to conjecture that the \textit{exact} Quantum Einstein
Gravity dynamically generates the $\phi_{\text{MOND}}$-term \eqref{1.10}
or something similar as a quantum effect in the Newtonian regime. Clearly
a proof of this conjecture is extremely difficult; but on the other hand
it is quite easy to understand why our above \textit{approximate} analysis is
not precise enough to reveal the comparatively subtle $M^{-1/4}$-dependence.

The RG trajectory, computed exactly or in a truncation, is a universal object
which does not ``know'' about the concrete system under consideration. In
particular the function $G = G (k)$ is the same for all galaxies and
independent of $M$ therefore. A $M$-dependence can enter only via the cutoff
identification $k = k_{M} (r)$ which makes $G_{M} (r) \equiv G \bigl(
k = k_{M} (r) \bigr)$ a potentially $M$-dependent function of the distance.
For the cutoff $k = \xi / d_{g}$ this is indeed the case in general
because $g_{\mu \nu}$ depends on $M$, but in the Newtonian limit this
$M$-dependence drops out and we are left with $k \propto 1/r$, implying a
$M$-independent $G (r)$ and $\ln r$-term. In the present approach we can hope to find
the Tully-Fisher law only once we have gained a much better
understanding of the physical cutoff process \cite{h1} and the resulting
refined cutoff identification which, most probably, will deviate from
$k \propto 1/r$ at subleading order.

It always had appeared problematic to construct a relativistic generalization
of MOND theory \cite{mond}. Leaving the Tully-Fisher issue aside, we can
say that the improved action approach based upon a power law running of $G$
does indeed provide a relativistic generalization of this kind.

We shall return to the discussion of galaxies in the RG framework in
Subsection \ref{s7.7}.
%
%
%
%
%
%
%
%
\section{\label{7}The modified field equation in the Newtonian regime}
In the previous sections we constructed exact vacuum solutions of the modified
Einstein equation and we found that, \textit{in the Newtonian regime already},
those solutions display features which suggest that the observed
rotation curves could be interpreted as a renormalization effect of $G$. For
this reason we are now going to analyze the Newtonian limit of the modified
Einstein equation itself rather than looking at special solutions. As a
by-product, this
discussion will make it clear how the running-$G$ scenario relates to the
``standard speculation'' about the origin of the flat rotation curves: dark
matter.

Contrary to the previous sections we shall now allow for the presence of 
``ordinary'' matter, described by an energy-momentum tensor $T_{\mu \nu}$
of the perfect fluid type. It will not be traceless in general so that
the results of this section cannot be obtained using the Weyl technique.

Except for staticity we shall not impose any symmetry requirements. Apart
from questions related to the cutoff identification the
discussion will also apply to systems which are not spherically
symmetric.
\subsection{\label{s7.1}Background, metric ansatz, small quantities}
Newton's constant is assumed to be time-independent but to have a weak
position dependence. We use the parameterization
\begin{align}
G (\mathbf{x}) = \overline{G} \bigl[ 1 + \mathcal{N} (\mathbf{x}) \bigr]
\label{7.1}
\end{align}
and assume that $\mathcal{N} (\mathbf{x}) \ll 1$. Being interested in small
isolated systems, the cosmological constant is assumed to vanish,
$\Lambda (\mathbf{x}) \equiv 0$.

As in the standard discussion of the Newtonian limit we make an ansatz for
the metric of the form
\begin{align}
\text{d} s^{2} =
- \bigl[ 1 + 2 \, \phi (\mathbf{x}) \bigr] \, \text{d} t^{2}
+ \bigl[ 1 - 2 \, \omega (\mathbf{x}) \bigr] \, \text{d} \mathbf{x}^{2}
\label{7.2}
\end{align}
where it is understood that $\phi (\mathbf{x}) \ll 1$ and 
$\omega (\mathbf{x}) \ll 1$. Here and in the following we use cartesian
coordinates $x^{\mu} \equiv (t,x^{i}) \equiv (t,\mathbf{x})$ and write
$\text{d} \mathbf{x}^{2} \equiv \delta_{ij} \, \text{d} x^{i} \text{d} x^{j}$.

When evaluating the field equations we are going to assume that $\phi$,
$\omega$, and $\mathcal{N}$ are of the same order of magnitude. We retain
terms of first order in $\phi,\omega,\mathcal{N} \ll 1$ and their
derivatives, but discard higher powers ($\phi^{2}$, $\cdots$) and cross terms
($\phi \, \mathcal{N}$, $\cdots$).

In the spacetime \eqref{7.2}, test particles with nonrelativistic
velocities $\mathbf{v}^{2} \equiv \left( \text{d} \mathbf{x} / 
\text{d} t \right)^{2} \ll 1$ obey a Newtonian equation of motion. Retaining
only terms linear in $\phi, \omega, \mathbf{v}^{2}$, the geodesic 
equation boils down to
\begin{align}
\frac{\text{d}^{2}}{\text{d} t^{2}} \, \mathbf{x} (t) =
- \nabla \phi \bigl( \mathbf{x} (t) \bigr).
\label{7.3}
\end{align}
Obviously $\phi$ is the ``modified Newtonian potential'', denoted
$\phi_{\text{mN}}$ in the previous sections.

To first order in $\phi$ and $\omega$ the Ricci tensor of the metric
\eqref{7.2} is given by
\begin{align}
R_{00} = \nabla^{2} \phi, \quad
R_{0i} = 0, \quad
R_{ij} = \partial_{i} \partial_{j} \, \left( \omega - \phi \right)
+ \delta_{ij} \, \nabla^{2} \omega.
\label{7.4}
\end{align}
Here $\nabla^{2} \equiv \delta^{ij} \, \partial_{i} \partial_{j}$
denotes the ordinary Laplacian on flat 3-space. Likewise the components of
the Einstein tensor are
\begin{align}
G_{00} = 2 \, \nabla^{2} \omega, \quad
G_{0i} = 0, \quad
G_{ij} = \Bigl[ \partial_{i} \partial_{j} 
- \delta_{ij} \, \nabla^{2} \Bigr] \, \left( \omega - \phi \right).
\label{7.5}
\end{align}
%
%
%
\subsection{\label{s7.2}The tensors $\boldsymbol{\theta_{\mu \nu}}$ and
$\boldsymbol{\Delta T_{\mu \nu}}$}
It is a particularly attractive feature of the modified Einstein equation
that, in the Newtonian limit, it becomes completely independent of the
tensor $\theta_{\mu \nu}$ whose form was not completely fixed by general
principles.
In fact, every tensor $\theta_{\mu \nu}$ which is at least quadratic in the
derivatives $\partial_{\mu} G = \overline{G} \, \partial_{\mu}
\mathcal{N}$ is of order $\mathcal{N}^{2}$ and can be neglected therefore.
For example, $\theta_{\mu \nu} = \varepsilon \, \theta_{\mu \nu}^{\text{ BD}}$
is indeed bilinear in $\partial_{\mu} G$, see eq.\ \eqref{2.16}. Since it is
impossible to satisfy the consistency condition identically with a tensor
which has less than two factors of $\partial_{\mu} G$ we conclude that
$\theta_{\mu \nu} = 0 + \textit{higher orders}$.

Upon evaluating \eqref{2.7} for the metric \eqref{7.2} we obtain for the
components of $\Delta t_{\mu \nu}$:
\begin{align}
\Delta t_{00} = - \nabla^{2} \mathcal{N}, \quad
\Delta t_{0i} = 0, \quad
\Delta t_{ij} = - \Bigl[ \partial_{i} \partial_{j} 
- \delta_{ij} \, \nabla^{2} \Bigr] \, \mathcal{N}.
\label{7.7}
\end{align}
The corresponding components of $\Delta T_{\mu \nu}$ are
\begin{subequations} \label{7.8}
\begin{align}
\Delta T_{00} & = - \left( 8 \pi \, \overline{G} \, \right)^{-1} \,
\nabla^{2} \mathcal{N},
\label{7.8a}
\\
\Delta T_{0i} & = 0,
\label{7.8b}
\\
\Delta T_{ij} & = - \left( 8 \pi \, \overline{G} \, \right)^{-1} \,
\Bigl[ \partial_{i} \partial_{j} 
- \delta_{ij} \, \nabla^{2} \Bigr] \, \mathcal{N}.
\label{7.8c}
\end{align}
\end{subequations}
The expressions \eqref{7.7} and \eqref{7.8} are correct up to terms of order
$\mathcal{N}^{2}$, $\mathcal{N} \, \phi$, and $\mathcal{N} \, \omega$.
\subsection{\label{s7.3}Content of the consistency condition}
The field equations whose Newtonian limit we are interested in consist of
the modified Einstein equation with $\Lambda =0$ and $\theta_{\mu \nu} =0$,
\begin{align}
G_{\mu \nu} = 8 \pi G \, T_{\mu \nu} + \Delta t_{\mu \nu},
\label{7.9}
\end{align}
coupled to the consistency condition which is most conveniently analyzed
in the version of \eqref{2.9-1}. In fact, since
$D^{\mu} G = \mathcal{O} \left( \mathcal{N} \right)$,
$\Delta t_{\mu \nu} = \mathcal{O} \left( \mathcal{N} \right)$, and
$R_{\mu \nu} = \mathcal{O} \left( \omega, \phi \right)$, its first term
$\propto D^{\mu} G \, \bigl( \Delta t_{\mu \nu} - R_{\mu \nu} \bigr)$
is of higher order and can be neglected. The same is true for the second
because $\vartheta_{\mu \nu} = \mathcal{O} \left( \mathcal{N}^{2} \right)$,
and the third one vanishes since $\Lambda =0$. What remains is the condition
$8 \pi \, \overline{G} \, T_{\nu}^{~\mu} \, \partial_{\mu} \mathcal{N} =
\mathcal{O} \left( \mathcal{N}^{2}, \mathcal{N} \, \phi,
\mathcal{N} \, \omega \right)$. Since $\mathcal{N}$ is time-independent
but an arbitrary function of $\mathbf{x}$ this condition requires that
$T_{\nu}^{~i} = \mathcal{O} \left( \mathcal{N}^{2}, \mathcal{N} \, \phi,
\mathcal{N} \, \omega \right)$. In the metric \eqref{7.2} the energy-momentum
tensor of a perfect fluid has the structure
$T_{\nu}^{~\mu} = \text{diag} \left[ - \rho, p, p, p \right]$. Imposing
$T_{\nu}^{~i} = 0 + \textit{higher orders}$ implies
$p = 0 + \textit{higher orders}$ and leaves $\rho$ unconstrained.

Thus we conclude that in the Newtonian limit only pressureless matter can
be coupled consistently. Henceforth we shall employ
the energy-momentum tensor of dust,
$T_{\nu}^{~\mu} = \rho \, \delta_{\nu}^{0} \, \delta_{0}^{\mu}$.
Note, however, that we are not neglecting the pressure terms in
$\Delta T_{\mu \nu}$.
\subsection{\label{s7.4}From Einstein's to Poisson's equation}
When we insert \eqref{7.5} and \eqref{7.7} into the modified Einstein
equation \eqref{7.9} we see that it decomposes into the $(00)$-component
equation
\begin{subequations} \label{7.10}
\begin{align}
\nabla^{2} \left( \omega + \tfrac{1}{2} \, \mathcal{N} \right) =
4 \pi \, G (\mathbf{x}) \, \rho
\label{7.10a}
\end{align}
and the $(ij)$-components
\begin{align}
\Bigl[ \partial_{i} \partial_{j} - \delta_{ij} \, \nabla^{2} \Bigr] \,
\left( \omega - \phi + \mathcal{N} \right) =0
\label{7.10b}
\end{align}
\end{subequations}
where $i,j=1,2,3$. The $(0i)$-equations are satisfied identically to the 
relevant order.

The system \eqref{7.10} can be solved in a very elegant way.
Eq.\ \eqref{7.10b} tells us that $\omega - \phi + \mathcal{N}$ must be
constant, and since we would like to recover the standard result
$\omega = \phi$ if $\mathcal{N} =0$, this constant must be zero:
$\omega - \phi + \mathcal{N} =0$. This allows us to express $\omega$
in terms of the (unknown) field $\phi$ and the (prescribed) function
$\mathcal{N}$:
\begin{align}
\omega = \phi - \mathcal{N}.
\label{7.11}
\end{align}
Inserting \eqref{7.11} into \eqref{7.10a} we obtain a modified Poisson
equation for $\phi$:
\begin{align}
\nabla^{2} \left[ \phi - \tfrac{1}{2} \, \mathcal{N} \right] =
4 \pi \, G (\mathbf{x}) \, \rho.
\label{7.12}
\end{align}
This equation differs from that of classical gravity in two ways: on its
LHS the potential $\phi$ is replaced with $\phi - \mathcal{N}/2$, and on
the RHS $G$ is allowed to depend on $\mathbf{x}$.

The equations \eqref{7.11} and \eqref{7.12} show that the metric of the
modified Newtonian limit can be found in the following way. Given the
position dependence of Newton's constant, $\mathcal{N} (\mathbf{x})$, and
the matter energy density $\rho (\mathbf{x})$ one first solves the Poisson
equation
\begin{align}
\nabla^{2} \widehat \phi = 4 \pi \, G (\mathbf{x}) \, \rho
\equiv 4 \pi \, \overline{G} \, \bigl( 1 + \mathcal{N} (\mathbf{x}) \bigr) \,
\rho.
\label{7.13}
\end{align}
Then, knowing the auxiliary potential $\widehat \phi (\mathbf{x})$, one
computes the metric components as
\begin{subequations} \label{7.14}
\begin{align}
\phi & = \widehat \phi + \tfrac{1}{2} \, \mathcal{N},
\label{7.14a}
\\
\omega & = \widehat \phi - \tfrac{1}{2} \, \mathcal{N}.
\label{7.14b}
\end{align}
\end{subequations}
The resulting line element reads
\begin{align}
\text{d} s^{2} =
- \left[ 1 + 2 \, \widehat \phi + \mathcal{N} \right] \,
\text{d} t^{2}
+ \left[ 1 - 2 \, \widehat \phi + \mathcal{N} \right]\,
\text{d} \mathbf{x}^{2}.
\label{7.15}
\end{align}

Eqs.\ \eqref{7.13} and \eqref{7.14a} are the most important result of this
section. They show that the modified Newtonian potential is given by
\begin{align}
\phi (\mathbf{x}) & = \widehat \phi (\mathbf{x}) 
+ \tfrac{1}{2} \, \mathcal{N} (\mathbf{x}),
\label{7.16}
\end{align}
where $\widehat \phi$ is a solution to the ``improved Poisson equation''
\eqref{7.13} which involves a position dependent Newton constant. Adopting
natural boundary conditions,
\begin{align}
\widehat \phi (\mathbf{x}) = - \int \!\! \text{d}^{3} x^{\prime} ~
\frac{G (\mathbf{x}^{\prime}) \, \rho (\mathbf{x}^{\prime})}
{|\mathbf{x} - \mathbf{x}^{\prime}|}.
\label{7.17}
\end{align}

Eq.\ \eqref{7.16} is a generalization of the result \eqref{A.20a} which
we derived in Appendix \ref{A} for the spherically symmetric case using
a completely different method. Note that the spatial part of the line
element \eqref{7.2} differs from that of the linearized standard metric
\eqref{A.1} employed in the Appendix.

Up to terms of higher orders, the metric \eqref{7.15} admits the
factorization
\begin{align}
\text{d} s^{2} = \left( 1 + \mathcal{N} \right) \, \bigg \{
- \left[ 1 + 2 \, \widehat \phi \right] \, \text{d} t^{2}
+ \left[ 1 - 2 \, \widehat \phi \right] \, \text{d} \mathbf{x}^{2} \bigg \}.
\label{7.18}
\end{align}
This factorization is reminiscent of the Weyl transformation \eqref{4.2}.
In fact, if $\rho =0$ for $|\mathbf{x}| > 0$ then
$\widehat \phi (\mathbf{x}) = - \overline{G} M / |\mathbf{x}|$
with $\overline{G} \equiv G (\mathbf{x} =0)$. As a result, the solution
belongs to Class II, and \eqref{7.18} is the familiar Weyl rescaling
$\text{d} s^{2} = \bigl[ G (\mathbf{x}) / \, \overline{G} \, \bigr] \,
\text{d} s_{\gamma}^{2}$ applied to the linearized Schwarzschild metric.
For $\rho \neq 0$ the result \eqref{7.18} with \eqref{7.17} is new and
cannot be proven by means of the Weyl trick since $T_{\mu}^{~\mu} \neq 0$
in this case.

For a $\delta$-function source at the origin $\widehat \phi$ equals the
classical $1/r$-potential, and according to \eqref{7.16} the corresponding
modified Newtonian potential is
\begin{align}
\phi (\mathbf{x}) = - \frac{\overline{G} \, M}{|\mathbf{x}|} 
+ \frac{1}{2} \, \mathcal{N} (\mathbf{x}).
\label{7.19}
\end{align}

The generality and simplicity of this result and its generalization 
\eqref{7.16} is striking: The impact of the $\mathbf{x}$-dependent $G$ 
on the Newtonian limit consists of the additional term
$\mathcal{N} (\mathbf{x}) /2$ in the potential only. This is true for any
externally prescribed weakly position dependent Newton constant. The
function $G (\mathbf{x})$ is neither required to have any special symmetry
nor is it necessary that $G (\mathbf{x})$ derives from a running $G (k)$
via some cutoff identification $k = k (\mathbf{x})$.

While this is difficult in general \cite{h1}, it is fairly straightforward
to convert $G (k)$ to a position dependent $G$ if the system is spherically
symmetric. In the Newtonian limit this step simplifies considerably.
In the relevant order the cutoff identification $k = k (r)$ is simply
$k = \xi / r$, $r \equiv |\mathbf{x}|$, since when inserted into
$\mathcal{N} (k = \xi /r) \equiv \mathcal{N} (r)$ the difference between
$r$ and $d_{g} (r)$ amounts to a higher order term.

As a result, the potential of a point mass at the origin is given by
$\phi (r) = - \overline{G} \, M / r + \mathcal{N} (r) / 2$. The
renormalization effects generate an additional contribution
$- \mathcal{N}^{\prime} (r) /2$ to the classical force
$- \overline{G} \, M / r^{2}$. The new force is attractive if 
$\mathcal{N} (r)$, and hence $G (r)$, grows as a function of $r$.

In Sections \ref{5} and \ref{6} we encountered two examples already. There
we calculated the \textit{exact} metrics pertaining to the fixed point and
the power law running of $G (k)$, respectively,
and we saw that in certain parts of those
spacetimes a Newtonian interpretation is possible. By eqs.\ \eqref{5.13} and
\eqref{6.25b} the corresponding $\mathcal{N}$-functions are
\begin{subequations} \label{7.20}
\begin{flalign}
& & & & & 
\mathcal{N} (r) = \kappa^{2} \, r^{2} && \text{(fixed point)} \qquad
\label{7.20a}
\\
& & & & & 
\mathcal{N} (r) = q \, \ln (\kappa \, r) && \text{(power law)} \qquad
\label{7.20b}
\end{flalign}
\end{subequations}
Using these $\mathcal{N}$'s in eq.\ \eqref{7.19} we reproduce exactly the
potentials \eqref{5.14} and \eqref{6.27} extracted from the exact solutions.
In Sections \ref{5} and \ref{6} we had to assume that 
$\theta_{\mu \nu} = \theta_{\mu \nu}^{\text{ BD}}$. Now we see that the
modified potentials obtained there are actually independent of the
choice for the $\theta$-tensor.
\subsection{\label{s7.5}The importance of the vacuum pressure}
Using the explicit expression for $\Delta T_{00}$, eq.\ \eqref{7.8a},
we can rewrite the Poisson equation \eqref{7.12} in the following equivalent
way:
\begin{align}
\nabla^{2} \phi = 4 \pi \, G (\mathbf{x}) \, \bigl[ T_{00}
- \Delta T_{00} \bigr].
\label{7.21}
\end{align}
At first sight it seems quite puzzling why the \textit{negative} of
$\Delta T_{00}$, the ``vacuum'' energy density due to the 
$\mathbf{x}$-dependence of G, should be added to the matter energy density
$\rho = T_{00}$. This paradox is resolved if one recalls that even in 
the Newtonian limit of standard general relativity \cite{weinbook}, when
the pressure is not negligible, Newton's $T_{00}$ gets replaced by
$T_{00} + \sum_{i=1}^{3} T_{ii}$ (which equals the familiar $\rho + 3 \, p$
for a perfect fluid):
$\nabla^{2} \phi = 4 \pi \, \overline{G} \, \bigl[ T_{00} + 
\sum_{i=1}^{3} T_{ii}\bigr]$.

Above we neglected the pressure of the ordinary matter but the pressure terms
stemming from $\Delta T_{\mu \nu}$, i.\,e.\ $\Delta T_{ij}$, where fully taken
into account. The $\Delta T_{\mu \nu}$-components \eqref{7.8} obey
the following identity which is quite remarkable:
\begin{align}
\Delta T_{00} + \sum_{i=1}^{3} \Delta T_{ii} = - \Delta T_{00}.
\label{7.22}
\end{align}
Adding the pressure terms has precisely the effect of flipping the sign of
$\Delta T_{00}$. Therefore we may rewrite \eqref{7.21} in the form
\begin{align}
\nabla^{2} \phi = 4 \pi \, G (\mathbf{x}) \, \bigl[ T_{00} + \Delta T_{00} +
\sum_{i=1}^{3} \Delta T_{ii}\bigr].
\label{7.23}
\end{align}
The source term of this Poisson equation has exactly the expected
``$\rho + 3 \, p$''-structure and we understand
that there is nothing wrong with the difference $T_{00} - \Delta T_{00}$
appearing in \eqref{7.21}. In fact, this relative minus sign resulting
from the inclusion of the vacuum pressure
is crucial in order to obtain a physically sensible Newton limit.
If we naively use the sum of the energy densities
$T_{00} + \Delta T_{00}$ as the source for $\phi$, the ensuing modified
Newtonian potential is given by \eqref{7.19} with
$\mathcal{N} (\mathbf{x})$ replaced by $- \mathcal{N} (\mathbf{x})$. The
consequence would be that a $G (r)$ growing with $r$ gives rise to a 
\textit{repulsive} component in the gravitational force. This is inconsistent
with the antiscreening nature of gravity and indeed is
not what the theory predicts.
\subsection{\label{s7.6}Effective energy density vs.\ dark matter}
For the interpretation of the RG improved theory it is helpful to
rewrite the modified Poisson equation in the style of a classical Poisson
equation involving a modified energy density, though. Defining the effective
density
\begin{align}
\rho_{\text{eff}} \equiv \left[ G (\mathbf{x}) / \, \overline{G} \, \right] \,
\left( \rho - \Delta \rho \right),
\label{7.24}
\end{align}
eq.\ \eqref{7.21} assumes the traditional form
\begin{subequations} \label{7.25}
\begin{align}
\nabla^{2} \phi = 4 \pi \, \overline{G} \, \rho_{\text{eff}}.
\label{7.25a}
\end{align}
\end{subequations}
Above we introduced the vacuum energy density
\begin{align}
\Delta \rho \equiv \Delta T_{00} =
- \left( 8 \pi \overline{G} \right)^{-1} \, \nabla^{2} \mathcal{N}
\label{7.26}
\end{align}
in analogy with $T_{00} = \rho$ (to lowest order). Since $\Delta \rho$ is of
order $\mathcal{N}$ already, \eqref{7.24} is equivalent to
\begin{align}
\rho_{\text{eff}} = \rho + \mathcal{N} \, \rho
+ \left( 8 \pi \overline{G} \right)^{-1} \, \nabla^{2} \mathcal{N}.
\tag{\ref{7.25}b} \label{7.25b}
\end{align}
What is the physical significance of $\rho_{\text{eff}}$?

In astronomy a standard method for measuring mass distributions proceeds,
schematically, as follows. One monitors the trajectories
$\mathbf{x} (t)$ of as many ``test particles'' as possible, assumes they
satisfy $\ddot{\mathbf{x}} = - \nabla \phi (\mathbf{x})$, then reconstructs
$\phi$ from the trajectories, and finally computes the matter density
responsible for this potential as $\left( \nabla^{2} \phi \right) / 4 \pi \,
\overline{G}$, taking the Newtonian field equation for granted.

In the Newtonian regime, the RG improved theory of gravity proposed in this
paper consists of two parts: the (modified) field equation \eqref{7.25}
and the (standard) particle equation of motion \eqref{7.3}.
Let us assume both of these laws are correct, and that they do indeed govern
the dynamics of ``test particles'' in a galaxy, say. Then, when an astronomer
tries to measure the mass density inside the galaxy by applying the method
described above, he or she will come to the conclusion that this mass density
is given by $\rho_{\text{eff}}$ rather than the density $\rho$ of ordinary
matter. In particular, in a halo with $\rho \equiv 0$ the measurement would
determine $\rho_{\text{eff}} = - \Delta \rho$.

The most popular speculation about the origin of the flat galaxy rotation
curves is the assumption that there exists a certain amount of dark matter
in galaxies, with density $\rho_{\text{DM}}$, which generates an appropriate
potential in the conventional way, i.\,e.\ via
$\nabla^{2} \phi = 4 \pi \overline{G} \, \left( \rho + \rho_{\text{DM}}
\right)$. Comparing this equation to \eqref{7.25} we see that in the RG
approach it is essentially $- \Delta \rho$ which takes the place of 
$\rho_{\text{DM}}$. In fact, typically the $\mathcal{N} \, \rho$-term
in \eqref{7.25b} is tiny compared to $\rho$ so that
$\rho_{\text{eff}} \approx \rho - \Delta \rho$.

Thus we conclude that a scale- and, as a consequence,
$\mathbf{x}$-dependent Newton constant can mimic the presence of dark
matter. In the Newtonian
limit the effective density of the dark matter substitute is given by
$+ \bigl( 8 \pi \, \overline{G}^{2} \bigr)^{-1} \, \nabla^{2}
G (\mathbf{x})$.
\subsection{\label{s7.7}A simple model of a galaxy}
Let us consider a spherically symmetric, static system so that $\rho$ and
$\mathcal{N}$ depend on $r \equiv |\mathbf{x}|$ only. Then the effective
density is given by
\begin{align}
\rho_{\text{eff}} (r) =
\bigl[ 1 + \mathcal{N} (r) \bigr] \, \rho (r)
+ \Bigl( 8 \pi \, \overline{G} \, r^{2} \Bigr)^{-1} \,
\frac{\text{d}}{\text{d} r} \, 
\left( r^{2} \, \frac{\text{d} \mathcal{N}}{\text{d} r} \right).
\label{7.27}
\end{align}
The Poisson equation \eqref{7.25a} is solved by integrating
\begin{align}
\frac{\text{d}}{\text{d} r} \, \phi (r) =
\frac{\overline{G} \, \mathcal{M}_{\text{eff}} (r)}{r^{2}}
\label{7.28}
\end{align}
with $\mathcal{M}_{\text{eff}} (r) \equiv 4 \pi \,
\int \limits_{0}^{r} \! \text{d} r^{\prime} ~
{r^{\prime}}^{2} \, \rho_{\text{eff}} (r^{\prime})$, or explicitly,
\begin{align}
\mathcal{M}_{\text{eff}} (r) = \mathcal{M} (r)
+ \frac{r^{2}}{2 \, \overline{G}\,} \, 
\frac{\text{d}}{\text{d} r} \, \mathcal{N} (r).
\label{7.29}
\end{align}
Here $\mathcal{M} (r) \equiv 4 \pi \,
\int \limits_{0}^{r} \! \text{d} r^{\prime} ~
{r^{\prime}}^{2} \, \bigl( 1 + \mathcal{N} (r^{\prime}) \bigr) \,
\rho (r^{\prime})$ is essentially the mass of the ordinary matter contained in
a ball of radius $r$, centered at the origin. Since $\mathcal{N} \ll 1$,
the factor $\left( 1 + \mathcal{N} \right)$ is very close to unity and plays
no role at the qualitative level; we shall ignore it in the
following. The effective mass inside the ball, $\mathcal{M}_{\text{eff}}$,
consists of the classical part $\mathcal{M}$, plus a term 
$\propto \text{d} \mathcal{N} / \text{d} r$ which is entirely due to the
renormalization effects. If Newton's constant is a growing function of $r$
then the nonclassical contribution is positive and
$\mathcal{M}_{\text{eff}} (r) > \mathcal{M} (r)$.

In the jargon of quantum field theory one would refer to $\mathcal{M}$
and $\mathcal{M}_{\text{eff}}$ as the ``bare'' and the ``dressed'' or
``renormalized'' mass, respectively. The antiscreening character of
Quantum Einstein Gravity \cite{mr} amounts to
$\text{d} \mathcal{N} / \text{d} r >0$ so that the quantum-gravitational
``dressing'' of a classical mass distribution increases the total mass.

From \eqref{7.28} with \eqref{7.29} we obtain the potential
\begin{align}
\phi (r) = \int \limits^{r} \!\! \text{d} r^{\prime} ~
\frac{\overline{G} \, \mathcal{M} (r^{\prime})}{{r^{\prime}}^{2}}
+ \frac{1}{2} \, \mathcal{N} (r).
\label{7.30}
\end{align}
The corresponding rotation curve $v^{2} (r) = r \, \phi^{\prime} (r)$
for particles orbiting in this potential is
\begin{align}
v^{2} (r) = \frac{\overline{G} \, \mathcal{M} (r)}{r}
+ \frac{1}{2} \, r \, \frac{\text{d}}{\text{d} r} \, \mathcal{N} (r).
\label{7.30-1}
\end{align}

On the basis of the potential \eqref{7.30} we can make a simple model of a 
spherically symmetric galaxy. We identify $\rho$ with the density of the
ordinary luminous matter. We model the luminous core of the galaxy by a ball
of radius $r_{0}$. The mass of the ordinary matter contained in the core is
$\mathcal{M} (r_{0}) \equiv \mathcal{M}_{0}$, the ``bare'' total mass
of the galaxy. Since, by assumption, $\rho =0$ and hence
$\mathcal{M} (r) = \mathcal{M}_{0}$ for $r > r_{0}$, the potential outside
the core is
\begin{flalign}
& & & & 
\phi (r) = - \frac{\overline{G} \, \mathcal{M}_{0}}{r}
+ \frac{1}{2} \, \mathcal{N} (r)
& & (r > r_{0}) \qquad
\label{7.31}
\end{flalign}
We shall refer to the region $r > r_{0}$ as the ``halo'' of the model
galaxy. For $r$ sufficiently large, the potential in the halo is
essentially $\mathcal{N} (r) / 2$.

Which $k$-dependence of $G (k)$ is phenomenologically favored? 
We saw that a power law behavior $G (k) \propto k^{-q}$
is, on the one hand, the simplest and most natural option theoretically
since it arises from a constant anomalous dimension.
On the other hand it was precisely this $k$-dependence which led to
flat rotation curves for point masses. Employing
the identification $k \propto 1 /r$ the $r$-dependence of $G$
is governed by a power law as well: $G (r) \propto r^{q}$. Only if $q$
is small and $r$ is not too large $\bigl( r \ll L \, \exp (1/q) \bigr)$ we may expand
$r^{q}$, obtaining $G (r) = \overline{G} \, \bigl[ 1 + \mathcal{N} (r) \bigr]$
with $\mathcal{N} (r) = q \, \ln (\kappa \,r)$.

Accepting the logarithmic $\mathcal{N}$-function as the most promising
candidate, the position dependence of $G$ generates the \textit{negative}
vacuum energy density
\begin{align}
\Delta \rho = - \frac{q}{8 \pi \overline{G} \, r^{2}\,}.
\label{7.32-1}
\end{align}
It also generates vacuum stresses (pressure terms) thanks to which the
effective energy density receives a \textit{positive} vacuum
contribution:
\begin{align}
\rho_{\text{eff}} (r) = \rho (r) + \frac{q}{8 \pi \overline{G} \, r^{2}\,}.
\label{7.33-1}
\end{align}
This formula is valid for all $r > r_{\text{tr}}$ where $r_{\text{tr}}$ is the
radius at which the nontrivial RG running gets ``switched on''.

For simplicity we assume that $r_{\text{tr}} < r_{0}$, i.\,e.\ that the
transition takes place inside the core. As we discussed already, the scale
$r_{\text{tr}}$ is comparatively unimportant: the renormalization effects
become strong, and visible, at the larger scale $r_{\text{nc}} =
2 \overline{G} \mathcal{M}_{0} / q$ only. Thus, everywhere in the halo,
\begin{align}
\phi (r) = - \frac{\overline{G} \, \mathcal{M}_{0}}{r}
+ \frac{q}{2} \, \ln (\kappa \, r).
\label{7.32}
\end{align}
Obviously \eqref{7.32} has the same form as the potential \eqref{6.27} for
a point mass at the origin. Strictly speaking \eqref{7.32}
is valid for $r_{0} < r \ll L \, \exp (1/q)$ only. However, most probably
the upper bound on $r$ is irrelevant from the practical point of view since
at too large distances the picture of a single isolated galaxy is unrealistic
anyhow. At a certain point the influence of other galaxies and of the
expansion of the Universe should be included.

Of course the most important experimental fact supporting the power law running of $G$
are the observed flat galaxy rotation curves. The power law scenario leads to the
prediction
\begin{align}
v^{2} (r) = \frac{\overline{G} \, \mathcal{M} (r)}{r} 
+ \frac{q}{2}
\label{7.33}
\end{align}
which is valid for $r > r_{\text{tr}}$, both inside and outside the core.
In the halo we have $v^{2} (r) = \overline{G} \mathcal{M}_{0} / r
+ q/2$ and the situation is the same as for the point mass in Section
\ref{6}. For $r \to \infty$, \eqref{7.33}
yields a constant velocity $v^{2}_{\infty} = q/2$. As we discussed
already, the observed values of $v_{\infty}$ suggest that
$q = \mathcal{O} \left( 10^{-6} \right)$. The smallness of this number
justifies the linearization with respect to $\mathcal{N}$. It also implies
that the variation of $G$ inside a galaxy is extremely small. The relative
variation of Newton's constant from some $r_{1}$ to $r_{2} > r_{1}$
is $\Delta G / G = q \ln (r_{2} / r_{1})$. As the radial extension of a
halo comprises only 2 or 3 orders of magnitude (from
$\unit[1]{kpc}$ to $\unit[100]{kpc}$, say) the variation between the
inner and the outer boundary of the halo is of the order
$\Delta G / G \approx q$, i.\,e.\ Newton's constant changes by one part in
a million only.

The $r$-dependence of the velocity \eqref{7.33} is in qualitative agreement with the
observations. For realistic density profiles, $\mathcal{M} (r) / r$ is an
increasing function for $r < r_{0}$, and it decays as
$\mathcal{M}_{0} / r$ for $r > r_{0}$. As a result, $v^{2} (r)$ rises steeply
at small $r$, then levels off, goes through a maximum at the boundary of
the core, and finally approaches the plateau from above:
$v^{2} (r > r_{0}) =\overline{G} \mathcal{M}_{0} /r + q/2$. Some galaxies
indeed show a maximum after the steep rise, but typically it is not very
pronounced, or is not visible at all. The prediction for the characteristic
scale where the plateau starts is $r_{\text{nc}} = 2 \overline{G}
\mathcal{M}_{0} / q$. With $q = 10^{-6}$ and
$\mathcal{M}_{0} = 10^{11} \, M_{\odot}$ one obtains
$r_{\text{nc}} = \unit[9]{kpc}$ which is just the right order of
magnitude.
\vspace*{2\baselinestretch pt}

We close this section with a set of further comments.

(a) As for deriving more accurate, quantitatively correct rotation curves the
limiting factor of our method is the cutoff identification $k = k (x)$.
Beyond spherical symmetry, for a disk say, there will be more than
one relevant length scale and there is no straightforward way of finding
out which function of those scales will take the place of the essentially
unique single-scale ansatz $k \propto 1/r$. In Subsection \ref{s6.5}
we explained already that for a very similar reason the Tully-Fisher
relation cannot be obtained easily. In principle it is clear how to solve these
problems: One computes $\Gamma = \Gamma_{k \to 0}$ and then solves the 
effective field equation $\delta \Gamma / \delta g_{\mu \nu} =0$. This is a
formidable task, though. Only under very special circumstances, strong
symmetry constraints for example, the improvement strategy is available as a
technically feasible ``short-cut'' to the full-fledged calculation.
(See \cite{h1} for a detailed discussion of this point.)

(b) The actual mass of the model galaxy,
including the renormalization effects, is
$\mathcal{M}_{\text{eff}}$. Integrating \eqref{7.33-1} we find for
this ``dressed'' mass
\begin{flalign}
& & & & 
\mathcal{M}_{\text{eff}} (r) = \mathcal{M}_{0}
+ \left( q / 2 \, \overline{G} \, \right) \, r
& & (r > r_{0}) \qquad
\label{7.34}
\end{flalign}
The first contribution, the ``bare'' mass $\mathcal{M}_{0}$, is the total
mass of the ordinary luminous matter contained in the galaxy, the second
is due to the running of $G$. The latter contribution increases with $r$
unboundedly so that, strictly speaking, the total dressed mass 
$\mathcal{M}_{\text{eff}} (\infty)$ is infinite. However, since it is clear that our
assumptions break down at some large value of $r$ the growth of
$\mathcal{M}_{\text{eff}}$ will stop at some point probably, but this
is outside the scope of the present model.

(c) In the dark matter scenario it is usually assumed that galaxies form in
two steps. In the first step, density perturbations which have grown
sufficiently lead to the formation of virialized, gravitationally bound
structures consisting of dark matter only. Then ordinary (baryonic)
matter will cool by radiating energy, sink to the centers of those dark
matter haloes, and thus form galaxies \cite{padman}.

In the physical picture we advocate in the present paper quantum fluctuations of the
gravitational field take the place of dark matter. If we try to keep
the essential features of the above structure formation scenario unaltered
it is natural to ask which objects could play the role of the pure
dark matter haloes which contain no ordinary matter yet. Among the exact
solutions which we found above there is an obvious 
candidate: the $\mathcal{M}_{0} \equiv M =0$ special case of the metric
$\text{d} s^{2} = w (\rho) \, \text{d} s_{\gamma}^{2}$. It is generated by
the Minkowski metric $\gamma_{\mu \nu} = \eta_{\mu \nu}$, whence
$g_{\mu \nu} = w (\rho) \, \eta_{\mu \nu}$ with the Weyl factor given by
\eqref{6.8} with $I_{f} (\rho) = \rho$. This
metric describes a ``solitonic'' excitation of the gravitational field in
absence of any matter. Contrary to the Schwarzschild spacetime which is a
vacuum solution, too, it owes its existence entirely to the renormalization
effects: for $q \to 0$ it approaches flat space.

Note that if one interprets the 4-momentum carried by the vacuum fluctuations
as due to some kind of matter, this matter is indeed ``dark'' in the sense
that it is not ``seen'' by the electromagnetic interaction, the reason being
that the solution $g_{\mu \nu}$ is conformally related to $\eta_{\mu \nu}$.

It seems conceivable that in the early Universe density fluctuations first
generate the ''solitons'' with $M=0$, thereby creating gravitational sinks which get
filled with ordinary matter later on. It would be interesting to see if the
current dark matter-based scenarios of structure formation can be
reinterpreted in a running-$G$ framework. This is beyond the scope of the
present paper, however.
%
%
%
%
%
%
%
%
\section{\label{8}Summary and conclusion}
As in any quantum field theory also in Quantum Einstein Gravity there is a 
standard way of investigating quantum effects both at small and large
distances: one computes the effective action 
$\Gamma \left[ g_{\mu \nu} \right]$ and then solves the effective field
equation $\delta \Gamma / \delta g_{\mu \nu} =0$ for $g_{\mu \nu}$, the
expectation value
of the microscopic metric. A sufficiently precise approximate,
let alone exact computation of $\Gamma \left[ g_{\mu \nu} \right]$
is an extremely difficult task, well beyond our present technical
abilities. However, if nonperturbative Quantum Einstein Gravity is
asymptotically safe, and there is growing evidence that this is actually
the case, then the problem is indeed more of a computational rather than
conceptual nature; many of the problems encountered in the traditional
approaches (perturbative nonrenormalizability etc.) would be
overcome then \cite{oliver2}.

Under certain particularly favorable physical conditions there exists a 
``short-cut'' on the way to the expectation value $g_{\mu \nu} (x)$ which
renders the calculation of $\Gamma$ unnecessary: renormalization group
improvement. The idea is to use a truncated form of the flow equation governing
the scale dependence of the effective average action
$\Gamma_{k} \left[ g_{\mu \nu} \right]$ in order to obtain running 
gravitational couplings $G (k)$, $\Lambda (k)$, $\cdots$ which then, in a 
second step, are converted to scalar functions on spacetime by means of
a ``cutoff identification'' $k = k (x)$. The most subtle issue of this approach
is finding such an identification which requires a detailed understanding of 
the relevant physical scales involved \cite{h1}. In ref.\ \cite{h1}
we proposed to use the scalars $G (x)$ and $\Lambda (x)$ thus constructed
for an improvement of the Einstein-Hilbert action (as opposed to its
field equation or solutions thereof).

While in ref.\ \cite{h1} we focused on general aspects of the resulting
Brans-Dicke-type
effective theory, as well as on applications to cosmology, the present paper
is dedicated to spherically symmetric systems. Our investigation of static
isotropic solutions to the effective field equation in Sections 
\ref{3} and \ref{4} has a variety of applications, only some of which we
already touched upon. In particular the methods developed there will be the
basis for a discussion of black holes in the improved action framework,
a topic to which we shall come back elsewhere.

In the present paper we analyzed how gravity gets modified at very large
distances if one assumes that Newton's constant has a nontrivial RG running
in the infrared, i.\,e.\ at astrophysical scales. As far as the pertinent RG
trajectories are concerned this issue is more speculative than the black
hole problem but it can be analyzed more easily since the essential physical
effects are of a quasi Newtonian nature. As an application we explored the
possibility that the flatness of the observed galaxy rotation curves, usually
attributed to the presence of dark matter, is actually due to a RG
running of $G$.

From the point of view of the present analysis, the ``RG trajectory'' $G (k)$
is a given function to be confirmed - or falsified - by future flow
equation studies. The ``IR fixed point scenario'' which we analyzed in Section
\ref{5} was inspired by recent work \cite{cosmo2,elo} which showed that
the $k$-dependence $G (k) \propto 1 / k^{2}$, at cosmological scales, leads
to a phenomenologically viable solution of the ``cosmic coincidence
problem''; it explains the approximate equality of the present matter and
vacuum energy density without a finetuning of parameters. We employed
the model trajectory $G (k) = \overline{G} + g_{*}^{\text{ IR}} / k^{2}$
which interpolates linearly between classical gravity in the laboratory
and the $1/k^{2}$-running at cosmological distances. We constructed exact
vacuum solutions for this $k$-dependence, analyzed their properties in various
regimes, and found that the modified Newtonian regime (realized at intermediate
distances) could perhaps be relevant to the physics of galaxies.
Here Newton's $1/r^{2}$-force is modified by a nonclassical contribution
linearly increasing with $r$. However, it turned out that this force is too
weak to explain the observations.

In Section \ref{6} we maintained the basic idea that Nature interpolates
between $G = const$ in the laboratory and (something similar to)
$G (k) \propto 1 / k^{2}$ in cosmology but we employed a different
interpolation. One could imagine that the anomalous dimension 
$\eta_{\text{N}} \equiv k \, \partial_{k} G / G$ interpolates smoothly between
$\eta_{\text{N}} =0$ and its fixed point value $-2$
with a very small rate of change, at least within some interval of $k$-values.
In particular we assumed, and this is the only assumption relevant to the 
present study, that $\eta_{\text{N}} = -q$ is approximately constant for
$k$ ranging over the 2 or 3 orders of magnitude which have an
impact on the structure of galaxies (corresponding to distances
between $\unit[1]{kpc}$ and $\unit[1]{Mpc}$, say). In this domain
Newton's constant scales as $G (k) \propto k^{-q}$ with a small positive
exponent $q$.

Also for the $k^{-q}$-trajectory we were able to find exact vacuum solutions generalizing
the Schwarzschild metric, and we analyzed their geodesics and the corresponding
rotation curves. We found that at intermediate distances there exists a
quasi Newtonian regime where Newton's $1/r$-potential is supplemented
by a term $q \, \ln r$ which is well known to produce a constant $v (r)$.
While this regime might indeed be relevant to the physics of galaxies, the full
nonlinear vacuum solutions are interesting also in their own right. They
are based upon a background independent, selfconsistent cutoff identification
$k \propto 1 / d_{g} (r)$ where, contrary to earlier approaches, the
proper distance $d_{g} (r)$ is computed from the solution $g_{\mu \nu}$ rather
than from a fixed background metric; as a result, the position dependence
$G = G (r)$ originating from a given $G = G (k)$ is \textit{computed}
rather than put in by hand. A particularly remarkable vacuum solution obtains 
by applying the solution-generating Weyl transformation to Minkowski space.
It represents a kind of gravitational ``soliton'' which can exist only thanks
to the running of $G$.

The ``phenomenologically'' most promising portion of the spacetimes constructed turned
out to be in the weak field regime. This motivated us, in Section 
\ref{7}, to investigate the Newtonian limit of the modified Einstein equation
in general, rather than of special solutions,
with matter included and without insisting on
spherical symmetry. One of the key results was that any arbitrary
$G (r)$ which increases with distance always gives rise to an
attractive nonclassical force which adds to Newton's
$1/r^{2}$-contribution. If $G = \overline{G} \, \bigl[ 1 
+ \mathcal{N} \bigr]$, a term $+ \mathcal{N}/2$ gets added to the
classical potential.

The upshot of all these investigations is that it is very
well conceivable that the approximately flat galaxy rotation curves one
observes in Nature are actually a renormalization effect of Newton's constant
rather than due to the presence of hitherto undetected dark matter.
If so, the $k$-dependence of $G (k)$ in the range of typical galaxy scales
is governed by a power law most probably. In Subsections \ref{s6.5} and
\ref{s7.7} we described in detail its consequences for the physics in
galactic haloes and the origin of the constant rotation curves in particular.
We also discussed various limitations of the present calculational scheme;
for instance, it will be difficult to go beyond spherical symmetry or
to describe higher order effects such as the Tully-Fisher relation.

Now that we know which type of RG trajectory we are looking for it will be
exciting to see whether future RG studies, employing improved truncations,
can confirm this picture. In a sense, it would replace dark matter with
the vacuum fluctuations which drive the RG evolution of the gravitational
parameters.
\vspace*{3\baselinestretch pt}
%
%
%
%
%
%
\appendix
\begin{flushleft}
{\Large \textbf{Appendix}}
\end{flushleft}
\section{\label{A}Linearized static isotropic field equations}
In this appendix we analyze the coupled system of equations \eqref{3.set}
in the modified Newtonian limit \cite{weinbook}. For the metric we make the
ansatz
\begin{align}
\text{d} s^{2} & = 
- \bigl[ 1 + 2 \, \phi (r) \bigr] \, \text{d} t^{2}
+ \bigl[ 1 - 2 \, \chi (r) \bigr] \, \text{d} r^{2}
+ r^{2} \, \text{d} \sigma^{2},
\label{A.1}
\end{align}
which amounts to
\begin{align}
A (r) = 1 - 2 \, \chi (r), 
\quad
B (r) = 1 + 2 \, \phi (r).
\label{A.2}
\end{align}
Deviations from flat Minkowski space are assumed tiny, and the functions
$\phi$ and $\chi$ are considered small compared to unity therefore. In the
following calculations we retain terms linear in $\phi$, $\chi$, and their
derivatives, but discard all higher powers or cross products of them. As a
result, Einstein's equation \eqref{3.10} consists of the linearized
$tt$-component
\begin{subequations} \label{A.3}
\begin{align}
\frac{\chi}{r^{2}} + \frac{\chi^{\prime}}{r} & =
- 4 \pi \, \overline{G} \, \rho^{\text{tot}},
\label{A.3a}
\end{align}
the $rr$-component
\begin{align}
\frac{\chi}{r^{2}} + \frac{\phi^{\prime}}{r} & =
4 \pi \, \overline{G} \, p^{\text{tot}}_{r},
\label{A.3b}
\end{align}
and the $\theta \theta / \varphi \varphi$-component
\begin{align}
\frac{\chi^{\prime}}{r} + \frac{\phi^{\prime}}{r} + \phi^{\prime \prime}
& =  8 \pi \, \overline{G} \, p^{\text{tot}}_{\perp}.
\label{A.3c}
\end{align}
\end{subequations}
The consistency condition assumes the form
\begin{align}
\tfrac{\text{d}}{\text{d}r} \, p^{\text{tot}}_{r}
+ \phi^{\prime} \, \Bigl( \rho^{\text{tot}} + p^{\text{tot}}_{r} \Bigr)
+ \frac{2}{r} \, \Bigl( p^{\text{tot}}_{r} - p^{\text{tot}}_{\perp} 
\Bigr) =0,
\label{A.4}
\end{align}
and the linearized form of the ordinary continuity equation
$D^{\mu} T_{\mu \nu} =0$ reads
\begin{align}
p^{\prime} + \left( \rho + p \right) \, \phi^{\prime} =0.
\label{A.5}
\end{align}

Next we must linearize the components of $\Delta T_{\mu}^{~\nu}$.
One finds
\begin{subequations} \label{A.6}
\begin{align}
\begin{split}
\Delta \rho & =
\frac{1}{8 \pi G} \, \left[ 2 \, \left( \frac{G^{\prime}}{G} \right)^{2}
- \frac{G^{\prime \prime}}{G} 
- \frac{2}{r} \, \left( \frac{G^{\prime}}{G} \right) \right]
\\ & \phantom{{==}}
+ \frac{1}{4 \pi G} \, \chi \, 
\left[ 2 \, \left( \frac{G^{\prime}}{G} \right)^{2}
- \frac{G^{\prime \prime}}{G} 
- \frac{2}{r} \, \left( \frac{G^{\prime}}{G} \right) \right]
- \frac{1}{8 \pi G} \, \chi^{\prime} \, \left( \frac{G^{\prime}}{G} \right)
+ \cdots,
\end{split}
\label{A.6a}
\\
\Delta p_{r} & = \frac{1}{4 \pi G} \, \frac{1}{r} \, 
 \left( \frac{G^{\prime}}{G} \right)
+ \frac{1}{2 \pi G} \, \frac{\chi}{r} \,  \left( \frac{G^{\prime}}{G} \right)
+ \frac{1}{8 \pi G} \, \phi^{\prime} \,  \left( \frac{G^{\prime}}{G}
\right) + \cdots,
\label{A.6b}
\\
\begin{split}
\Delta p_{\perp} & = 
\frac{1}{8 \pi G} \, \left[ - 2 \, \left( \frac{G^{\prime}}{G} \right)^{2}
+ \frac{G^{\prime \prime}}{G} 
+ \frac{1}{r} \, \left( \frac{G^{\prime}}{G} \right) \right]
\\ & \phantom{{==}}
+ \frac{1}{4 \pi G} \, \chi \, 
\left[ - 2 \, \left( \frac{G^{\prime}}{G} \right)^{2}
+ \frac{G^{\prime \prime}}{G} 
+ \frac{1}{r} \, \left( \frac{G^{\prime}}{G} \right) \right]
\\ & \phantom{{===}}
+ \frac{1}{8 \pi G} \, \chi^{\prime} \,  \left( \frac{G^{\prime}}{G} \right)
+ \frac{1}{8 \pi G} \, \phi^{\prime} \,  \left( \frac{G^{\prime}}{G} \right)
+ \cdots.
\end{split}
\label{A.6c}
\end{align}
\end{subequations}
From the expansion of
$\theta_{\mu}^{~\nu} = \varepsilon {\theta^{\text{ BD}}}_{\mu}^{~\nu}$ one
finds analogously:
\begin{subequations} \label{A.7}
\begin{align}
\rho_{\theta} & = 
- \frac{3 \, \varepsilon}{32 \pi G} \, \left( \frac{G^{\prime}}{G} \right)^{2}
- \frac{3 \, \varepsilon}{16 \pi G} \, \chi \, \left( \frac{G^{\prime}}{G} \right)^{2}
+ \cdots, 
\label{A.7a}
\\
p_{\theta,r} & =
- \frac{3 \, \varepsilon}{32 \pi G} \, \left( \frac{G^{\prime}}{G} \right)^{2}
- \frac{3 \, \varepsilon}{16 \pi G} \, \chi \, \left( \frac{G^{\prime}}{G} \right)^{2}
+ \cdots, 
\label{A.7b}
\\
p_{\theta,\perp} & =
\frac{3 \, \varepsilon}{32 \pi G} \, \left( \frac{G^{\prime}}{G} \right)^{2}
+ \frac{3 \, \varepsilon}{16 \pi G} \, \chi \, \left( \frac{G^{\prime}}{G} \right)^{2}
+ \cdots.
\label{A.7c}
\end{align}
\end{subequations}
Inserting the above leading order approximations into Einstein's equation
\eqref{A.3} we obtain the $tt$-component
\begin{subequations} \label{A.8}
\begin{align}
\begin{split}
& 2 \, \chi \, \left[ 
\frac{1}{r^{2}} 
- \frac{2}{r} \, \left( \frac{G^{\prime}}{G} \right)
- \frac{G^{\prime \prime}}{G}
+ \frac{1}{4} \, \left( 8 - 3 \, \varepsilon \right) \, 
\left( \frac{G^{\prime}}{G} \right)^{2} \right]
+ \chi^{\prime} \, \left[
\frac{2}{r} - \frac{G^{\prime}}{G} \right]
\\ & \phantom{{=}} = 
- \Lambda - 8 \pi G \, \rho
+ \frac{2}{r} \, \left( \frac{G^{\prime}}{G} \right)
+ \frac{G^{\prime \prime}}{G}
- \frac{1}{4} \, \left( 8 - 3 \, \varepsilon \right) \, 
\left( \frac{G^{\prime}}{G} \right)^{2},
\end{split}
\label{A.8a}
\end{align}
the $rr$-component
\begin{align}
\begin{split}
& 2 \, \chi \, \left[ 
\frac{1}{r^{2}} 
- \frac{2}{r} \, \left( \frac{G^{\prime}}{G} \right)
+ \frac{3 \, \varepsilon}{4} \left( \frac{G^{\prime}}{G} \right)^{2} \right]
+ \phi^{\prime} \, \left[ 
\frac{2}{r} - \frac{G^{\prime}}{G} \right]
\\ & \phantom{{=}} =
- \Lambda + 8 \pi G \, p
+ \frac{2}{r} \, \left( \frac{G^{\prime}}{G} \right)
- \varepsilon \, \frac{3}{4} \left( \frac{G^{\prime}}{G} \right)^{2},
\end{split}
\label{A.8b}
\end{align}
and the $\theta \theta / \varphi \varphi$-component
\begin{align}
\begin{split}
& 2 \, \chi \, \left[ 
- \frac{1}{r} \, \left( \frac{G^{\prime}}{G} \right)
- \frac{G^{\prime \prime}}{G}
+ \frac{1}{4} \, \left( 8 - 3 \, \varepsilon \right) \, 
\left( \frac{G^{\prime}}{G} \right)^{2} \right]
\\ & \phantom{{==}}
+ \chi^{\prime} \, \left[ 
\frac{1}{r} - \frac{G^{\prime}}{G} \right]
+ \phi^{\prime} \, \left[ 
\frac{1}{r} - \frac{G^{\prime}}{G} \right]
+ \phi^{\prime \prime}
\\ & \phantom{{=}} =
- \Lambda + 8 \pi G \, p
+ \frac{1}{r} \, \left( \frac{G^{\prime}}{G} \right)
+ \frac{G^{\prime \prime}}{G} 
- \frac{1}{4} \, \left( 8 - 3 \, \varepsilon \right) \, 
\left( \frac{G^{\prime}}{G} \right)^{2}.
\end{split}
\label{A.8c}
\end{align}
\end{subequations}
For the consistency condition we find in the same fashion:
\begin{align}
\begin{split}
& 2 \, \chi \, \left[
\frac{1}{r} \, \left( 2 - 3 \, \varepsilon \right) 
\left( \frac{G^{\prime}}{G} \right)^{2}
- \frac{3 \, \varepsilon}{2} 
\, \left( \frac{G^{\prime} G^{\prime \prime}}{G^{2}} \right)
+ \varepsilon \, \frac{3}{2} \, \left( \frac{G^{\prime}}{G} \right)^{3}
\right]
\\ & \phantom{{=}}
+ \chi^{\prime} \, \left[ 
\frac{2}{r}
- \frac{3 \, \varepsilon}{2} \, \left( \frac{G^{\prime}}{G} \right)
\right] \, \left( \frac{G^{\prime}}{G} \right)
+ \frac{\phi^{\prime}}{2} \, \left(
2 - 3 \, \varepsilon \right) \, \left( \frac{G^{\prime}}{G} \right)^{2}
- \phi^{\prime \prime} \, \left( \frac{G^{\prime}}{G} \right)
\\ & \phantom{{==}}
+ \left[
\frac{1}{r} \, \left( 2 - 3 \, \varepsilon \right) \, 
\left( \frac{G^{\prime}}{G} \right)
- \frac{3 \, \varepsilon}{2} 
\, \left( \frac{G^{\prime \prime}}{G} \right)
+ \frac{3 \, \varepsilon}{2} \, \left( \frac{G^{\prime}}{G} \right)^{2}
\right] \,\left( \frac{G^{\prime}}{G} \right) 
\\ & \phantom{{===}}
- \Lambda^{\prime} + 8 \pi \, G^{\prime} \, p
=0.
\end{split}
\label{A.9}
\end{align}

For $\phi$ and $\chi$ to be small it is necessary that the sources
of the gravitational field are weak. Likewise we are going to assume that the
variability of $G (r)$ is small since it is obvious from the above equations
that also the position dependence of $G$ acts as a source. We use the
parameterization
\begin{align}
G (r) = \overline{G} \, \bigl[ 1 + \mathcal{N} (r) \bigr] 
\label{A.10}
\end{align}
and assume that $\mathcal{N} \ll 1$ is of the same order as $\phi$ and $\chi$.
We retain terms linear in $\phi$, $\chi$, $\mathcal{N}$, and their
derivatives, but neglect higher powers and products. Hence, to
leading order, $G^{\prime} / G = \mathcal{N}^{\prime} + \cdots$,
$\left( G^{\prime} / G \right)^{2} = 0 + \cdots$,
$G^{\prime \prime} / G = \mathcal{N}^{\prime \prime} + \cdots$,
and $G^{\prime} G^{\prime \prime} / G^{2} = 0 + \cdots$.
Furthermore, we neglect the cosmological constant, $\Lambda (r) \equiv 0$,
because it is not important in the present context. As a result, the
$tt$-component \eqref{A.8a} boils down to
\begin{subequations} \label{A.11}
\begin{gather}
\frac{\chi}{r^{2}\,}  + \frac{\chi^{\prime}}{r} = 
- 4 \pi \, G (r) \, \rho
+ \frac{{\mathcal N}^{\prime}}{r}
+ \frac{{\mathcal N}^{\prime \prime}}{2}
\label{A.11a}
\end{gather}
and the $rr$-component \eqref{A.8b} becomes
\begin{gather}
\frac{\chi}{r^{2}\,} + \frac{\phi^{\prime}}{r} = 
4 \pi \, G (r) \, p
+ \frac{{\mathcal N}^{\prime}}{r}.
\label{A.11b}
\end{gather}
\end{subequations}

Rather than the $\theta\theta/\varphi\varphi$-component, we consider the
consistency condition along with the continuity equation the independent
equation. Linearizing the consistency condition \eqref{A.9} yields
the statement that, up to higher orders,
$8 \pi \, \overline{G} \, \mathcal{N}^{\prime} \, p =0$.
Since, by assumption, $\mathcal{N}^{\prime} \neq 0$, we obtain the condition
$p=0$, up to higher orders. This shows that metrics of the type \eqref{A.1}
can describe the modified Newtonian limit at most for pressureless
matter (``dust'').

The last equation we must satisfy is the continuity equation \eqref{A.5}.
It does not contain $\mathcal{N}$, and coincides with the one appearing in the
standard Newtonian limit therefore. For $p=0$ it says that 
$\rho \, \phi^{\prime} =0$ up to terms $\propto \phi^{2}, \chi^{2},
\phi \, \chi$. As in the standard case, we consider the product
$\rho \, \phi^{\prime}$ negligible, formally of order $\phi^{2}$, so that
\eqref{A.5} is satisfied with sufficient accuracy. This is the Newtonian
approximation of ``quasistatic dust'' \cite{rindler}.

It remains to solve \eqref{A.11a} coupled to \eqref{A.11b} with $p=0$.
The most elegant method is to make an ansatz
\begin{subequations} \label{A.13}
\begin{align}
\phi (r) & = \widehat \phi (r) + \tfrac{1}{2} \, \mathcal{N} (r)
\label{A.13a}
\\
\chi (r) & = \widehat \chi (r) + \tfrac{1}{2} \, r \, 
\mathcal{N}^{\prime} (r)
\label{A.13b}
\end{align}
\end{subequations}
with two new functions, $\widehat \phi$ and $\widehat \chi$,
yet to be determined. Inserting this ansatz into \eqref{A.11a} and
\eqref{A.11b} for $p=0$ yields, respectively,
\begin{subequations} \label{A.14}
\begin{gather}
\frac{\widehat \chi}{r^{2}\,}  + \frac{\widehat \chi^{\prime}}{r} = 
- 4 \pi \, G (r) \, \rho,
\label{A.14a}
\\
\frac{\widehat \chi}{r^{2}\,} + \frac{\widehat \phi^{\prime}}{r} = 
0.
\label{A.14b}
\end{gather}
\end{subequations}
The equations (\ref{A.14}\,a,b) have a strikingly simple interpretation.
They are nothing but the $tt$-and the $rr$-component, respectively, of the
field equation
\begin{align}
G_{\mu \nu} \bigl( \widehat g \bigr) =
8 \pi \, G (r) \, T_{\mu \nu}
\label{A.15}
\end{align}
evaluated for the following metric analogous to \eqref{A.1}:
\begin{align}
\text{d} \widehat s^{\,2} 
& \equiv \widehat g_{\mu \nu} \, \text{d} x^{\mu} \text{d} x^{\nu} =
- \bigl[ 1 + 2 \, \widehat \phi (r) \bigr] \, \text{d} t^{2}
+ \bigl[ 1 - 2 \, \widehat \chi (r) \bigr] \, \text{d} r^{2}
+ r^{2} \, \text{d} \sigma^{2},
\label{A.16}
\end{align}
Obviously \eqref{A.15} is precisely the Einstein equation the 
\textit{improved equation} approach is based upon: it obtains from the
classical $G_{\mu \nu} = 8 \pi \, \overline{G} \, T_{\mu \nu}$
by the substitution $\overline{G} \to G (r)$. While $\phi$ and $\chi$
determine the metric in the \textit{improved action} setting,
$\widehat \phi$ and $\widehat \chi$ play the analogous role when
the field equation is improved.

The system (\ref{A.14}\,a,b) is easily solved. \eqref{A.14b} tells us that
$\widehat \chi$ is known once we know $\widehat \phi$:
$\widehat \chi = - r \, \widehat \phi^{\prime}$.
Inserting this into \eqref{A.14a} yields the one-dimensional Poisson equation
\begin{align}
\nabla^{2}_{\text{rad}} \widehat \phi
= 4 \pi \, G (r) \, \rho (r),
\label{A.18}
\end{align}
where
$\nabla^{2}_{\text{rad}} \equiv \frac{\text{d}^{2}}{\text{d} r^{2}}
+ \frac{2}{r} \, \frac{\text{d}}{\text{d} r}
= \frac{1}{r} \, \frac{\text{d}^{2}}{\text{d} r^{2}} \, r$
is the radial part of the Laplacian on flat space.

Let us summarize.
If the matter is quasistatic dust of density $\rho$, the modified Newtonian
limit implied by the improved action functional is given by the metric
\eqref{A.1} with
\begin{subequations} \label{A.20}
\begin{align}
\phi (r) & = \widehat \phi (r) + \tfrac{1}{2} \, \mathcal{N} (r)
\label{A.20a}
\\
\chi (r) & = - r \, \widehat \phi^{\prime} (r) + \tfrac{1}{2} \, r \, 
\mathcal{N}^{\prime} (r)
\label{A.20b}
\end{align}
\end{subequations}
where $\mathcal{N} (r)$ parameterizes the variation of Newton's constant
via \eqref{A.10}, and $\widehat \phi$ is to be obtained by solving
the ``improved Poisson equation'' \eqref{A.18}. Note that the final result
does not depend on $\varepsilon$, i.\,e.\ on the $\theta$-tensor chosen.

The possibility of generating solutions of the modified field equations
from solutions of a much simpler equation, here the improved Poisson equation,
is reminiscent of the Weyl transformation technique. However, the relations
\eqref{A.20} just discovered apply to matter with
$T_{\mu}^{~\nu} = \text{diag} [-\rho,0,0,0]$. Hence $T=-\rho \neq 0$, so
that the solutions are not in Class II or IIIa where the Weyl method
is available.

It is instructive to study the special case of vacuum solutions
in the classical limit. The solution to Poisson's equation,
for $r \neq 0$, is Newton's potential then:
\begin{align}
\widehat \phi (r) = - \frac{\overline{G} \, M}{r} \equiv \phi_{\text{N}}
(r)
\label{A.21}
\end{align}
In this case $- r \, \widehat \phi^{\prime} = \widehat \phi
= \phi_{\text{N}}$ so that
$\chi = \phi_{\text{N}} + \tfrac{1}{2} \, r \, \mathcal{N}^{\prime}$ and the
metric assumes the form
\begin{align}
\text{d} s^{2} 
& =
- \bigl[ 1 + 2 \, \phi_{\text{N}} (r) + \mathcal{N} (r) \bigr] 
\, \text{d} t^{2}
+ \bigl[ 1 - 2 \, \phi_{\text{N}} (r) - r \, \mathcal{N}^{\prime} (r) \bigr] 
\, \text{d} r^{2}
+ r^{2} \, \text{d} \sigma^{2}.
\label{A.22}
\end{align}

The special case \eqref{A.22} does indeed belong to Class II, and it can be
obtained by a Weyl transformation. In the Newtonian limit
$g_{\mu \nu}$ can be read off straightforwardly from
$\text{d} s^{2} = \left[ G (\rho) / \, \overline{G} \, \right] \,
\text{d} s^{2}_{\gamma}$ because the difference between a selfconsistent
cutoff based upon $d_{g} (\rho)$ and the naive one, employing the
Minkowskian $d_{\gamma} (\rho) = \rho$, is of higher order and negligible
therefore. Hence $G (\rho) = G (k = \xi / \rho) \equiv
\overline{G} \, \bigl[ 1 + \mathcal{N} (\rho) \bigr]$ is independent of
$g_{\mu \nu}$ so that the explicit solution reads, in the $\rho$-coordinate
system,
\begin{align}
\text{d} s^{2} = \bigl[ 1 + \mathcal{N} (\rho) \bigr] \,
\Bigl( - f (\rho) \, \text{d} t^{2}
+ f (\rho)^{-1} \, \text{d} \rho^{2}
+ \rho^{2} \, \text{d} \sigma^{2} \Bigr)
\label{A.23}
\end{align}
with $f (\rho) = 1 + 2 \, \phi_{\text{N}} (\rho)$.
Retaining only terms
linear in $\phi_{\text{N}}$ and $\mathcal{N}$, this line element can easily
be cast into the $AB$-form. Upon defining the new radial coordinate
\begin{align}
r = \rho \, \sqrt{1+ \mathcal{N}\,} =
\rho \, \bigl[ 1 + \tfrac{1}{2} \, \mathcal{N} +
\mathcal{O} \left( \mathcal{N}^{2} \right) \bigr]
\label{A.24}
\end{align}
one finds precisely eq.\ \eqref{A.22} as the final result. It amounts to the
solution 
\begin{subequations} \label{A.25}
\begin{align}
A (r) & = 1 + \frac{2 \, \overline{G} \, M}{r} - r \, \mathcal{N}^{\prime} (r),
\label{A.25a}
\\
B (r) & = 1 - \frac{2 \, \overline{G} \, M}{r} + \mathcal{N} (r),
\label{A.25b}
\\
G (r) & = \overline{G} \, \bigl[ 1 + \mathcal{N} (r) \bigr],
\label{A.25c}
\end{align}
\end{subequations}
which is valid for an arbitrary (but small) function $\mathcal{N} (r)$.
%
%
%
%
%
%
%
%
\section{\label{B}Geodesics}
In this appendix we study the trajectories of test particles in spacetimes
with a metric of the type
\begin{align}
\text{d} s^{2} =
w (\rho) \, \biggl[ 
- f (\rho) \text{d} t^{2}
+ f (\rho)^{-1} \, \text{d} \rho^{2}
+ \rho^{2} \, \Bigl( \text{d} \theta^{2}
+ \sin^{2} \theta \, \text{d} \varphi^{2} \Bigr)
\biggr]
\label{B.1}
\end{align}
where $w (\rho)$ and $f (\rho)$ are arbitrary functions for the time being.
Geodesics $x^{\mu} (\lambda)$, parametrized by the parameter $\lambda$, can
be obtained from the Lagrangian
\begin{align}
L = - w (\rho) \, f (\rho) \, \dot t^{2}
+ w (\rho) \, f (\rho)^{-1} \, \dot \rho^{2}
+ w (\rho) \, \rho^{2} \,
\Bigl( \dot \theta^{2} + \sin^{2} \theta \, \dot \varphi^{2} \Bigr).
\label{B.2}
\end{align}
In this appendix, a dot denotes the derivative with respect to $\lambda$,
and a prime a derivative with respect to $\rho$. The Euler-Lagrange equations
implied by \eqref{B.2} for $t$, $\varphi$, $\rho$, and $\theta$, respectively,
read
\begin{subequations} \label{B.3}
\begin{gather}
w \, f \, \dot t = const \equiv \alpha,
\label{B.3a}
\\
w \, \rho^{2} \, \sin^{2} \theta \, \dot \varphi = const \equiv \beta,
\label{B.3b}
\\
2 \, \frac{w}{f} \, \ddot \rho + \left( \frac{w}{f} \right)^{\prime} \,
\dot \rho^{2} = - \left( w \, f \right)^{\prime} \, \dot t^{2}
+ \Bigl( w \, \rho^{2} \Bigr)^{\prime} \,
\Bigl( \dot \theta^{2} + \sin^{2} \theta \, \dot \varphi^{2} \Bigr),
\label{B.3c}
\\
w \, \rho^{2} \, \ddot \theta + \Bigl( w \, \rho^{2} \Bigr)^{\prime} \,
\dot \theta^{2} = 
w \, \rho^{2} \, \sin \theta \, \cos \theta \, \dot \varphi^{2}.
\label{B.3d}
\end{gather}
\end{subequations}

Without loss of generality we may assume that the motion takes place in the
plane $\theta = \pi / 2$. Eq.\ \eqref{B.3d} is satisfied identically by the
ansatz $\theta (\lambda) = \pi / 2 = const$, and the remaining equations are
\begin{subequations} \label{B.4}
\begin{gather}
\dot t = \alpha / \left( w \, f \right),
\label{B.4a}
\\
\dot \varphi = \beta / \Bigl( w \, \rho^{2} \Bigr),
\label{B.4b}
\\
2 \, \frac{w}{f} \, \ddot \rho
+ \left( \frac{w}{f} \right)^{\prime} \, \dot \rho^{2} =
- \left( w \, f \right)^{\prime} \, \dot t^{2}
+ \Bigl( w \, \rho^{2} \Bigr)^{\prime} \, \dot \varphi^{2}.
\label{B.4c}
\end{gather}
\end{subequations}

We are not going to embark on a general analysis of the equations
\eqref{B.4} here. Instead we focus on the question whether or not the
spacetime admits, for massive test particles, closed \textit{circular}
orbits at $\rho = const$. If they exist, $w \, f$ and $w \, \rho^{2}$ are
constant for a given orbit, and the equations \eqref{B.4a} and \eqref{B.4b}
simply tell us that $t (\lambda) \propto \lambda$ and 
$\varphi (\lambda) \propto \lambda$. Inserting \eqref{B.4a} and \eqref{B.4b}
into \eqref{B.4c} with $\dot \rho = \ddot \rho =0$ we see that circular
orbits do exist if constants $\alpha$ and $\beta$ can be found
such that
\begin{align}
\frac{\Bigl( w \, \rho^{2} \Bigr)^{\prime}}{\rho^{4}}
- \left( \frac{\alpha}{\beta} \right)^{2} \,
\frac{\left( w \, f \right)^{\prime}}{f^{2}} =0.
\label{B.5}
\end{align}
As expected, there are no solutions in Minkowski space ($w=f=1$), but there
exist circular orbits in the Schwarzschild geometry
($w=1$, $f (\rho) = 1 - 2 \overline{G} M / \rho$), say.
If one specializes for the setting of Section \ref{4}, $f (\rho)$ is fixed to be
of the form \eqref{4.3b} so that it is the Weyl factor $w (\rho)$,
related to the $\rho$-dependence of $G$, which then decides about the existence
of circular orbits.

The angular velocity referring to $\lambda$, $\dot \theta$, as well as
the angular velocity defined in terms of the coordinate time $t$,
$\text{d} \varphi / \text{d} t = \dot \varphi / \dot t$, are constant for
a given circular orbit, but in general they will depend on $\rho$, of
course. From \eqref{B.4c} we obtain
\begin{align}
\left( \frac{\text{d} \varphi}{\text{d} t} \right)^{2} =
\frac{\left( w \, f \right)^{\prime}}{\Bigl( w \, \rho^{2} \Bigr)^{\prime}}
= \frac{w^{\prime} \, f + w \, f^{\prime}}{w^{\prime} \, \rho^{2}
+ 2 \, \rho \, w}.
\label{B.6} 
\end{align}
This quantity is indeed positive precisely if \eqref{B.5} is satisfied.
It determines the angular velocity for which the centrifugal force balances
the gravitational attraction. The period of an orbit at coordinate radius
$\rho$, measured in units of the coordinate time $t$, is given by
$T (\rho) = 2 \pi / \left( \text{d} \varphi / \text{d} t \right)$, or
\begin{align}
T (\rho) = 2 \pi \, \rho \,
\left[ \frac{\frac{2}{\rho} + \frac{w^{\prime}}{w}}
{f^{\prime} + \frac{w^{\prime}}{w} \,f} \right]^{1/2}.
\label{B.7}
\end{align}
In particular for $f (\rho)$ given by \eqref{4.3b},
\begin{align}
T (\rho) = 2 \pi \, \rho \, 
\left[ \frac{1 + Y (\rho)}{\left( \, \overline{G} \, M / \rho \right)
+ f (\rho) \, Y (\rho)} \right]^{1/2},
\label{B.8}
\end{align}
with the abbreviation
\begin{align}
Y (\rho) \equiv \frac{\rho}{2} \, \frac{w^{\prime} (\rho)}{w (\rho)}.
\label{B.9}
\end{align}
In the classical case, $Y=0$, and we recover Kepler's law
$T^{2} \propto \rho^{3}$.
%
%
%
%
%
%
%

\end{document}